\documentclass[onecolumn]{aastex631}

\newcommand{\K}{\ensuremath{\mathrm{K}}}
\newcommand{\gcc}{\ensuremath{\mathrm{g}\,\mathrm{cm}^{-3}}}
\newcommand{\cms}{\ensuremath{\mathrm{cm}\,\mathrm{s}^{-1}}}

\newcommand{\ergs}{\ensuremath{\mathrm{erg}\,\mathrm{s}^{-1}}}

\newcommand{\Ms}{~{\ensuremath{M_{\odot}}}}

\newcommand{\cutt}[1]{\textcolor{blue}{}}

\newcommand{\code}[1]{{\tt{#1}}}

\newcommand{\N}{{\ensuremath{^{14} \mathrm{N}}} }

\newcommand{\FIGFF}[2]{{\ref{fig:#2}{#1}}}

\newcommand{\FIG}[2]{{Figure~\FIGFF{#1}{#2}}}
\newcommand{\Fig}[1]{{\FIG{}{#1}}}

\newcommand{\Sectff}[1]{{\ref{sec:#1}}}
\newcommand{\Sect}[1]{{\S~\Sectff{#1}}}

\newcommand{\eq}[2]{\begin{equation} \label{eq:#1} #2 \end{equation}}

\newcommand{\CASTRO}{\texttt{CASTRO}}
\newcommand{\OPAL}{\texttt{OPAL}}

\newcommand{\MESA}{\texttt{MESA}}
\newcommand{\KEPLER}{\texttt{KEPLER}}

\renewcommand{\fig}[4]{\begin{figure*}\centering\includegraphics[width=#4\textwidth]{#2}\caption{#3}\label{fig:#1}\end{figure*}}
\newcommand{\sfig}[4]{\begin{figure}\centering\includegraphics[width=#4\textwidth]{#2}\caption{#3}\label{fig:#1}\end{figure}}

\DeclareRobustCommand{\dl}{\bgroup\markoverwith{\textcolor{red}{\rule[.5ex]{2pt}{0.4pt}}}\ULon}
\usepackage{amsmath}

\begin{document}

\title{Multidimensional Radiation Hydrodynamics Simulations of Supernova 1987A Shock Breakout}


\author[0009-0002-3816-4732]{Wun-Yi Chen}
\affiliation{Institute of Astronomy and Astrophysics, Academia Sinica, Taipei 10617, Taiwan}

\affiliation{National Taiwan University, Graduate Institute of Astrophysics \\
Department of Physics/Center for Condensed Building R401, No.1, Sec. 4, Roosevelt Rd, \\
Taipei 10617, Taiwan}


\author[0000-0002-4848-5508]{Ke-Jung Chen}
\affiliation{Institute of Astronomy and Astrophysics, Academia Sinica, Taipei 10617, Taiwan}

\author[0000-0002-0603-918X]{Masaomi Ono}
\affiliation{Institute of Astronomy and Astrophysics, Academia Sinica, Taipei 10617, Taiwan}
\affil{Astrophysical Big Bang Laboratory (ABBL), RIKEN Cluster for Pioneering Research, \\
2-1 Hirosawa, Wako, Saitama 351-0198, Japan}

\begin{abstract}

Shock breakout is the first electromagnetic signal from supernovae (SNe), which contains important information on the explosion energy and the size and chemical composition of the progenitor star. This paper presents the first two-dimensional (2D) multi-wavelength radiation hydrodynamics simulations of SN 1987A shock breakout by using the \CASTRO\ code with the opacity table, \OPAL\, considering eight photon groups from infrared to X-ray. To investigate the impact of the pre-supernova environment of SN 1987A, we consider three possible circumstellar medium (CSM) environments: a steady wind, an eruptive mass loss, and the existence of a companion star. In sum, the resulting breakout light curve has an hour duration and its peak luminosity of $\sim 4\times 10^{46}\,\rm{erg\,s^{-1}}$ then following a decay rate of $\sim 3.5\,\rm{mag\,hour^{-1}}$ in X-ray. The dominant band transits to UV around 3 hours after the initial breakout, and its luminosity has a decay rate of $\sim 1.5\,\rm{mag\,hour^{-1}}$ that agrees well with the observed shock breakout tail. The detailed features of breakout emission are sensitive to the pre-explosion environment. Furthermore, our 2D simulations demonstrate the importance of multidimensional mixing and its impacts on shock dynamics and radiation emission. The mixing emerging from the shock breakout may lead to a global asymmetry of SN ejecta and affect its later supernova remnant formation.
\end{abstract}


\section{Introduction} \label{sec:intro}

The electromagnetic signals from supernovae (SNe) of massive stars begin when the explosion shock breaks out of the stellar surface. This is the so-called {\it shock breakout}, a luminous phenomenon that has only been detected through a few serendipitous observations \citep{doi:10.1126/science.1160456, 2015ApJ...804...28G}. The peak luminosity and duration of breakout emission can be used to probe the explosion energy, the radius of a progenitor star, and its circumstellar environment \citep{2017hsn..book..967W}.

Previous theoretical studies of shock breakout of massive stars usually assumed an adiabatic shock propagating in a spherical stellar envelope with a power-law density profile \citep{1999ApJ...510..379M, 2017hsn..book..967W}. However, modeling the realistic SN shock breakout requires a deeper understanding of the shock propagation in a clumpy stellar atmosphere and the circumstellar medium (CSM) formed before the star explodes \citep{2011ApJ...729L...6C, rsgwind}, and the emission from the shock typically ranges from hard X-ray to UV and infrared wavelengths \citep{10.1093/mnras/sts577, Katz_2010}.

Previous one-dimensional (1D) shock breakout simulations \citep{2017ApJ...845..103L} have provided insights into the physics of SN shock breakout. However, these simulations have limitations, as they fail to accurately model fluid instabilities, resulting in nonphysical features such as a distinct density spike (thin shell). Limited fluid instabilities in 1D simulations often concentrate a significant amount of shocked ejecta into a small region, moving at the same velocity, and lead to the formation of a nonphysical thin shell \citep[see, for example, discussion in][]{Ken2023ApJ}. This thin shell, where most of the electromagnetic emission originates, may underestimate the peak luminosity and the duration of the breakout emission. Furthermore, 2D hydrodynamics shock breakouts by \cite{2011ApJ...727..104C} revealed a longer light curve (LC) duration due to more efficient mixing in 2D. The complex interplay of radiative shock among the structured stellar atmosphere and CSM can only be properly modeled through the multidimensional radiation hydrodynamics (RHD) \citep{LEVINSON20201}, which has been utilized to model the transport of neutrinos and photons in supernovae \citep{Ott_2008, Utrobin2021ApJ, Ken2023ApJ}.

2D frequency-integrated RHD simulations of shock breakout \citep{suzuki2016, suzuki2021} have revealed the emergence of bipolar explosion shocks on the stellar surface as a possible source of long-lived bright emission. 3D RHD simulations on red supergiant (RSG) envelopes have also shown density fluctuations in the stellar surface may extend the LC duration \citep{2022ApJJiang}.

However, these simulations, only containing a single energy group of photons known as {\it grey} approximation, fail to capture the shock heating and cooling processes that generate X-ray and UV emissions accurately. The corresponding grey opacity falls short in producing a cooling shell during the free expansion phase of the ejecta. To investigate the shock breakout, one has to follow the propagation of the forward shock into regions from optically thick to optically thin regions and follow the resulting fluid instabilities. 
Therefore, 2D RHD simulations, combined with multi-group radiation transport spanning hard X-rays to near-infrared radiation, are required to properly model shock breakout. This paper presents the first results of multidimensional multi-group RHD simulations of SN shock breakout and provides more realistic observational signatures. 

To conduct simulations incorporating CSM alterations and facilitating comparisons with 1D calculations, we employ 2D Multi-Group Flux Limited Diffusion (MGFLD) in {\CASTRO} \citep{ 2010ApJ...715.1221A,Almgren2020}, utilizing the \code{OPAL} opacity table \citep{1996ApJ...464..943I}.

We use SN models for SN~1987A for our shock breakout study. SN~1987A stands out among well-studied supernova targets. With plenty of observations and detailed investigations, including shock breakout estimations \citep{Imshennik1988, 1992ApJ...393..742E,  2017ApJ...845..103L}, SN~1987A makes itself an ideal candidate for probing multi-dimensional supernova shock breakout phenomena.

The progenitor star of SN~1987A has been known as a compact blue supergiant (BSG) \citep{1987ApJ...321L..41W,1987A&A...177L...1W} with a radius of (2--4) $\times$ 10$^{12}$ cm \citep{1989ARA&A..27..629A}. Recent theoretical studies \citep{2019MNRAS.482..438M,2020ApJ...888..111O,2020A&A...636A..22O,Utrobin2021ApJ} with progenitor models based on a binary merger scenario for SN~1987A \citep{2017MNRAS.469.4649M,2018MNRAS.473L.101U} have overall supported this scenario including modeling of LCs after the shock breakout \citep{2019MNRAS.482..438M,Utrobin2021ApJ}. Several related studies on SN~1987A with the binary merger progenitor models have also been reported \citep{2019ApJ...882...22A,2021ApJ...908L..45G,2022ApJ...931..132G,2022MNRAS.514.3941N,2023ApJ...949...97D,Ono2023,2024ApJ...961L...9S}.

We conduct simulations to calculate luminosity variation curves (LCs) by considering different wavelength bands and viewing angles for SN~1987A (see Table \ref{tab:RG} for details). Although recent studies have shown the binary merger progenitors better describe the optical \citep{2019MNRAS.482..438M,Utrobin2021ApJ} and X-ray \citep{2020A&A...636A..22O} LCs as partly mentioned, to compare our results with those of the previous study \citep{2017ApJ...845..103L}, we adopt the same $15\Ms$ single star progenitor (a BSG) for SN~1987A and are motivated by the asymmetries in the nebula of SN~1987A \citep[e.g.][]{2005ApJ...627..888S} to incorporate inhomogeneity in stellar envelopes and CSM, including perturbations, a ring-like structure caused by an eruptive mass loss, and a binary companion star \citep{Tsai_2023} to investigate the impact of variety of CSM environment. Furthermore, we also explore the CSM resulting from a steady stellar wind of varying strengths spanning two orders of magnitude.

We introduce the {\CASTRO} code, progenitor models, and methodologies in \Sect{NM}. To verify RHD simulations, we then examine the numerical effects of simulation resolutions and initial perturbations in \Sect{test}. Our simulation results are presented in \Sect{uni}, \Sect{clum}, and \Sect{bi}. The significance and implications of our work are discussed in \Sect{Dis}. Finally, we conclude our findings in \Sect{conclusions}.

\section{Numerical Method}
\label{sec:NM}
 
\subsection{\CASTRO}
{\CASTRO} \citep{2010ApJ...715.1221A,Almgren2020} is a compressible RHD code that solves hyperbolic radiation fluid and frequency space advection with the higher-order Godunov scheme and solves photon diffusion and source-sink terms by first-order backward Euler method. It uses co-moving frame Multi-Group Flux Limited Diffusion approximation (MGFLD) RHD \citep{2011ApJS..196...20Z, 2013ApJS..204....7Z} with Poisson gravity \citep{2016ApJ...819...94K}. \CASTRO\ is equipped with highly parallel Adaptive Mesh Refinement (AMR) to resolve turbulent fine structures formed in explosive phenomena.

We perform 2D simulations of the shock breakout using the cylindrical coordinates ($r-z$) covering direction angles $\theta =0^\circ-90^\circ$ and $\phi=0^\circ-360^\circ$ with a symmetry along $z$, which essentially represents an axis-symmetric half-sphere of the shock breakout. The lower and upper boundaries of the $r$ and $z$ axes are set as reflective and outflow boundaries, respectively. Our simulations are conducted in a box with the size of $10^{14}$ cm $\times$ $10^{14}$ cm for the 2D single star SN~1987A simulations, employing a resolution of $1024^2$. For simulations of an explosion with a nearby companion star, we use a box with the size of $4\times 10^{13}$ cm $\times 4\times 10^{13}$ cm with a higher grid resolution of $2048^2$ in Cartesian coordinates. Single simulation with $1024^2$ resolution, 8 groups MGFLD, and \code{OPAL} takes around a thousand CPU hours on the Kawas Cluster at the Academia Sinica Institute of Astronomy and Astrophysics (ASIAA).

\subsubsection{Gas Physics}
We adopt a $\gamma$ law equation of state for the shock breakout problem due to the wide range of gas density ($10^{-15} \, \gcc$ to $10^{-3}\,\gcc$) and temperature ($10^4 \, \K$ to $10^7\,\K$) in our simulations. The ideal gas with $\gamma = 5/3$ is appropriate for this density and temperature range. As our simulations commence long after the completion of explosive nucleosynthesis, nuclear burning networks are omitted.

\subsubsection{Radiation Hydrodynamics}

The RHD module of {\CASTRO} \citep{2011ApJS..196...20Z} independently evolves the temperatures of gas and radiation. The introduction of radiation cooling and heating terms is represented by $\partial_t E_{\text{gas}} \propto c(aT^4 - E_{\text{r}})$, which captures the balance between the gas energy $E_{\text{gas}}$ and radiation energy $E_{\text{r}}$, where $c$ is the speed of light, $a$ is the radiation constant, and $T$ is the gas temperature. This equation characterizes the local radiative heating and cooling process. As the shock propagates, the radiation temperature and gas temperature of a cell can vary, allowing for a non-thermalized radiation spectrum.

{\CASTRO} MGFLD module \citep{2013ApJS..204....7Z} enables multi-group radiation transport and can evaluate the color evolution of shock breakout. {\CASTRO} MGFLD employs a closure relation and flux approximate limiter between the radiation flux and the radiation energy to ensure isotropic diffusion in optically thick regions, and that in the free streaming region, the radiation flux adopts the form $||F_r|| = c E_r$, where $F_r$ represents the radiation flux, $E_r$ is the radiation energy, and $c$ denotes the speed of light. FLD approximation aligns effectively with the shock breakout scenario, where radiation flux may follow different directions relative to the gas flow, depending on the breakout environments.

In our 2D simulations, we consider photon frequencies ranging from $10^{13}$ Hz to $10^{19}$ Hz, dividing them into 8 groups with a logarithmic scale spacing. These resulting photon groups span from mid-infrared to X-rays. The specific range of each group is outlined in Table \ref{tab:RG}. At the simulation's outset, the radiation energy of each cell is set to $E_r=aT^4$. This value is based on the gas temperature $T$ from the progenitor stars, assuming local thermal equilibrium. The initial radiation energy of each radiation group is determined by integrating the Planck function within the corresponding band, and the following color evolution is governed by the closure relation, e.g., the dominant photon group can evolve depending on the interaction between gas and radiation.

We calculate the LCs by tallying the photons passing through the virtual observer point close to the boundary of the simulation box. The luminosity of the breakout can be expressed as
\begin{equation}
L(\nu,\theta)= 4\pi r_{s}^2 \times F({\nu,\theta}), 
\label{Lumi}
\end{equation}
where $L(\nu,\theta)$ is the luminosity within the frequency band of $\nu$ at the viewing angle $\theta$. $r_{s}$ is the radius of the sphere for calculating the total luminosity that is set close to the boundary of the simulation box, and $F({\nu,\theta})$ is the corresponding radiation flux. We collect lab frame fluxes calculated by \CASTRO\ at the virtual observer (VO) point. Furthermore, we calculate the LCs for distant observers who only see photons traveling along the line of sight from the SN emission, much closer to a point source for an extra-galactic SN. Therefore, we calculate such LCs by considering the light projected along the real observer (RO) point with a correction of light travel time. The comparison of VO, RO, and uncorrected RO will be discussed in \Sect{test}

\begin{deluxetable}{cccc}
\tabletypesize{\scriptsize}
\tablewidth{0pt} 
\tablecaption{Radiation Groups \label{tab:RG}}
\tablehead{
\colhead{Band} & \colhead{Central Frequency (Hz)} & \colhead{Band Width (Hz)} & \colhead{Central Wavelength (\AA)}
} 
\colnumbers
\startdata 
{I}  &$2.37 \times 10^{13}$  & $4.62 \times 10^{13}$ & $1.26\times 10^{5}$ \\
{II }&$1.33 \times 10^{14}$  & $2.59 \times 10^{14}$ & $2.25\times 10^{4}$ \\
{III}&$7.49 \times 10^{14}$  & $1.46 \times 10^{15}$ & $4.00\times 10^{3}$ \\
{IV} &$4.21 \times 10^{15}$  & $8.22 \times 10^{15}$ & $7.12\times 10^{2}$ \\
{V}  &$2.37 \times 10^{16}$  & $4.62 \times 10^{16}$ & $1.26\times 10^{2}$ \\
{VI}  &$1.33 \times 10^{17}$  & $2.59 \times 10^{17}$ & $2.25\times 10$ \\
{VII}&$7.49 \times 10^{17}$  & $1.46 \times 10^{18}$ & $4.00$ \\
{VIII}&$4.21 \times 10^{18}$  & $8.22 \times 10^{18}$ & $7.12\times 10^{-1}$ \\
\enddata
\tablecomments{The table lists the information of eight bands of radiation groups used in the simulations. It contains the central frequency, bandwidth, and central wavelength for each group.}
\end{deluxetable}

\subsubsection{Opacity}
\label{subsec:opacity}
We adopt opacity from \code{OPAL} \citep{1996ApJ...464..943I}. Instead of approximating constant opacity, \code{OPAL}\ can better model the total opacity in the cooling envelope and the regions with turbulent structures after shock passing. \code{OPAL} takes the inputs of the gas density, electron density, hydrogen fraction, helium fraction, metallicity, and temperature from each cell and yields the opacity for each photon group. In our simulations, we only consider the stellar envelope and CSM, which contains mainly hydrogen and helium and a tiny fraction of heavy elements. Therefore, we neglect the opacity contributed by individual heavy elements. Assuming the gas is fully ionized, we adopt the hydrogen fraction, helium fraction, and metallicity \footnote{Note: the values of the binary merger progenitor model \citep{2018MNRAS.473L.101U} of 1987A adopted, e.g., in the previous studies \citep{2020ApJ...888..111O,2020A&A...636A..22O}, are $X=0.639$; $Y=0.355$; $Z=0.006$.} as $X=0.47$, $Y=0.52$, and $Z\sim 0.01$, respectively. It is noted that the differences in the metallicity by a few factors slightly affect the consequent shock breakout due to the dominance of electron scattering opacity. 

\OPAL\ provides an effective opacity, including absorption and scattering. However, {\CASTRO} RHD calculation requires absorption ($\kappa_{\nu, \mathrm ab}$) and scattering ($\kappa_{\nu, \mathrm sc}$) opacities, independently. Therefore, the absorption opacity is obtained from the three dominant absorption opacity sources following \cite{2017ApJ...845..103L} including bound-free ($\kappa_{\nu, bf}$), free-free ($\kappa_{\nu,ff}$) and consider Compton scattering from a tiny fraction of electron scattering ($\epsilon\times \kappa_{\mathrm es}$, where $\epsilon=10^{-4}$) assuming the fraction of photon energy can be transferred to the gas during photon-electron collisions. These opacities depend on the gas temperature ($T$) and density ($\rho$) in the form:
\eq{kb}{\kappa_{\nu,bf} \approx \rho^2  T^{-0.5}  \nu^{-3} \, Z(1+X),}
\eq{kf}{\kappa_{\nu,ff} \approx \rho^2  T^{-0.5}  \nu^{-3} \, \frac{Z^2}{\mu_e\mu_I},}
\begin{equation}
\begin{aligned}
\kappa_{\mathrm es} = \rho_e\, K_{\mathrm sc}\, \left[1+\left(\frac{T}{4.5\times10^8 \ {\mathrm K}}\right)^{0.86}\right]^{-1} \\ \times \ \left[1+2.7\times10^{11} \, \frac{\rho}{T^2}\right]^{-1}.
\end{aligned}
\label{kes}
\end{equation}
$\mu_e$ is the mean molecular weight of electrons, $\mu_I$ is the mean molecular weight of ions, and $\rho_e$ is the electron density. We also consider partial degeneracy at a high-temperature regime for Thomson electron scattering opacity $K_{\mathrm sc}=0.4$ \citep{1976ApJ...210..440B}. Then, we obtain the scattering opacity with the effective opacity from {\OPAL} $\kappa_{\nu, {\OPAL}}$ and absorption opacity $\kappa_{\nu,\mathrm ab}$
\begin{equation}
\label{nusc}
\kappa_{\nu, \mathrm sc} = \kappa_{\nu, {\OPAL}} - \kappa_{\nu, \mathrm ab},
\end{equation}
\begin{equation}
\label{nuab}
\kappa_{\nu, \mathrm ab} = \kappa_{\nu,ff}+\kappa_{\nu,bf}+ \epsilon \kappa_{\mathrm es}.
\end{equation}
We take one of our 2D simulations and plot its 1D opacity profiles in \Fig{opacityks}. The total opacity behind the stellar radius of $\sim 3.2\times 10^{12}$ cm is higher than pure electron scattering and is consistent with the findings in \cite{2017ApJ...845..103L}. We adopt the value of specific opacity from {\OPAL}. In the innermost regions of \Fig{opacityks}, the scattering opacity exceeds the electron scattering opacity because \OPAL\ includes additional opacity from bound-bound transitions and photoionizations. Since we cannot distinguish the component of resonant scattering in \OPAL, we simply assume the opacities from bound-bound transitions and photoionizations contribute to scattering. This may overestimate the scattering opacity and underestimate the absorption opacity and the associated energy degradation during the radiative cascade processes. If the effective opacity is unavailable from {\OPAL} for a certain temperature and density region, for example, in the outer low-density regions of \Fig{opacityks}, we assume $\kappa_{\nu, \mathrm ab}= \epsilon \kappa_{\mathrm es}$ and $\kappa_{\nu, \mathrm sc}=(1-\epsilon) \kappa_{\mathrm es}$, with $\epsilon = 10^{-4}$ to fill the opacity gaps in {\OPAL}. This simplified treatment is effective for dilute ionized gas with $\rho < 10^{-12} \gcc$ and $10^8 > T > 10^4$ K \citep{2011Paxton,2024Farag}, in which electron scattering dominates. The $\epsilon$ is required to $\ll 1$, but its exact value can be adjusted for numerical stability without considering the detailed physics for the absorption opacity. This setup provides a realistic frequency-dependent opacity in our 2D simulations. Nevertheless, \OPAL\ remains incomplete, and the simulation can benefit from better treatments of the opacity calculation; we will discuss this limitation and possible improvement in \Sect{Dis}.
\fig{opacityks}{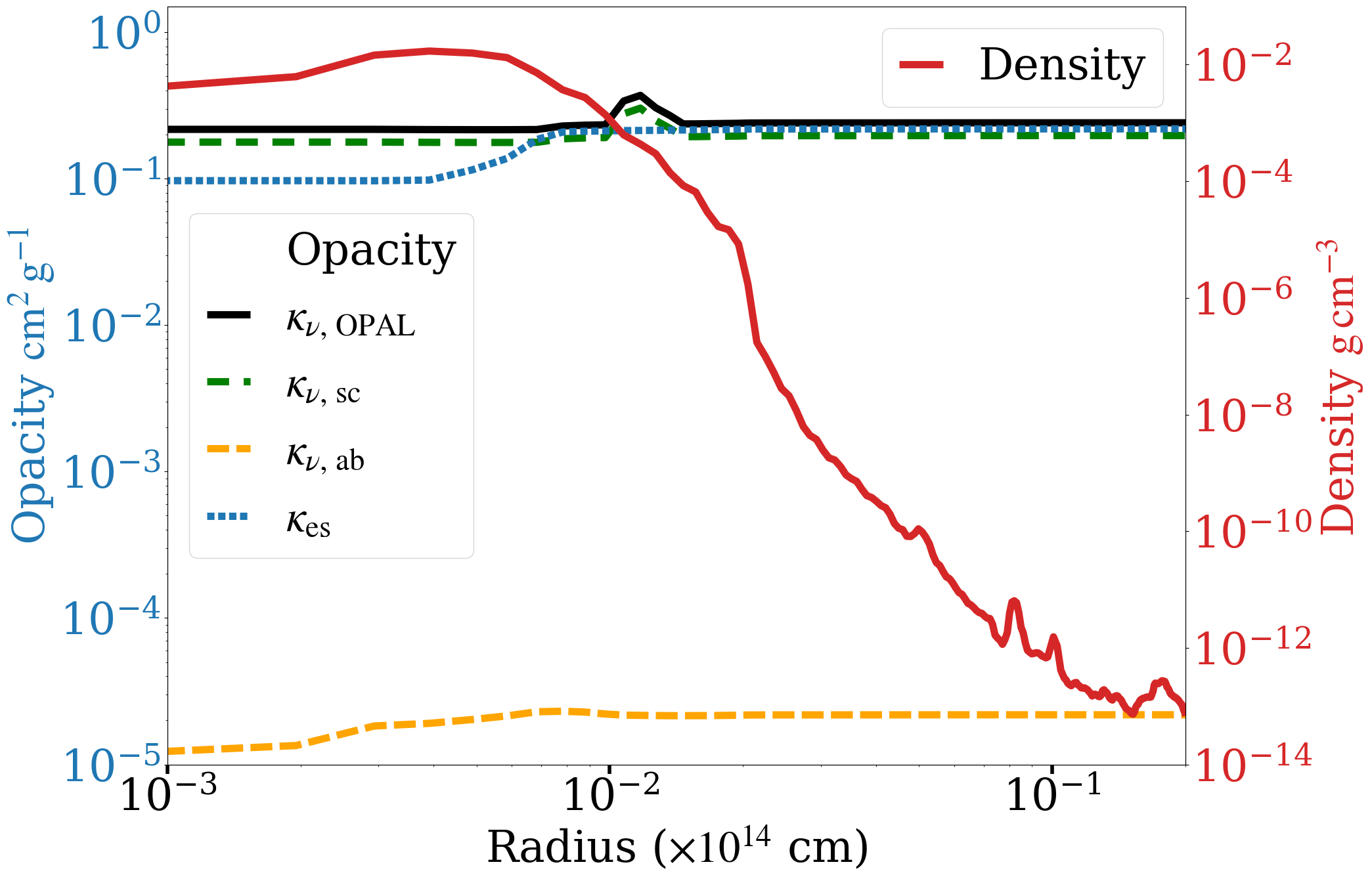}{1D density and opacity profiles for Model \texttt{T} when the shock is about to break out. Profiles show various opacity components from the scattering opacity, and absorption opacity from {\OPAL}
for the band VIII, and the electron scattering. The \OPAL\ opacity in the region within a stellar radius, $r=3.2\times 10^{12}$ cm, is higher than electron scattering. The overlapping lines are slightly shifted for legibility.}{0.7}

\subsection{Progenitor Star Models of SN~1987A}

\subsubsection{Single-Star Progenitor Model}
We adopt the single-star progenitor of SN~1987A from \cite{2017ApJ...845..103L}, which was computed using the stellar evolution code, \KEPLER\ \citep{1978ApJ...225.1021W}. 
During the iron core collapse phase, a thermal bomb of $2.3\times 10^{51}\,\mathrm{erg}$ is injected at the outer boundary of the iron core to initiate a successful explosion with a kinetic energy of $\sim2\times 10^{51}$ erg at the breakout. This process generates a strong shock driven by radiation, eventually leading to the complete destruction of the star. Subsequently, we map the 1D \KEPLER\ explosion profile onto the 2D {\CASTRO} $r-z$ coordinate system at the moment when the shock reaches $r \approx 1\times10^{12}$ cm and a shock speed of $4\times 10^{8}\, \mathrm{cm}\, \text{s}^{-1}$.

\subsubsection{Explosion with a nearby Companion Star}

We examine the shock breakout in a close binary by assuming the existence of a companion star during the explosion of SN 1987A. Recent studies \citep{2019MNRAS.482..438M,2020ApJ...888..111O,2020A&A...636A..22O,Utrobin2021ApJ} have suggested that SN 1987A progenitor may have formed from a binary merger and result in a single BSG progenitor before the explosion. Nevertheless, observations of massive stars \cite{Sana2012} show the majority of them are in binary, and some companion stars likely exist during the explosion of donor stars. Therefore, studying this scenario remains scientifically intriguing.
To examine the shock breakout in this hypothetical binary system, we place a companion star near the SN~1987A progenitor star, excluding the wind from the companion star. The companion star is a $13\Ms$ main-sequence star with $r \approx 5\times 10^{11}$ cm, located at a distance of $7\times10^{12}$ cm from the SN~1987A progenitor star based on the distance of interacting binaries suggested by \cite{Tsai_2023}. 
We place the companion star at a location along $\theta = 45^\circ$ to avoid boundary effects and use 2D Cartesian grids to prevent the nonphysical torus structure for the companion star. The 2D simulation box is considered a sliced view through the centers of the SN and the companion star. This setup is to investigate the fluid instabilities of the shock collision on a companion star. However, we recognize that only 3D simulations can properly model the binary interactions, and we may overestimate the impacts of the companion star due to the limitations in 2D. The differences in coordinate systems raise difficulties in quantitative comparison with other 2D models, therefore we focus on qualitative discussions for this Model \texttt{B}.

\subsection{Stellar Surface and Circumstellar Medium}
\label{subsec:ini}

The simulation box is extended to accommodate the shock interaction with the stellar surface and the CSM. For our single-star progenitor, the box size extends to $1\times10^{14}$ cm on each side, while for the binary system, it is set to $4\times 10^{13}$ cm for resolving the companion star. We fill the medium between the stellar surface and inner CSM from $3.2\times 10^{12}$ cm to $2\times 10^{13}$ cm using an interpolation scheme in \cite{2017ApJ...845..103L}. 

The regions beyond $2\times 10^{13}$ cm are filled with CSM based on a steady wind profile of \eq{constant}{\rho(r)=\frac{\dot{M}}{4\pi r^2 v_{\mathrm w}},} where the mass loss rate $\dot{M}\approx8\times10^{-6} \,M_{\odot}\, \mathrm{yr^{-1}}$, stellar wind velocity $v_{\mathrm w}\approx 450\, \mathrm{km}\, \mathrm{}{\mathrm s}^{-1}$, and the radius $r$. These values are based on the observational constraint and the BSG wind profiles \citep{1987Natur.328...44C, 2005ApJ...627..888S}. Moreover, we create various stellar wind profiles by adjusting the mass loss rate via $\alpha\dot{M}$, where $\alpha=[0.1, 0.2, 0.5, 1.0, 2.0, 5.0, 10]$. These profiles serve as the foundation for our investigations into the influence of the CSM on observable aspects of shock breakouts. The models with only the steady wind in the CSM are denoted as \texttt{N} for 1D simulation with 1024 resolution and \texttt{T} for 2D simulation. For all the models in Table \ref{tab:name}, $\alpha$ is set to 1.0 as the fiducial value; the comparison of different $\alpha$ values are presented in \S \ref{subsec:amplitude}.

The observed asymmetries in the nebula of SN~1987A \citep[e.g.,][]{2005ApJ...627..888S} inspire us to investigate the gas dynamics of SN explosions within inhomogeneous CSM environments. Such environments may naturally arise from stellar evolution and the formation of CSM. In addition to considering the binary companion, we also independently account for perturbations and an eruptive mass loss in the implemented BSG progenitor star and the steady wind CSM system.

\subsubsection{Sources of gas inhomogeneity}
\label{subsec:pert}

The envelopes of massive stars can be highly convective and may not be aware of the collapsing core before the shock arrives \citep{CHEN201370, Mao_2015}. To assess the effects of density inhomogeneities that may be introduced in such environments, we initialize density perturbations in the envelopes to evaluate their consequences on shock breakout. We introduce 2D density perturbations from the base of helium layer where the shock emerges to the stellar radius ranging over $6.2\times10^{10} \,\mathrm{cm} \leq r \leq 3.2 \times 10^{12} \,\mathrm{cm}$ to mimic the inhomogeneities: $\rho(r,\theta) = \rho_0(r)  [1+s\cos (m\theta)]$. To conserve momentum, we encompass sinusoidal perturbations $s\cos (m\theta)$ with $m=32$ and $s=1\%,\,3\%,\,5\%,\,10\%,\,15\%,\,20\%,\,25\%,\, \mathrm{and}\,30\%$, $s$ represents the amplitude of perturbed density $\approx \delta \rho/<\rho>$. Since seeding initial perturbations aim to trigger inhomogeneities, perturbations are only added to the steady wind CSM, which has a homogeneous density. The models with the perturbation strength $s=5\%$ are denoted as \texttt{P}.

\subsubsection{Circumstellar Medium Structure from Eruptive mass loss}
\label{subsec:CSM}

Mass loss from massive stars at their final stage of evolution may form a CSM torus and hourglass shape cavity \citep{ring2012}, SN~1987A is also believed to have an hourglass shape CSM formed during its BSG phase. The shock breakout likely occurs within the inner part of the hourglass structures and the well-known inner equatorial and outer double rings in the nebula of SN 1987A. We first assume constant mass loss wind in our simulations without considering the variation of mass loss rate as in Models \texttt{T} and \texttt{P}. However, years before the explosion, observations suggest that the progenitor star can have an enhanced mass loss \citep{2007Natur.450..390W, 2014ApJ...792...44C, 2014ARA&A..52..487S, 2021ApJ...923...41L, Hiramatsu_2023}. The origin of such mass loss might be eruptive mass loss caused by unknown mechanisms. To investigate the possible impact of eruptive mass loss on the shock breakout, we assume the eruptive mass loss forms a ring-like dense object located at $5.0\times 10^{13}$ cm from the progenitor star, with an elliptical cross-section of major axis $10^{13}$ cm and minor axis $10^{12}$ cm. The density of the ring is set to be 100 times higher than the surrounding CSM previously formed from a steady wind of $\alpha = 1.0$.
Such structure corresponds to an eruptive mass loss $\sim 10$ days before the explosion, assuming the velocity of the eruptive mass loss is of the same order of a BSG wind, $\sim 500\,\mathrm{km\,s^{-1}}$. The eruptive mass loss may be related to the final nuclear burning stage, but the exact mechanism remains elusive \citep{2016Quataert,2018Fuller,2019Ouchi&Maeda}. For this numerical experiment, we place the ring-like structures at $\theta = 45^\circ$ to avoid boundary effects. The ring model is denoted as \texttt{R}. Hereafter, we summarize our models and their naming with descriptions in Table \ref{tab:name}.

\begin{deluxetable*}{llcc}
\tabletypesize{\scriptsize}
\tablewidth{0pt}
\tablecaption{Summary of model setup and naming\label{tab:name}}
\tablehead{
\colhead{Model} & \colhead{Description} & \colhead{Wind}& \colhead{$\alpha$}
} 
\colnumbers
\startdata 
{ \texttt{N}}& 1D normal spherical simulation. &steady wind& 1.0 \\
{ \texttt{T}}& 2D test simulation. &steady wind & 1.0 \\
{ \texttt{P}}& 2D simulation with perturbation strength 5\%. &steady wind& 1.0 \\
{ \texttt{R}}& 2D simulation with a ring-like CSM.&eruptive mass loss wind& 1.0 \\
{ \texttt{B}}& 2D simulation with a companion star.& steady wind& 1.0 \\
\enddata
\tablecomments{The summary of model names and their setups. \texttt{N} stands for {``(N)ormal"}, \texttt{T} stands for {``(T)est"}, \texttt{P} stands for {``(P)erturbation"}, \texttt{R} stands for {``(R)ing-like"} CSM, and \texttt{B} stands for {``(B)inary"}. For the above-mentioned models, the strength of steady-wind of $\alpha = 1$ is applied unless specified with $*$. 
The detailed setting of each model can be found in \S \ref{sec:NM}.
}
\end{deluxetable*}

\section{Verification of Radiation Hydrodynamics with Resolutions and Perturbations}
\label{sec:test}
\subsection{Resolution}
To assess the influence of resolution on the results, we compare the density structures from simulations at grid resolutions of $512^2$, $1024^2$, and $2048^2$ in \Fig{2DResolution}.

\fig{2DResolution}{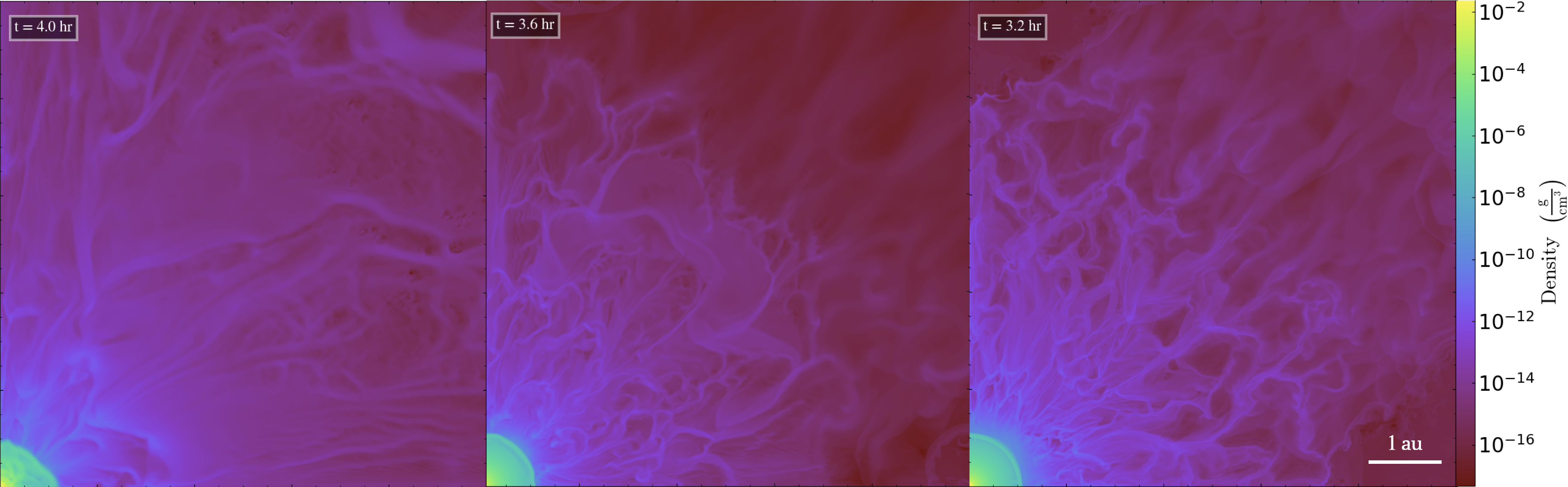}{Resolution comparison of Model \texttt{R}. \textbf{Left}: Resolution of $512^2$. \textbf{Middle}: Resolution of $1024^2$. \textbf{Right}: Resolution of $2048^2$. After the shock breakout, the shocked SN ejecta and CSM deviate from the spherical symmetry. The filaments and layers of the shock front are reasonably resolved. The difference between $ 512^2$ and $1024^2$ is the more detailed filamentary structures. Although $2048^2$ resolves the details of fine filamentary structures, it looks similar to $1024^2$. These tests suggest the optimal resolution is close to $1024^2$.}{1.00}

We examine density structures in two regions to assess resolution at different distances from the star: the spherically expanding remnant and the filamentary post-shock region. Their density structures converge around the resolution of $1024^2$. In \Fig{Resolution_LC}, their LC duration\footnote{Note: LC duration is defined as the time interval between the nearest instances when the luminosity reaches 10\% of its peak value before and after the major peak.}, peak luminosity, and LCs fluctuations across angles are similar, indicating that the physics of shock breakout is well resolved in our simulations with a resolution of $512^2$ or higher. However, for $512^2$, the angular-dependent LCs from Model \texttt{R} in \Fig{Resolution_LC} show a different result and imply the unresolved structures that impact the breakout emission. Therefore, we adopt a resolution of $1024^2$ for our simulations presented in later sections to optimize the scientific outputs and reduce the computing expense. Although LCs of the Model \texttt{N} in the left panel of \Fig{Resolution_LC} shows converging LC duration of $\sim 0.2$ hours and peak luminosity of $\sim10^{45}-10^{46} \ergs$, their LC fluctuates with resolutions significantly due to the nonphysical density spike in 1D. We use the resolution of 1024 for the later 1D and 2D comparison.
\sfig{Resolution_LC}{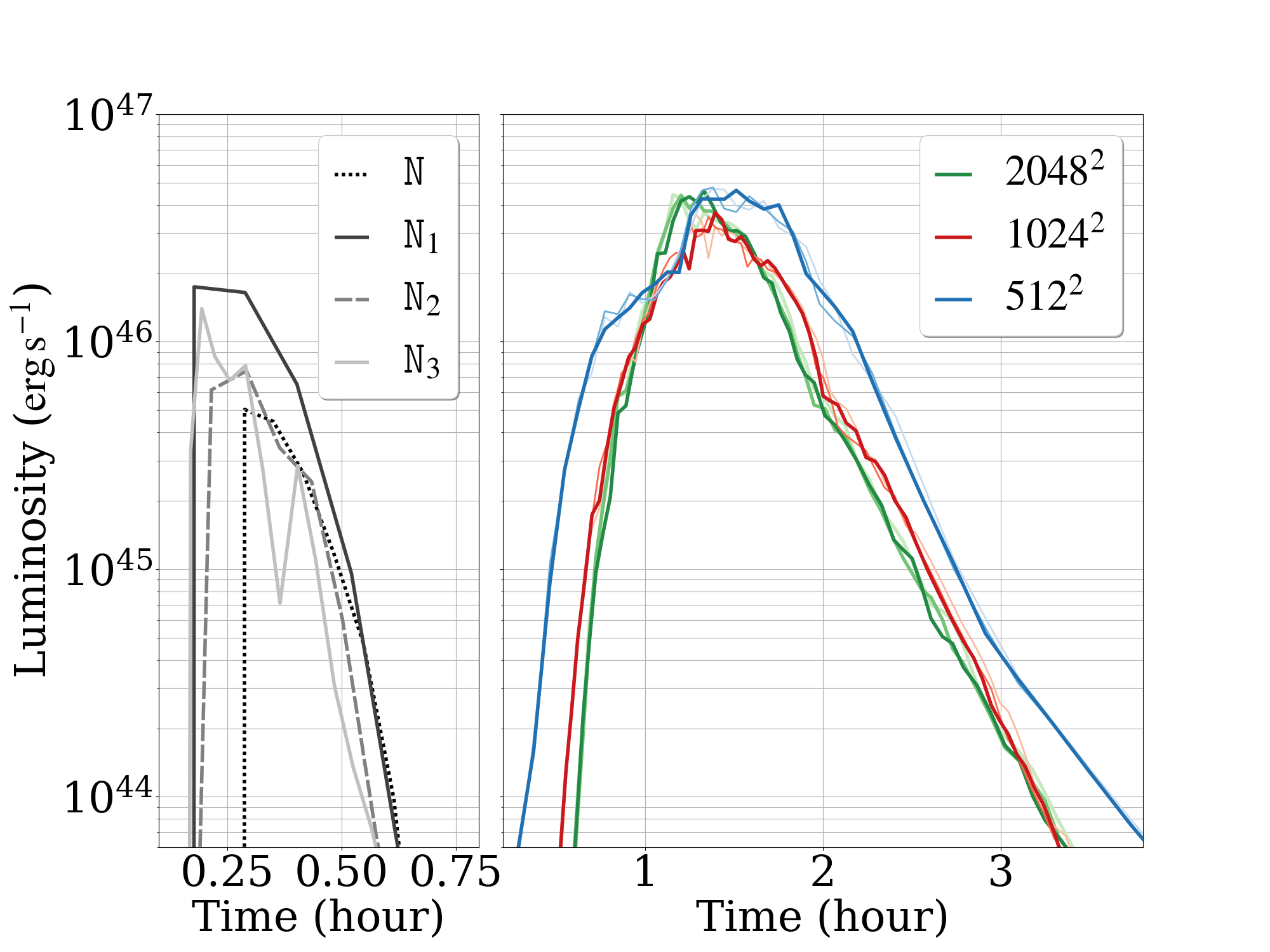}{LCs from simulations with different resolution. \textbf{Left}: LCs from the Model \texttt{N} with resolutions: \texttt{N}$_1$ $4096$, \texttt{N}$_2$ $10240$, and \texttt{N}$_3$ $102400$. The 1D LC duration is $\sim 0.2-0.3$ hours with peak luminosity of $\sim 10^{45}-10^{46}\ergs$ \textbf{Right}: LCs from Model \texttt{R} with resolutions at different viewing angles: greens for $2048^2$, reds for $1024^2$, and blues for $512^2$. The 2D LC duration is $\sim 2-3$ hours with peak luminosity of $\sim 2-4\times10^{46}\ergs$. Although the LC duration and peak luminosity from current resolutions are similar, the results from $512^2$ deviate more from $2048^2$ and $1024^2$, indicating insufficient resolution can affect the shock breakout emissions.}{0.7}
\subsection{Effect of Initial Perturbations}

To assess the impact of the perturbations, we examine their influence on the LC durations and peak luminosities for different perturbation amplitudes in \Fig{PS}. We examine the sinusoidal perturbations with the amplitudes of $s=1\%$, $3\%$, $5\%$, $10\%$, $15\%$, $20\%$, $25\%$, and $30\%$ in the gas density. \Fig{PS} shows the LC durations and peak luminosities converge around $5-6\times 10^{46} \ergs$ and $1.2-1.5$ hours, respectively. The higher peak luminosity usually accompanies the shorter LC duration. For $s\geq 10\%$, depending on the perturbation strength, LC duration and peak luminosity seem to fluctuate more than that of $s\leq 5\%$. The mixing from different perturbation strengths is shown in \Fig{gasPS}, and their final evolved structures show similar patterns. We focus on small perturbations because multi-dimensional stellar evolution calculations have suggested that the density fluctuations can be at most 10\% \citep{2007ApJ...667..448M, 2011ApJ...741...33A}, larger density perturbations may originate from effects that are difficult to take into account in current stellar models \citep[e.g.,][]{Mao_2015}. Therefore, we adopt $5\%$ initial density perturbations for Model \texttt{P}.

\sfig{PS}{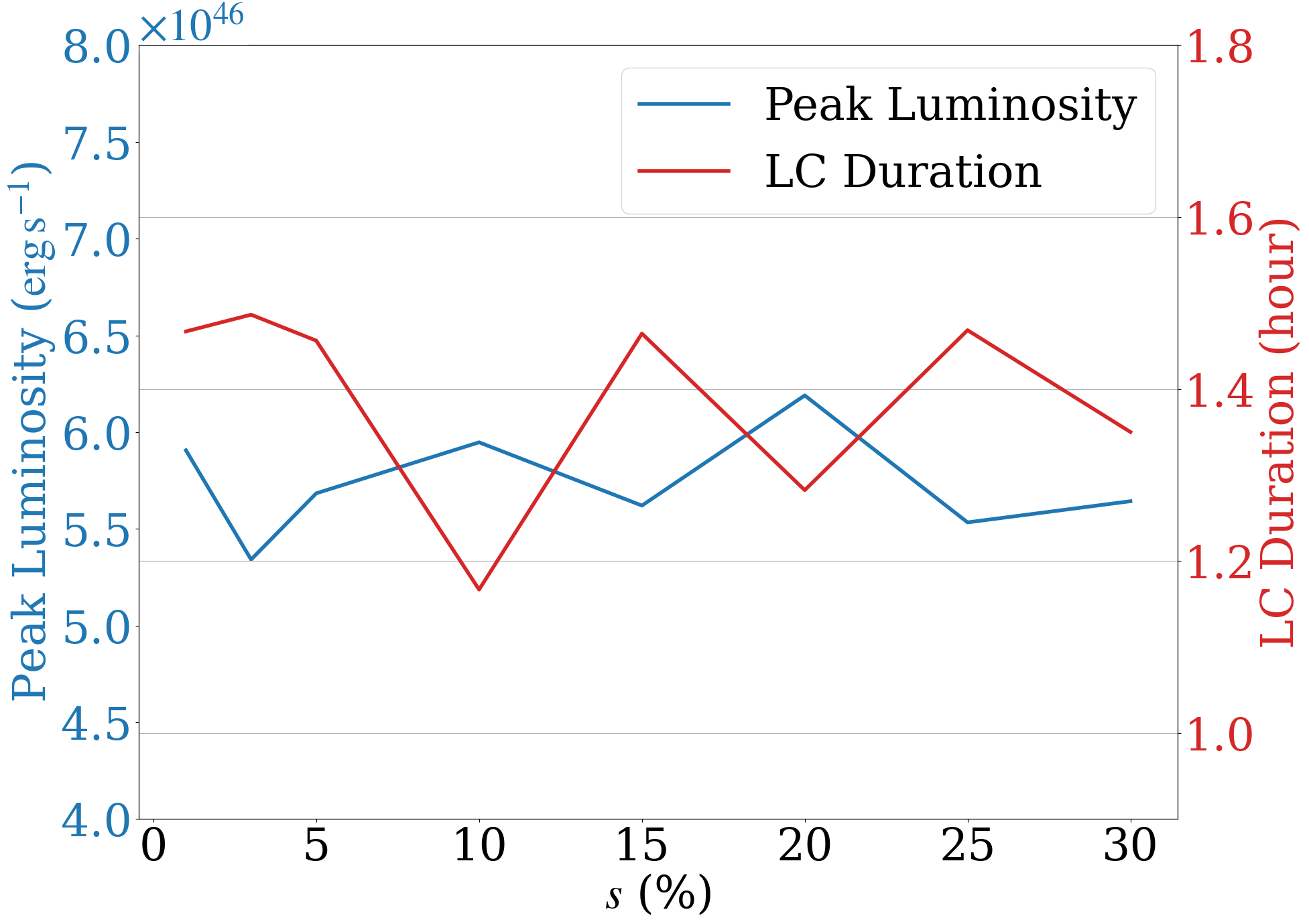}{LC peak and duration of Band VIII plot with different perturbation strengths (\texttt{P}; $s=1\%$, $3\%$, $5\%$, $10\%$, $15\%$, $20\%$, $25\%$, and $30\%$). Models of different perturbation strengths show roughly consistent peak luminosity of $5-6\times 10^{46} \ergs$ and LC duration of $1.2-1.5$ hours. In general, the higher peak luminosity corresponds to the shorter duration.}{0.7}
\sfig{gasPS}{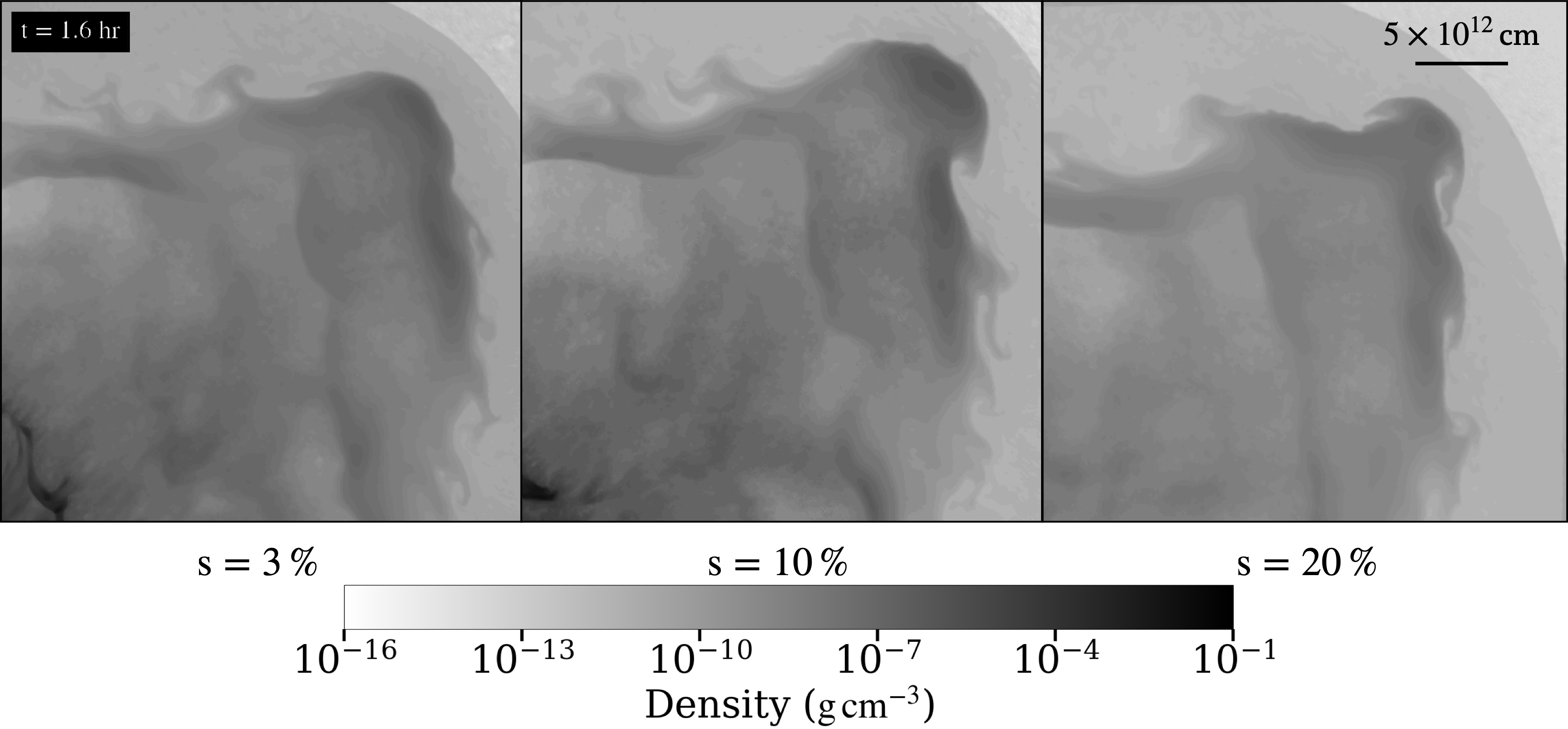}{Zoom-in of gas density structures. Panels from left to right show the gas density for models with perturbation strengths of $s$ $=$ 3, 10, 20\%, respectively. All panels clearly show the development of Rayleigh-Taylor instabilities, which are less sensitive to the strengths of perturbations.}{0.800}

\subsection{RHD and Radiation Precursor}

RHD is essential for modeling SN shock breakout, where radiation transport significantly affects gas dynamics as the gas becomes optically thin, $\tau \approx 1$. \Fig{RMS} displays density, radiation-to-gas pressure ratio profiles, and optical depths in Model \texttt{T}. The region around the forward shock is becoming optically thin, with its radiation-to-gas pressure ratio exceeding $1000$.

Before the shock breakout peaks to its maximum luminosity in LCs, the shock front communicates with pre-shocked regions through a radiation precursor \citep{Epstein1981, Schawinski_2008, 2017hsn..book..967W}\footnote{radiation precursor is the escaping radiation emitted ahead of the shock front from the SN explosion, and it can heat up the pre-shocked gas and affect the dynamics of shock propagation.}, and the radiation temperature in the pre-shocked regions increases as shown in \Fig{cjshock}. Due to the interaction between the radiation precursor and the pre-shocked CSM gas, the forward shock feature becomes a smoother transition in temperature and density. Furthermore in 2D, the mixing from fluid instabilities can enhance the radiation precursor and boost the breakout emissions. Therefore, multi-dimensional RHD simulations are required to model the shock breakout properly. 
\sfig{RMS}{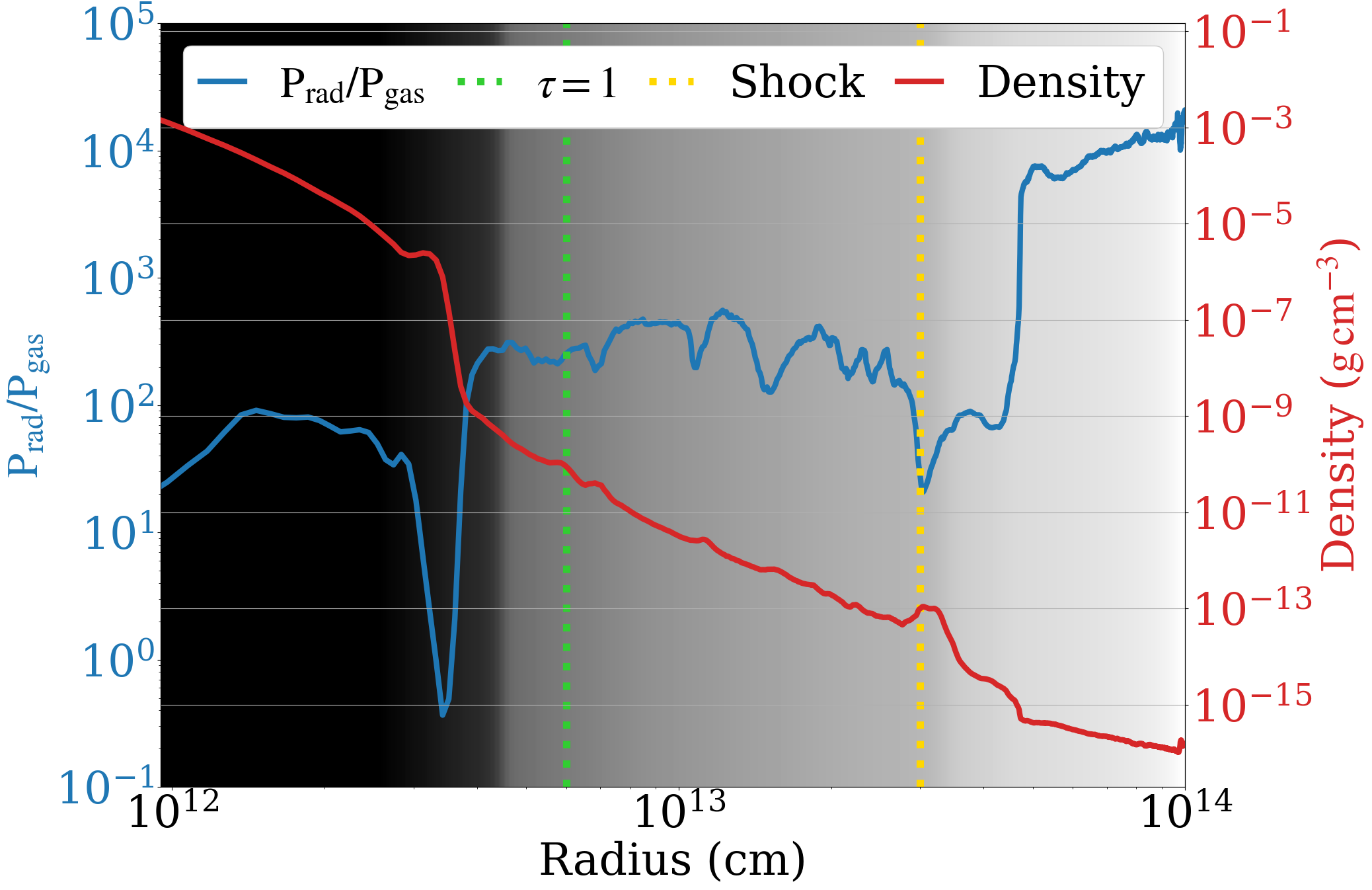}{The 1D radial profile of Model \texttt{T} when the shock travels to $3\times 10^{13}$ cm (yellow-dotted line). The background gray-scale color map shows the optical depth ($\tau$) as a function of $r$, and the green-dotted line shows the position where $\tau=1$. We show the radiation pressure to gas pressure ratio and gas density. The radiation pressure largely exceeds the gas pressure throughout the simulation regions.}{0.7}
\sfig{cjshock}{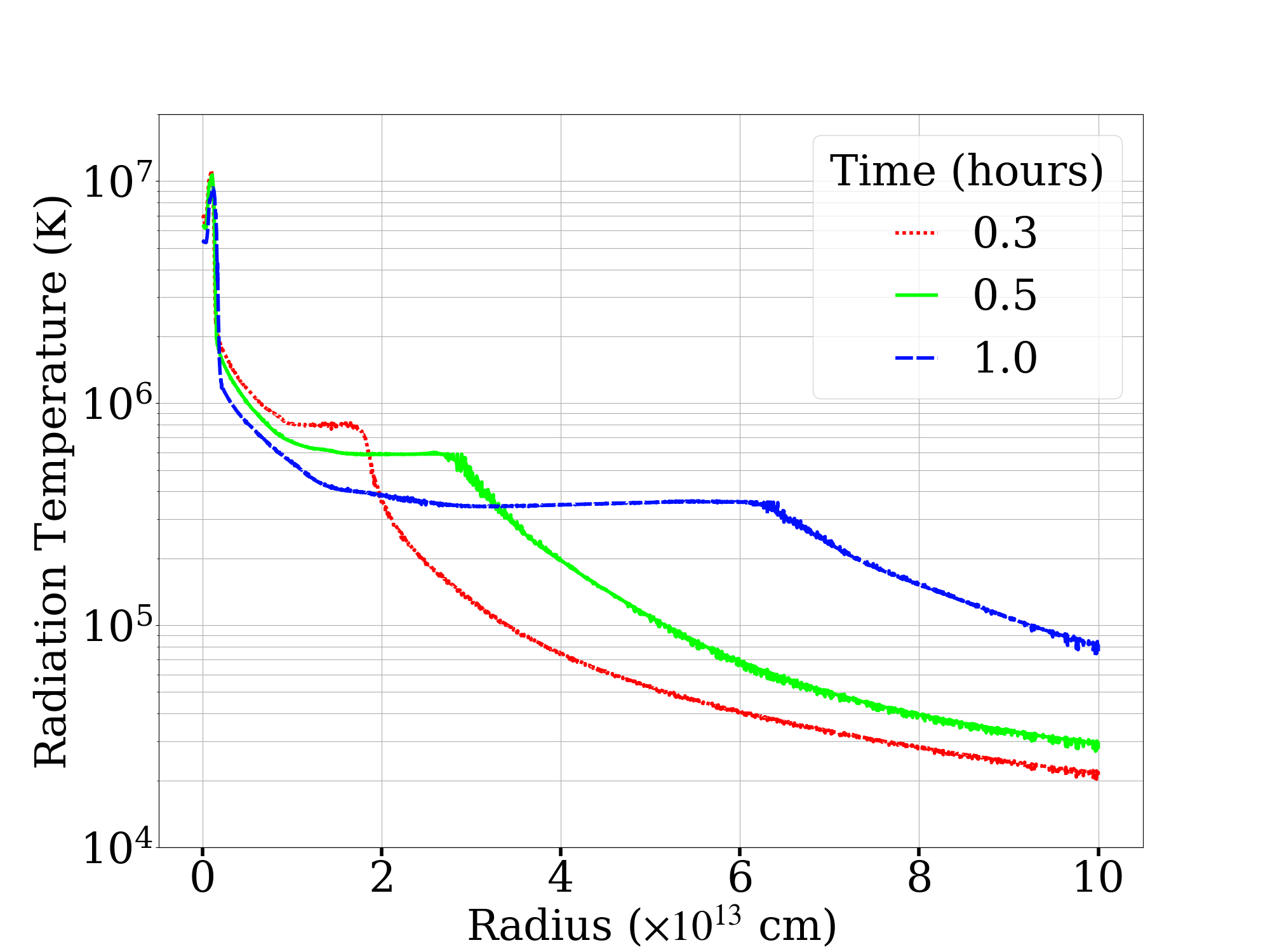}{The evolution of the radiation temperature profile for Model \texttt{T}. The shock is located at a turning point from flat to decay of the profile. As the shock propagates from $\sim 2 \times 10^{13}$ cm to $\sim 6 \times 10^{13}$ cm, the CSM gas beyond the shock front is heated up by the radiation precursor.}{0.700}

\subsection{LCs Calculation}
The approach of LC calculation is essential for properly capturing shock breakout emission in multi-dimension. For RO, we calculate the effective luminosity of shock breakout by integrating the elements of projected surface emission towards RO using Equation (\ref{LCintegrate}). \Fig{LTT} illustrates the viewing frame (blue) and lab frame (black) coordinates utilized for calculating RO LCs; the two frames share the same $x$-axis, and the former is obtained by rotating the latter clockwise by the viewing angle $\theta$ around the $x$-axis through rotation matrix $\mathbf{R_x}(-\theta)$; the coordinates in the viewing frame are denoted with a prime $``\,^{\prime}\,"$. \Fig{LTT} also demonstrates an element of emitting surface at the polar angle of $\beta^{\prime}$ and its distance difference from the observer between the elements at polar angles of $0$ and $\beta^{\prime}$: $\mathit{\Delta} d\,(\beta^{\prime}) = 2\,r_{s} \sin^2 (\beta^{\prime}/2)$ and the corresponding time difference, $\mathit{\Delta} t\,(\beta^{\prime}) = \mathit{\Delta} d\,(\beta^{\prime})/c$. For arbitrary $\theta$, the RO LCs at observer time $t$ are derived by integrating the contribution from each emitting surface over polar angles $0\leq \beta^{\prime} \leq \frac{\pi}{2}$ and azimuth angles around RO direction $0\leq \phi^{\prime} \leq 2\pi$, and $r_{s}=8\times 10^{13}$ cm for a half-sphere $S$:
\begin{equation}
\begin{aligned}
L_{\theta}(t) = \int_{S} F(t-\Delta t,\beta^{\prime},\phi^{\prime})\times(d\vec{ s} \,\cdot\, \hat{\boldsymbol{r}}_{\mathrm{RO}})= \int_{\phi^{\prime}=0}^{2\pi}\int_{\beta^{\prime} =0}^{\frac{\pi}{2}}{F(t-\Delta t,\beta^{\prime},\phi^{\prime})\,r_{s}^2\,\sin\beta^{\prime}\, \cos\beta^{\prime}\, d\beta^{\prime}\, d\phi^{\prime}},
\end{aligned}
\label{LCintegrate}
\end{equation}
where the integration is in the viewing frame, the projected surface element toward the RO is $d\vec{s} \,\cdot \,\hat{\boldsymbol r}_{\mathrm{RO}}=r_{s}^2\,\sin\beta^{\prime}\,\cos\beta^{\prime}\,d\beta^{\prime}\,d\phi^{\prime}$; $\hat{\boldsymbol r}_{\mathrm{RO}}$ is a unit vector directed to RO. While $F$ is cylindrical and reflective symmetric in the lab frame, we clock-wise rotate the coordinate to obtain $F$ in the viewing frame as shown in \Fig{LTT}:
\begin{equation}
\begin{aligned}
F(t-\Delta t, \beta^{\prime}, \phi^{\prime})=|\mathbf{R_x}(-\theta) \vec{F}(t-\Delta t, \beta, \phi)|, \quad
\mathbf{R_x}(-\theta)
=
\begin{pmatrix}
1 & 0 & 0 \\
0 & \cos \theta & \sin \theta \\
0 & -\sin \theta & \cos \theta
\end{pmatrix}
\end{aligned}
\label{Flux}.
\end{equation}
The integration of Equation (\ref{LCintegrate}) is done by summing all emitting surface elements of 1 square degree. \Fig{LCdis} shows the LCs for VO at different radii, and these LCs look similar, suggesting the VO's distance slightly affects the LCs. \Fig{LCt} and \Fig{LCb} show the LCs of Models \texttt{T} and \texttt{B} including the project w/o light traveling correction effects for RO. The corrected LCs for RO show a lower peak luminosity and longer duration than LCs for VO without corrections. The corrected LCs for RO are more realistic. Therefore, we adopt the RO approach to calculate and compare LCs among different models.
\sfig{LTT}{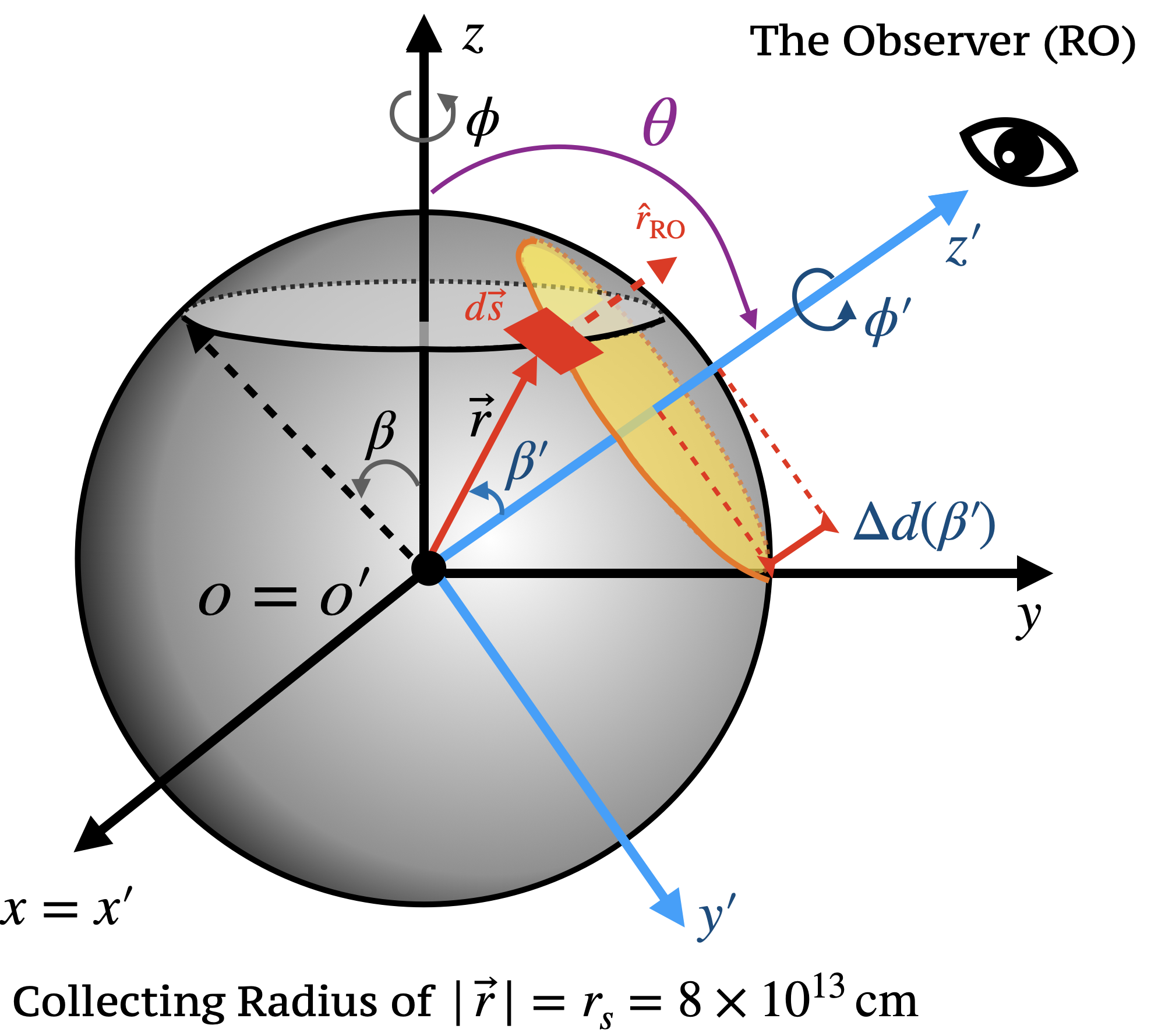}{Schematic diagram of observer and star's coordinates with the light traveling time correction. The black axis of ($x, y, z$) with its spherical coordinate of $(r,\beta,\phi)$ is for the star's frame. The blue axis of ($x^{\prime}, y^{\prime}, z^{\prime}$) with its spherical coordinate of $(r^\prime,\beta^\prime,\phi^\prime)$ is for the observer's viewing frame obtained by rotating star's frame along the x-axis by $\theta$. The red square in the direction of $\vec{r}$ represents the emitting surface elements on the sphere of collecting radius, $r_s$. The angle between $\vec{r}$ and  the viewing direction of $\hat{\boldsymbol r}_{\mathrm {RO}}$ or $z^\prime$ is $\beta^\prime$. The distance difference from the observer between the elements at polar angles of $0$ and $\beta^{\prime}$: $\mathit{\Delta} d\,(\beta^{\prime}) = 2\,r_{s} \sin^2 (\beta^{\prime}/2)$ and the corresponding light traveling time difference, $\mathit{\Delta} t\,(\beta^{\prime}) = \mathit{\Delta} d\,(\beta^{\prime})/c$. }{0.5}
\sfig{LCdis}{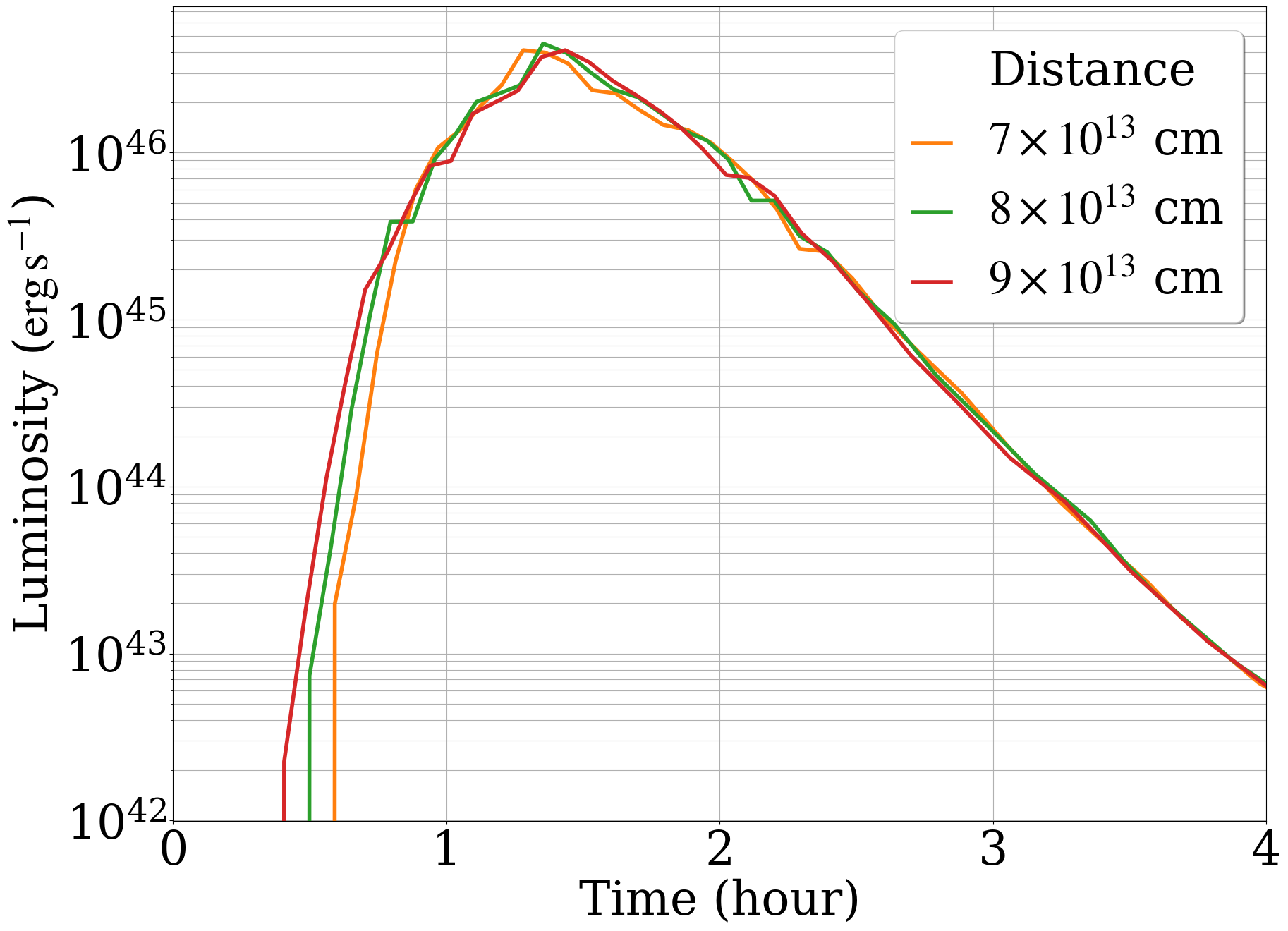}{The LCs calculated with VO at different distances $r = 7\times10^{13}$, $r = 8\times10^{13}$, and $r = 9\times10^{13}$ for Model \texttt{P}. The peak luminosities are 4.1, 4.5, and $4.1 \times 10^{46}\ergs$ with durations of 1.4, 1.3, and 1.4 hours, respectively. These angle-averaged LCs of different photon-collecting radius show minor differences.}{1.0}
\sfig{LCt}{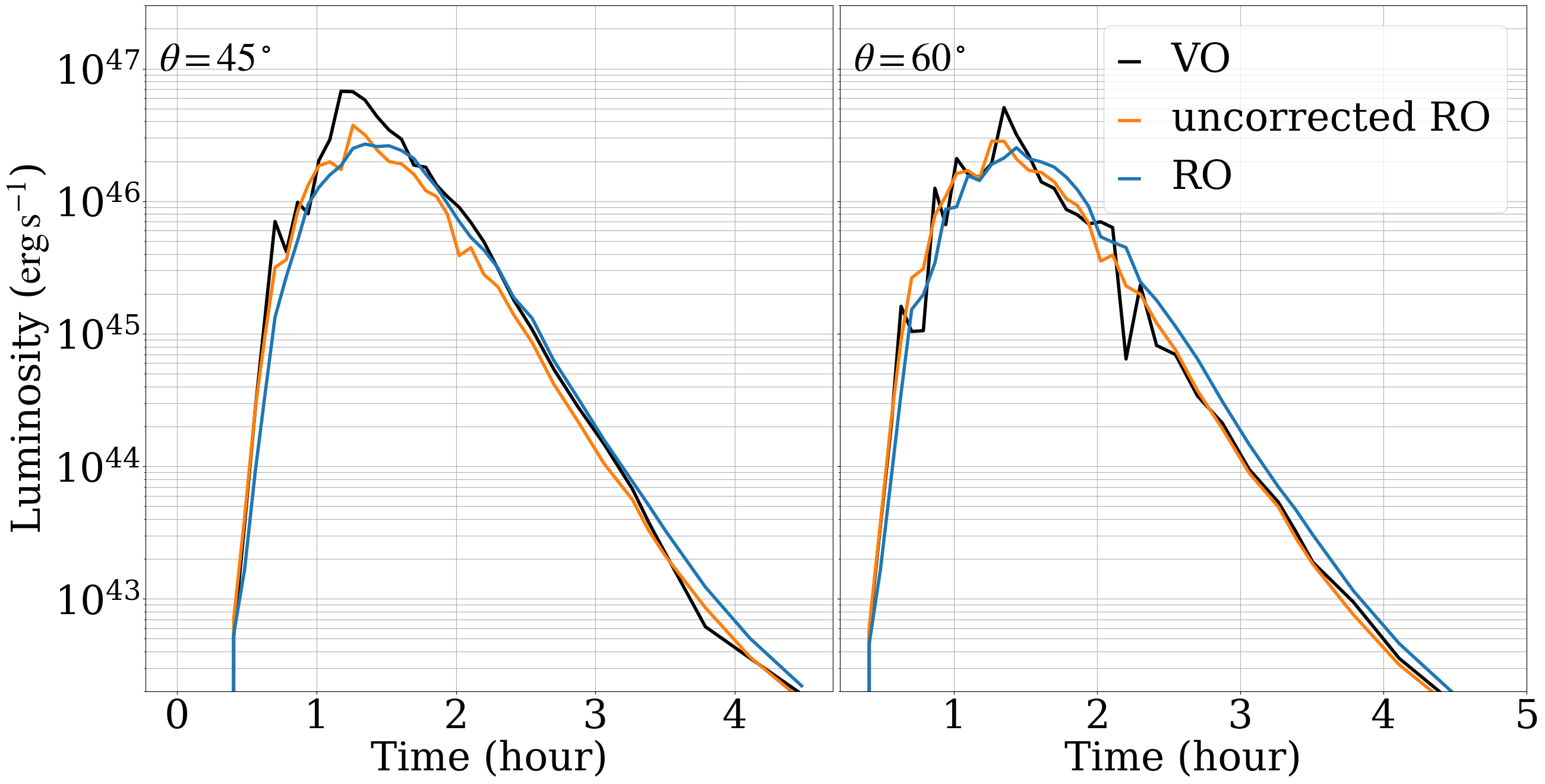}{Comparison of LCs in Model \texttt{T} with different methods: VO (black), uncorrected RO (orange), and RO (blue) for the viewing angles of $45^\circ$ and $60^\circ$. The RO LCs are smoother compared to VO and the uncorrected RO and have lower peak luminosities; the LCs for viewing angles of $45^{\circ}$ and $60^{\circ}$ are very similar. The angle-averaged LC durations for them are 1.34, 1.33, and 1.50 (hours), respectively. The corresponding peak luminosities are 4.5, 4.6, and 3.8 $\times 10^{46}\ergs$.}{1.0}
\sfig{LCb}{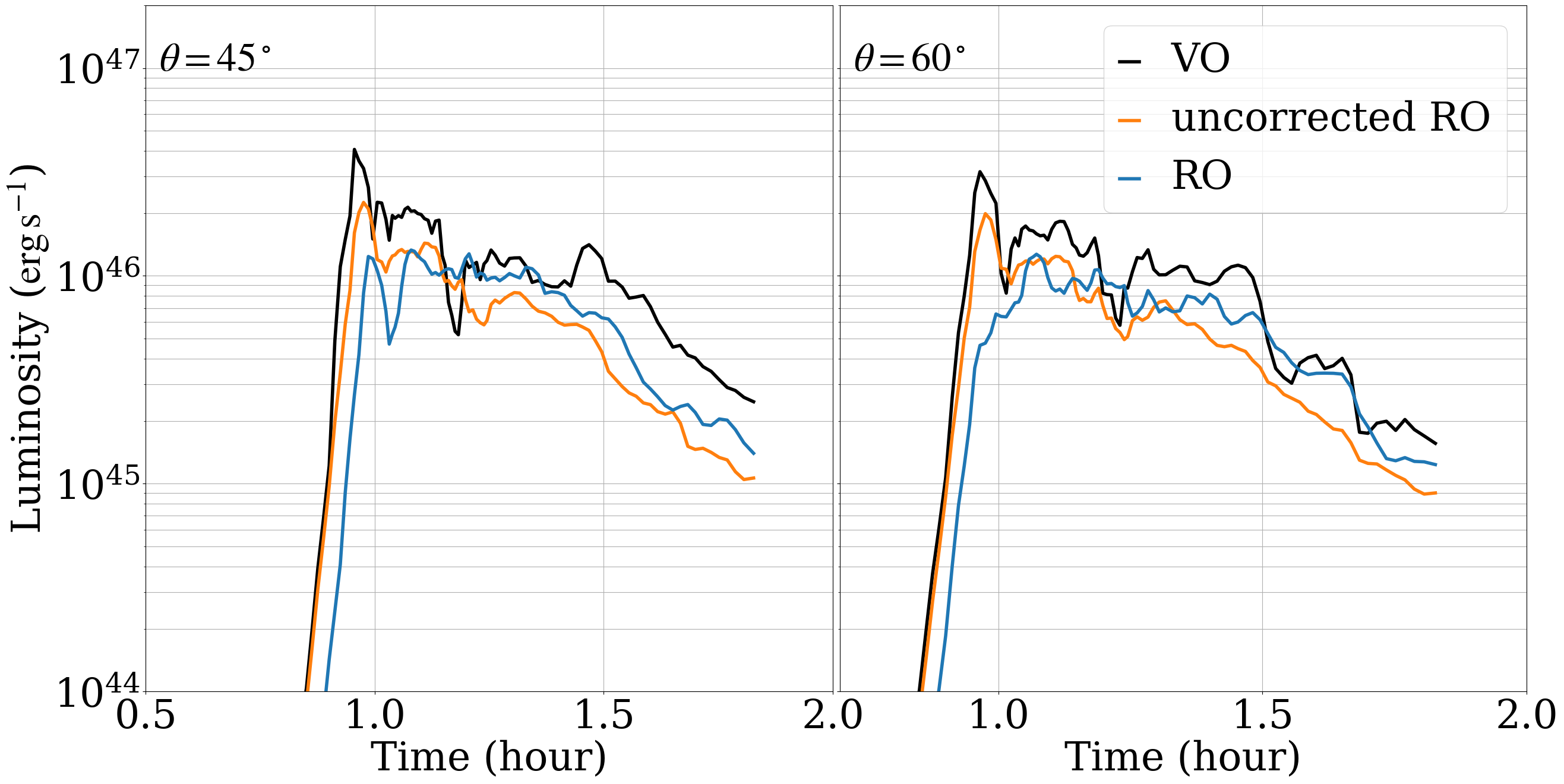}{Comparison of LCs in Model \texttt{B} with different methods: VO (black), uncorrected RO (orange), and RO (blue) for the viewing angles of $45^\circ$ and $60^\circ$. The RO LCs are smoother compared to VO and projection and have lower peak luminosities. The angle-averaged LC duration for them are 0.57, 0.70, and 0.83 (hours), respectively. The corresponding peak luminosities are 3.0, 2.1, and 1.3 $\times 10^{46}\ergs$.}{1.0}

\section{Shock Breakout with steady wind CSM}
\label{sec:uni}

Here, we present the results of the breakout simulation with steady wind CSM with zero perturbation (\texttt{T}) in \Fig{test}. The explosion shock propagates nearly spherically in the 2D space, while the reverse shock induces turbulence behind the forward shock colliding with the steady wind CSM. We use Model \texttt{T} as a reference for comparing shock structures, LC duration, and peak luminosity across different 2D models (\texttt{P}, \texttt{R}, and \texttt{B}). The selected LCs for comparison are from the band VIII of a wavelength: $\lambda=7.12 \times 10^{-1}$ {\AA}, which dominates the peak luminosity of shock breakout (see Table \ref{tab:RG}). In \Fig{LCAngle}, the LC duration and the peak luminosity in our 2D runs are $1.66\pm 0.07$ hours and $(3.24 \pm 0.26) \times 10^{46}\,\ergs$, respectively. Comparing with our 1D simulation (\texttt{N}) and previous 1D calculations \citep{1992ApJ...393..742E, 1999ApJ...510..379M,2017ApJ...845..103L}, our 1D LC is $\sim 3$ times longer and $\sim 3-6$ times brighter.

The major differences between 1D and 2D models are probably attributed to multi-dimensional effects that cause a broadening of density structures around the breakout shock, the consequence diffusion of photons, and the RHD interaction through radiation precursors. A detailed analysis of the multi-dimensional effects will be presented in Section \ref{sec:Dis}.

As shown in \Fig{test}, even without perturbations, A non-spherical shock appears along $\theta \approx 45^{\circ}$. The $45^{\circ}$ elongation may partly be due to numerical artifacts of the shock propagation during the breakout along $45^{\circ}$ because of the effectively lower resolution in the 2D grid cells, which can be improved with higher resolution. For other numerical artifacts, the Carbuncle instability may affect shock waves propagating along the symmetric axis and the grid structures, which arise independent of the resolution \citep{2009MNRAS.400.1283I}. Nevertheless, the hydro solver in {\CASTRO} is able to minimize this effect \citep[e.g.][]{Chen2014PPI,Ken2014PI, Ken2023ApJ}.

The non-spherical shock structure in Model \texttt{T} slightly affects the LCs from different viewing angles as shown in \Fig{LCAngle}; the peak luminosity for $0^{\circ}$ and $90^{\circ}$ is $3.53\times 10^{46}\ergs$ and $3.44\times 10^{46}\ergs$, respectively and are $\sim(3.64-3.81)\times 10^{46}\ergs$ for $(15^{\circ}-75^{\circ})$; the LC duration for $0^{\circ}$ and $90^{\circ}$ are $\sim 1.5$ hours and are $\sim(1.4-1.5)$ hours for $(15^{\circ}-75^{\circ})$. The angle-dependent LCs for $0^{\circ}$ and $90^{\circ}$ show slightly different behaviors compared to other angles, while the LC duration is relatively consistent among different angles. 

Although the simulation results are not sensitive to the initial perturbations, we adopt $s=5\%$ initial perturbations for Model \texttt{P}. We exclude the perturbations in models \texttt{R} and \texttt{B} because of the presence of asymmetry CSM and a companion star. 
\fig{test}{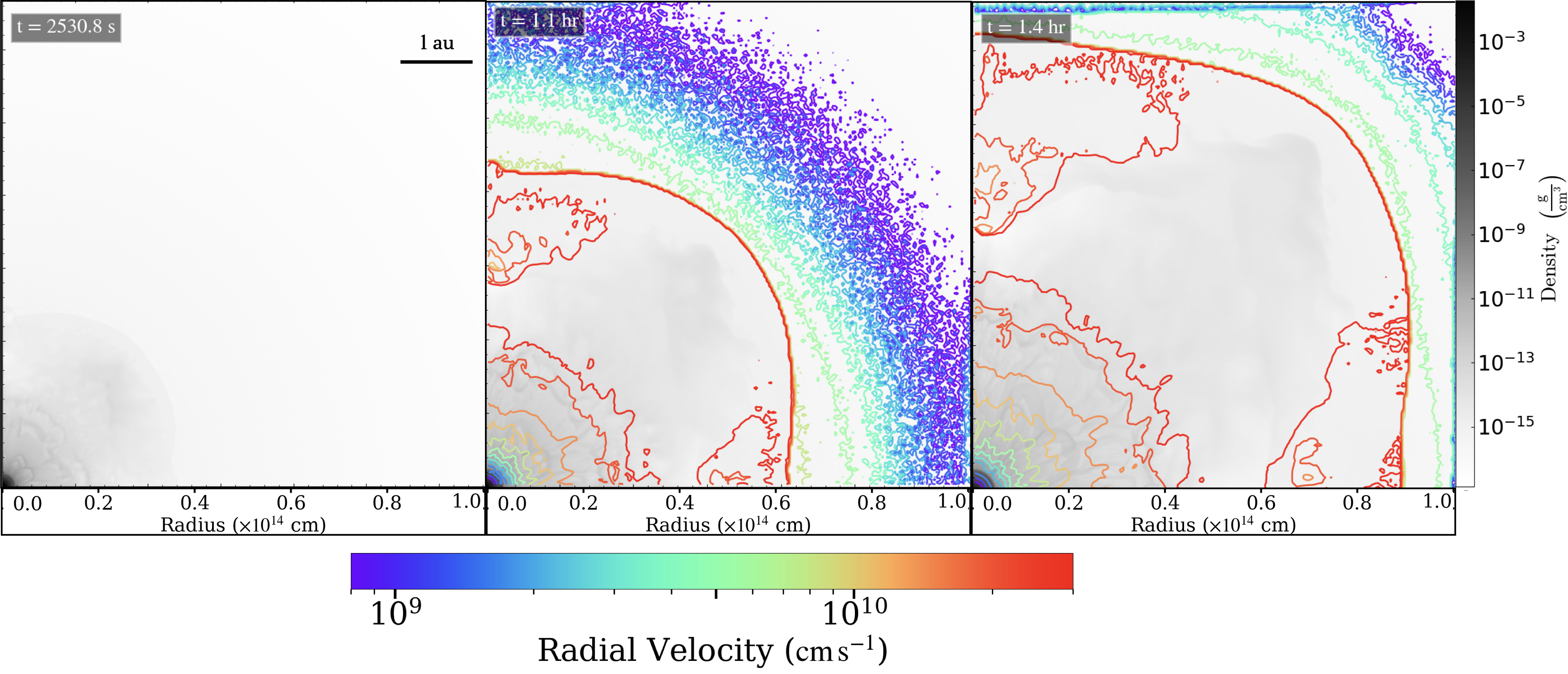}{The density and velocity evolution of Model \texttt{T}. The left panel is plotted without velocity contours to show the density structure clearly. The density and velocity distribution remains spherical around the shock front, However, the post-shock CSM becomes much turbulent.}{1.0}
\fig{LCAngle}{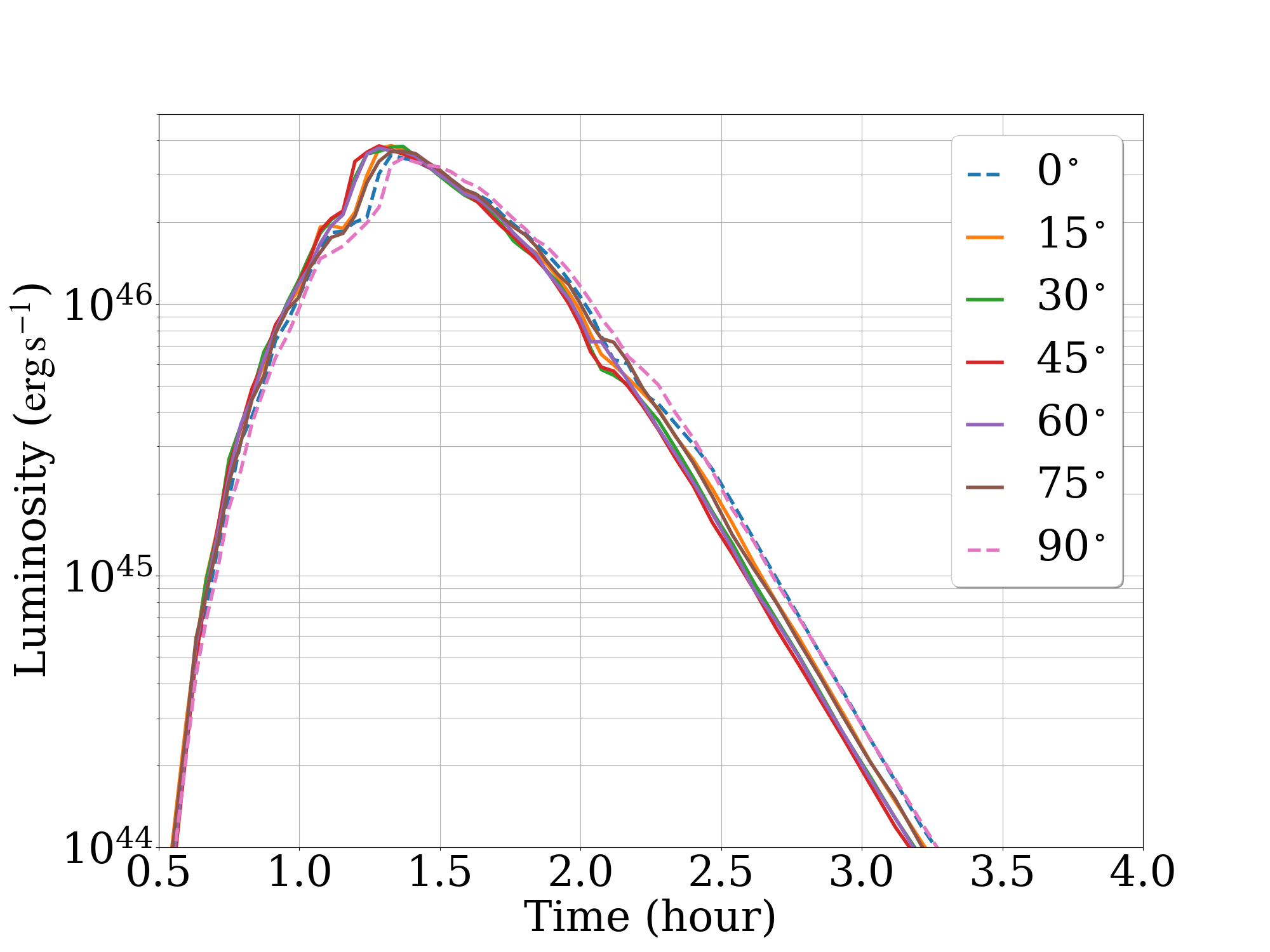}{RO LCs for Model \texttt{T} with different viewing angles. The dotted lines are for $0^{\circ}$ and $90^{\circ}$ for differentiating from other angles, and their peak luminosities are lower.}{1.0}
\subsection{Gas Dynamics}

\Fig{2D325} shows the evolution of the shock breakout with the CSM from the steady wind. During the breakout, the forward shock rapidly accelerates to a high velocity $v_r\sim \times 10^{10}\cms$, forming a blob of high-speed gas.  
When the shock travels roughly spherically to $r\sim 3.5\times 10^{13}$ cm, the initial perturbations start to grow within both the shock front ($r\sim 3.5\times 10^{13}$ cm) and the reverse shock ($r\sim 3\times 10^{13}$ cm) regions. The thickness of the radiating regions is affected by clumpy structures that can increase the radiation flux leakage and deviate the duration of breakout emission at different viewing angles.  
The middle panel in \Fig{2D325} shows a density bulb moving at high velocity because the shock first breaks out of the stellar surface around $\theta \sim 45 ^\circ$ ahead of the mixing zone. This occurs due to Rayleigh-Taylor (RT) instabilities that evolve from perturbations when the gas in the shocked region becomes unstable against the RT instability, characterized by different signs of density and pressure gradients \citep{1882Rayleigh,1950Taylor,1976ApJ...207..872C}. RT instability is less evident in Model \texttt{T} of zero perturbations. The RT instability driven by the formation of the reverse shock continues to enhance post-breakout mixing.

To quantitatively describe the gas dynamics of shock breakout, we plot the 1D velocity and density profiles of different viewing angles in \Fig{2D325_line} \citep[as in][]{sharma_ram_sachdev_1987}. The steep radial velocity jump is the forward shock front. The shock velocity is at $\sim 10^{10} \cms$, slightly varying among different viewing angles. This implies that shock breakouts occur at different times and can cause differences in LCs among viewing angles. Large density fluctuations in the region around $(1-5) \times 10^{13}$ cm imply the clumpy density structure by mixing.  

In \Fig{2D325_line}, small fluctuations ahead of the shock front are fluid instabilities driven by radiation precursors. Because the radiation precursor originates from turbulent regions behind the shock, the pre-shocked regions inherit these fluctuations. 
By comparing the initial density profile with the right panel in \Fig{2D325}, the post-breakout density at $8\times 10^{13}$ cm increases from $1.3\times 10^{-16}\gcc$ to $4\times 10^{-14}\gcc$. Nevertheless, only $2\times 10^{-4}\Ms$ is ejected from the star during the first hour of shock breakout. 
\fig{2D325}{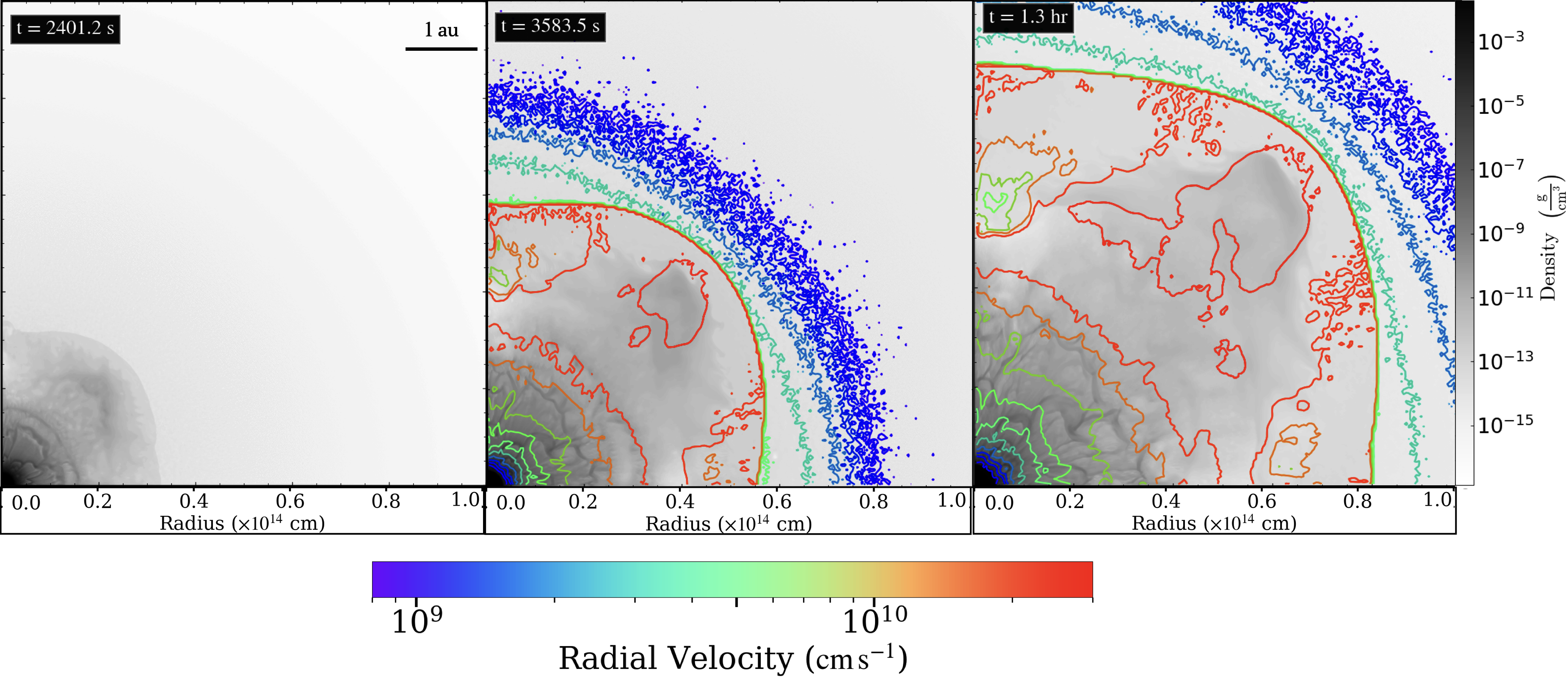}{
The density and velocity evolution of Model \texttt{P}. Panels show the evolution of SN and CSM mixing as the shock propagates outward. The shock velocity reaches $>10^{10} \cms$ and the post-shock region becomes highly turbulent.}{1.000}
\sfig{2D325_line}{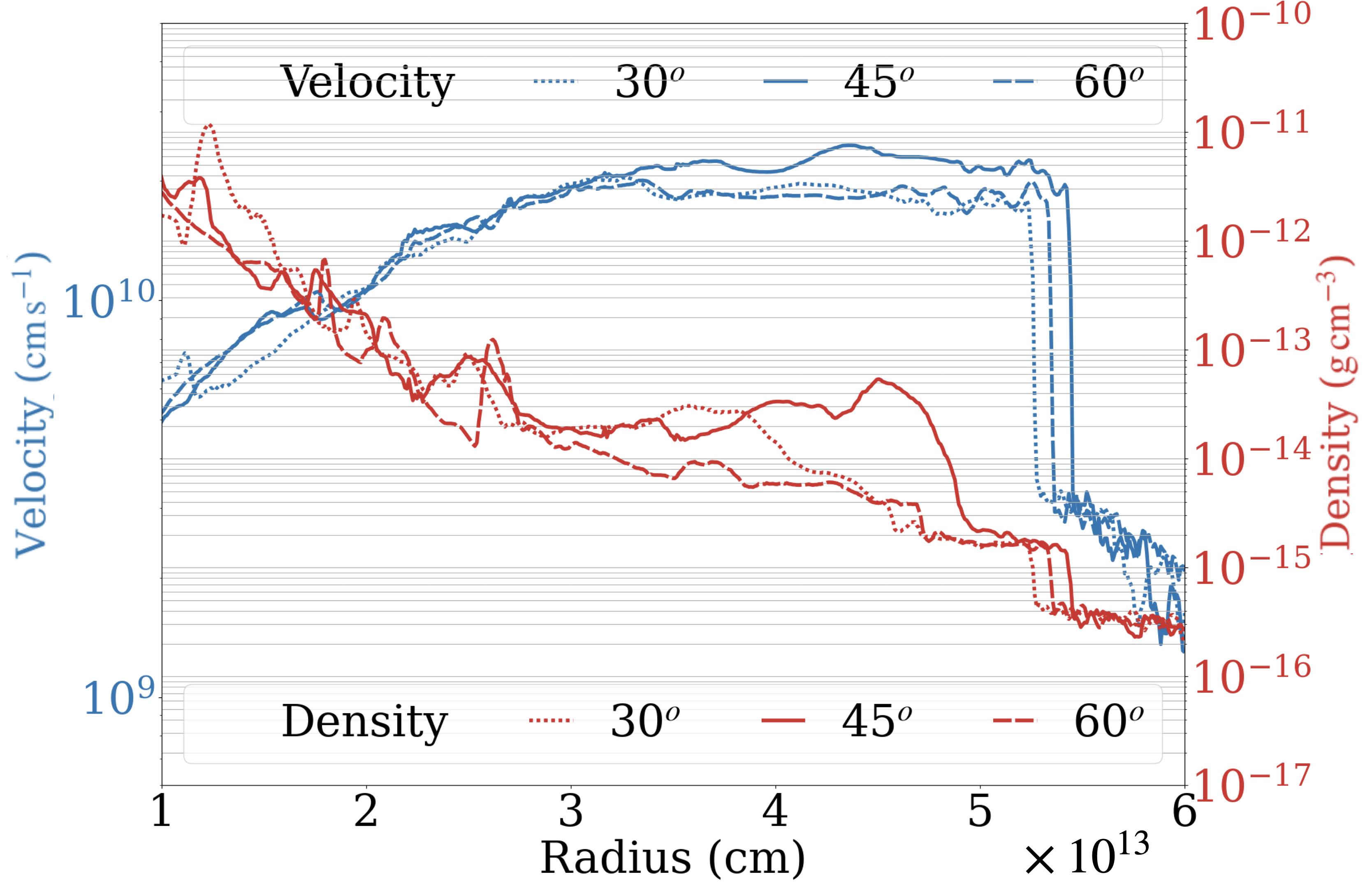}{1D velocity and density profiles at three different viewing angles for Model \texttt{P}. The shock front is shown as the velocity jumps at $\sim 5.3 \times 10^{13}$ cm. The density profiles vary across different viewing angles due to the mixing. Small fluctuations of density and velocity in regions $> 5.3 \times 10^{13}$ cm are caused by the radiation precursor.}{0.700}
\subsection{Light Curves}

In \Fig{2D325LC} and \Fig{2D325LC2}, we present the multi-color LCs with different viewing angles of Model \texttt{P}. The LC duration is $\approx 1.42\pm 0.04$ hours with a peak luminosity ($3.83\pm0.15) \times 10^{46}\,\ergs$ among viewing angles. The angular-dependent LCs are essential in capturing the emission characteristics of shock breakout in multidimensional simulations. During the rising phase of LC, the shock front initially has spherical symmetry, and its emission is roughly homogeneous and isotropic. However, the emission is shortly affected by the mixing of post-shock CSM where the photosphere is located. As time evolves, the mixing becomes large enough to result in angle-dependent LCs. Meanwhile, mixing also enhances the fluctuations of LC for each viewing angle at a late time, as shown in the radial profiles \Fig{2D325_line} and post-breakout LCs in \Fig{2D325LC}. The non-uniform and non-radial distribution of radiation flux vectors is shown in \Fig{325emission}.

\Fig{2D325LC2} shows multi-color LCs with the evolution of the photospheric temperatures of the shock breakout. The rising time of each color deviates slightly. The luminosity for each color is sensitive to the radiation temperature. A sudden increase in radiation temperature occurs at the rising phase of the luminosity. The timescale for the increase in radiation temperature is short compared to the breakout timescale. Therefore, the rising LCs are similar in \Fig{2D325LC2} among different colors. Besides the color evolution, the total luminosity increases during the initial shock breakout. The peak luminosity of bands I to VIII is as follows: $6.2\times 10^{39},\, 1.7\times 10^{41},\, 2.0\times 10^{42},\, 1.4\times 10^{43},\, 6.3\times 10^{43},\, 1.3\times 10^{45},\, 4.2\times 10^{45},\, \mathrm{and}\, 4.0\times 10^{46}$ \ergs, respectively. Their LC durations are 1.4, 1.3, 2.1, 1.8, 3.3, 2.0, 1.7, and 1.4 hours, respectively. Although all color LCs have a similar shape during the rising phase, longer wavelengths are more sensitive to viewing angles.

Post-breakout cooling becomes prominent after the LCs reach their peaks, and the luminosity then drops. For band VIII, the decline rate of luminosity is $\dot{L}=4\,\mathrm{mag}\,\mathrm{hour}^{-1}$. The dominant wavelengths shift towards longer wavelengths, and the decline rate of luminosity for band V is $\dot{L}=1.5\,\mathrm{mag}\,\mathrm{hour}^{-1}$, reflecting the expansion of the photosphere with a decreasing radiation temperature.
\sfig{2D325LC}{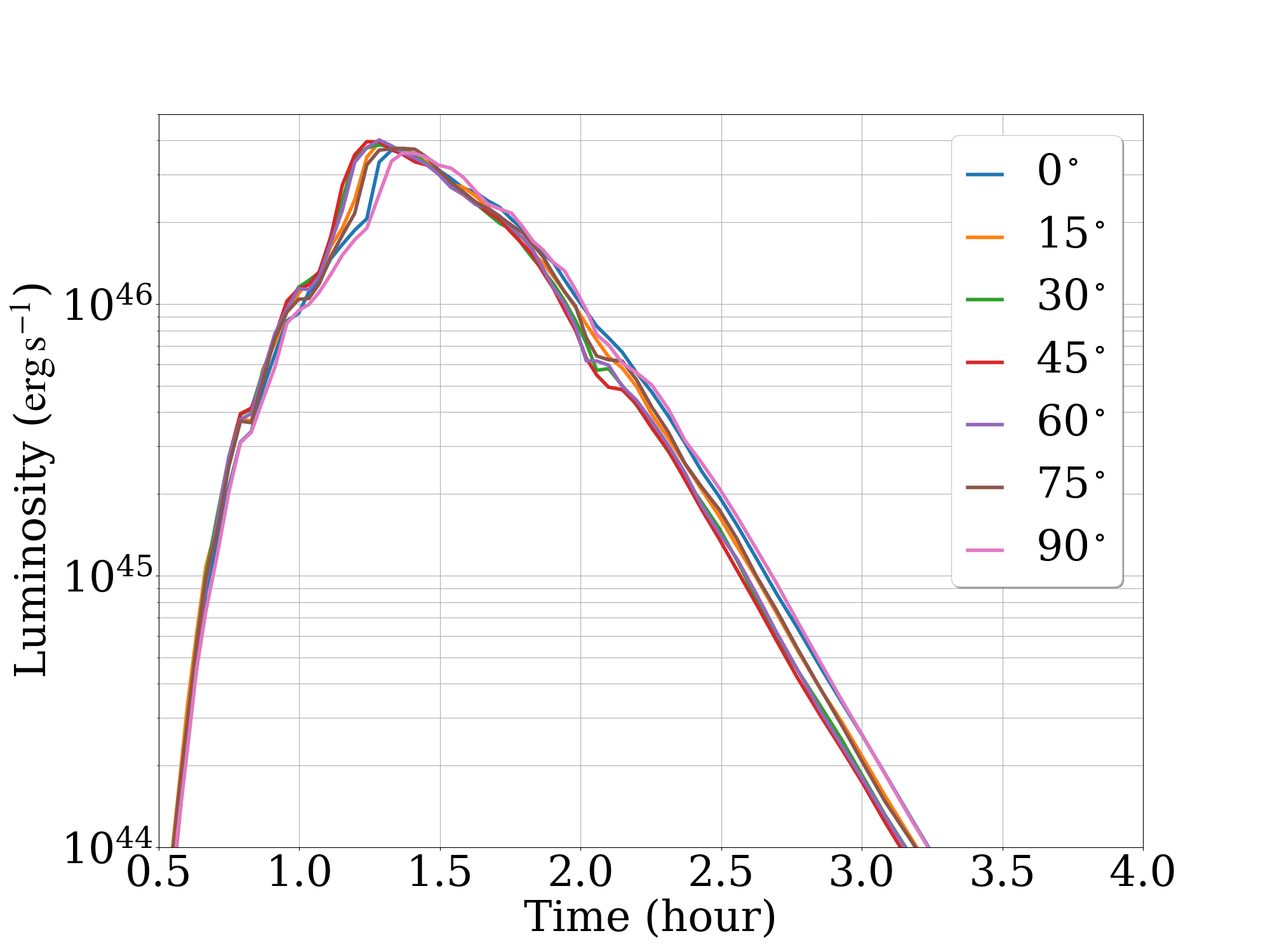}{ LCs from different viewing angles for band VIII in Model \texttt{P}. The LCs show small deviations during the rising phase, and the deviations slightly increase around the peak. }{0.700}
\sfig{325emission}{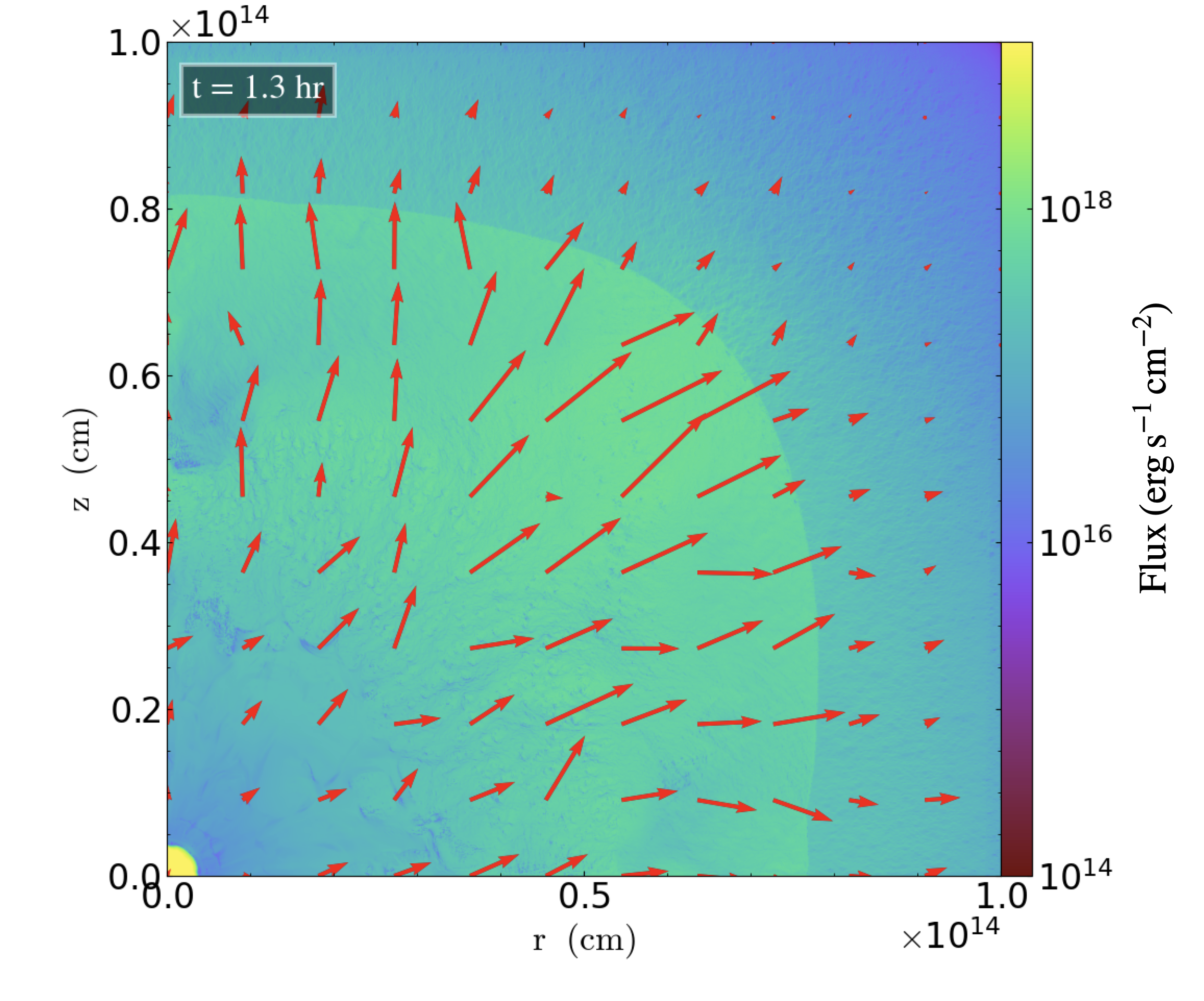}{The distribution of radiation flux vectors of band VIII in Model \texttt{P}. The flux magnitude is maximum at the emitting region between the shock and the photosphere. Due to mixing, the radiation flux vectors are misaligned with radial directions, and their magnitude is also non-uniform.}{0.7}
\sfig{2D325LC2}{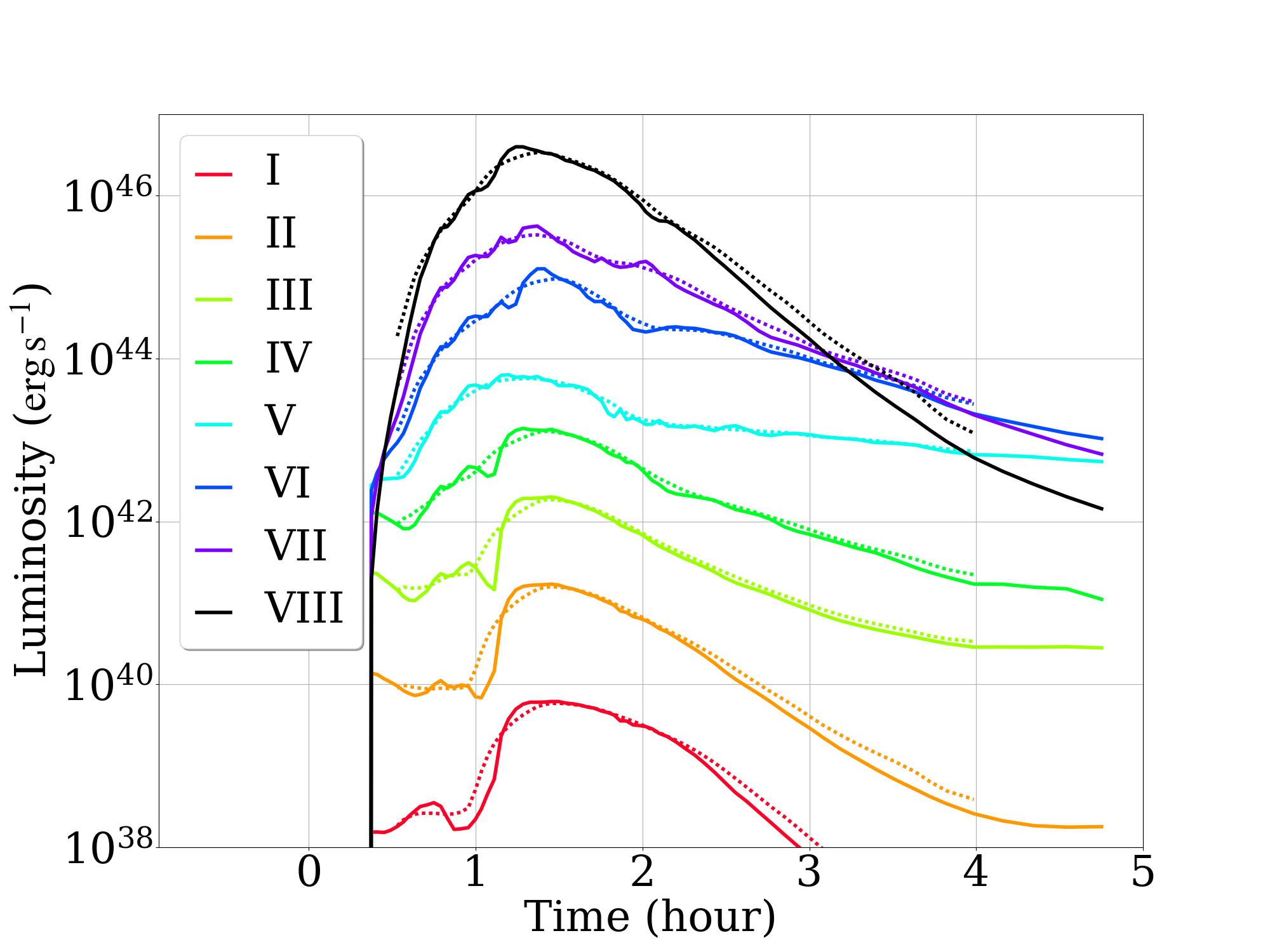}{Color LCs of Model \texttt{P} at the viewing angle of $45^{\circ}$, the dotted curves are smoothed by data binning to better identify the peak for longer wavelengths. The peak luminosity for the band VIII is $4\times10^{46}\ergs$, and the corresponding duration is $1.4$ hour. The dominant band VIII transits to lower frequency bands VII and VI at $\sim 3$ hours due to the cooling of the photosphere. Larger fluctuations appear around the peak, followed by bumpy features at lower wavelengths. }{0.700} 

\subsection{Effect of Steady Wind Strength}
\label{subsec:amplitude}
To examine the observational signatures of the shock breakout associated with the CSM formed prior to the explosion, we modify the density profiles by varying the mass loss rate as $\alpha\dot{M}$, where $\alpha=0.1,\, 0.2,\, 0.5,\, 1.0,\, 2.0,\, 5.0,\,\mathrm{and}\,10$, spanning two orders of magnitude as outlined in \Sect{NM}. The correlation is depicted in \Fig{MLR}. We observe that a higher $\alpha$ with a denser and more extensive CSM prolongs the LC duration and decreases its peak luminosity. Higher mass loss broadens the shock structures, resulting in longer LC durations and lower peak luminosities, as shown in \Fig{MLRdensity}. We also derive a fitting formula of the peak luminosity as a function of $\alpha$, $L\,= \alpha^{-0.16}\times 10^{46.7}$ \ergs, $\tau_{\rm shock}=0.17\, \log{\alpha}+1.48$ hours, where $\tau_{\rm shock}$ is the LC duration, and $t_{\rm peak}=0.1\, \log{\alpha}+1.17$ hours is the rising time to peak luminosity. When $\alpha$ increases, the breakout is delayed, the LC duration increases, and peak luminosity decreases. As discussed in the previous 1D studies on the RSG shock breakout \citep{delay2018}, the presence of the dense CSM can delay the breakout and extend the photon diffusion time, which increases the LC duration and delays the emergence of shock breakout signals. In our models, a higher mass loss rate ($\alpha$) increases CSM density and reduces luminosity due to higher optical depth. Comparing the density distributions at the shock breakout for the three mass loss rates in \Fig{MLRdensity}, higher $\alpha$ would lead to a smoother density distribution ahead of the shock front with larger photosphere radii and longer LC duration.
Considering the 2D effects and the relationship between shock breakout signals and CSM density in this study, we suggest that estimations of CSM density based solely on previous 1D calculations may be incorrect. Such estimations could either underestimate the LC duration or overestimate the density of the CSM. Therefore, our research demonstrates the importance of counting multi-dimensional effects when evaluating the properties of gas dynamics and radiative emission of shock breakout with a dense CSM.
\sfig{MLR}{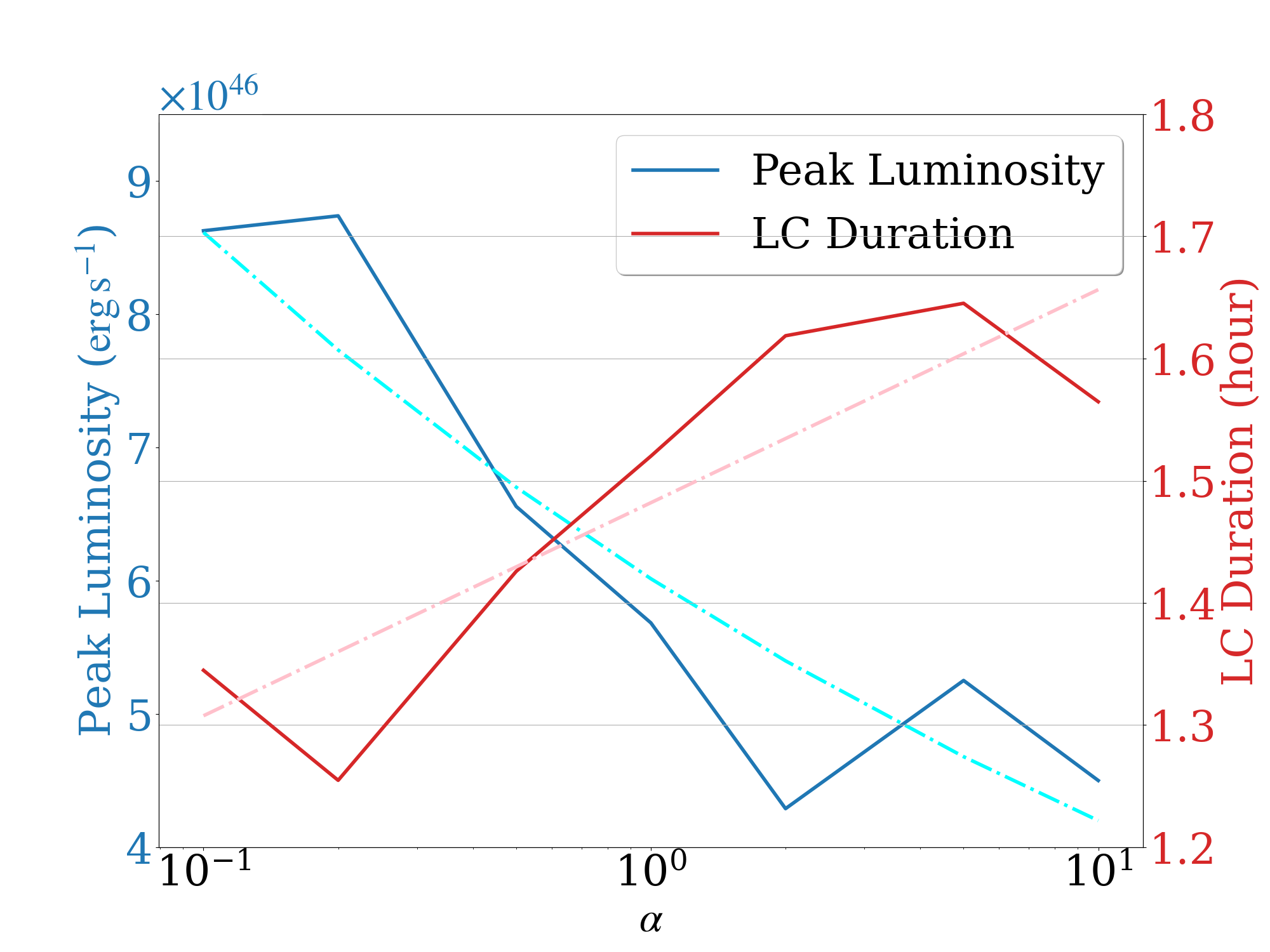}{The peak luminosity and LC duration as a function of the mass loss rate parameter (\texttt{P};  $\alpha=0.1,\, 0.2,\, 0.5,\, 1.0,\, 2.0,\, 5.0,\,\mathrm{and}\,10$). The dashed fitting curves are for the correlation between $\alpha$ and LC duration (pink) and peak luminosity (cyan). Higher values of $\alpha$ correspond to denser CSM profiles. We find that LC duration positively correlates with $\alpha$: $\tau_{\rm shock}=0.17\,\log{\alpha}+1.48$ hours. A dilute CSM tends to produce a higher peak luminosity with a shorter duration.}{0.700}
\sfig{MLRdensity}{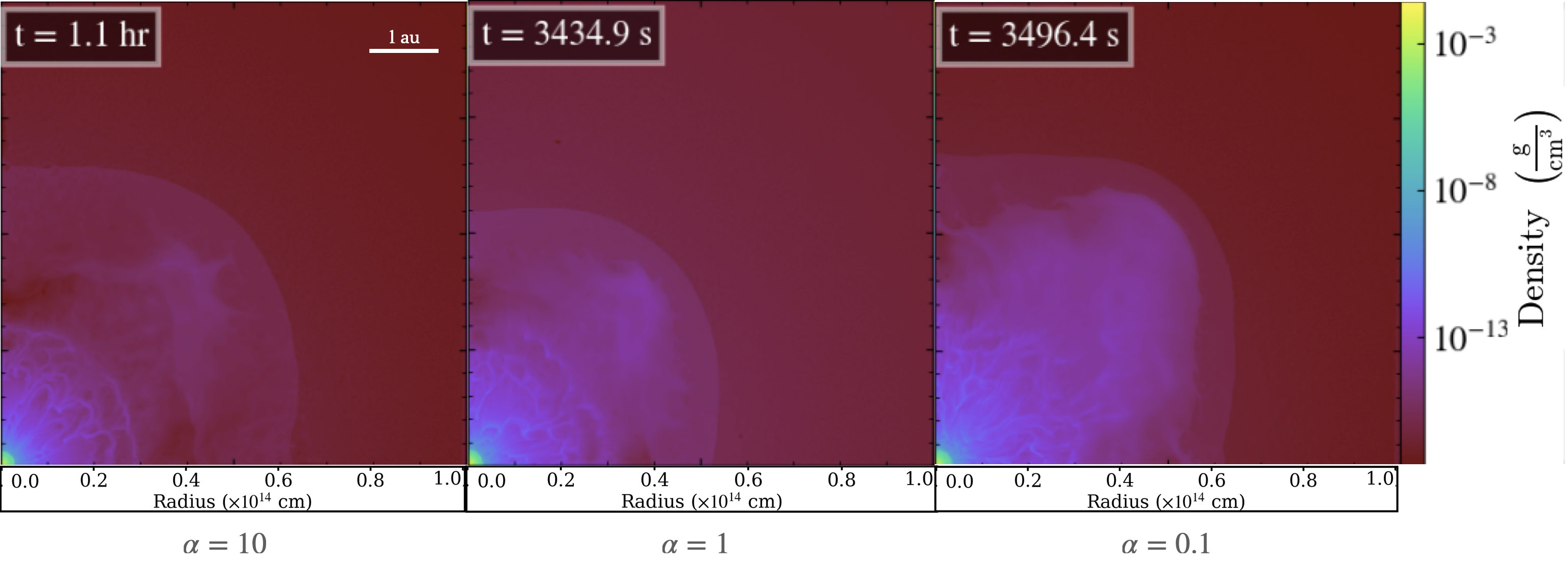}{Density snapshots from Model \texttt{P} with $\alpha=0.1,\, 1.0,\,\mathrm{and}\,10$. These snapshots are taken when the shock front arrives at $\sim 6\times 10^{13}$ cm. Higher $\alpha$ models form denser CSM profiles, and their breakout emission is delayed, but the shock front can still propagate to a comparable distance regardless $\alpha$ values.}{1.00}
\section{Shock Breakout with a ring-like CSM}
\label{sec:clum}
\subsection{Gas Dynamics}

We present the evolution of the shock breakout within the ring-like CSM (ring) based on Model \texttt{R} in \Fig{gas_ring}. In the left panel of \Fig{gas_ring}, the shock front travels to $4\times10^{13}$ cm, with radial velocity of $\sim 2\times 10^{10}\cms$. The thickness between the shock front and the reverse shock is $5\times 10^{12}$ cm. At this moment, the ring does not influence the shock propagation. In the middle panel, the collision of SN shock creates an oblique shock that wraps around the ring. The shock morphology is deformed during the collision, a strong reverse shock forms and propagates back to the ejecta and drives a vigorous mixing, as shown by the filamentary structures in the left-bottom corner of the middle panel in \Fig{gas_ring}. When the shock approaches the boundary of the simulation box, the ring is destroyed, and its debris starts to follow the forward shock at a velocity of $2\times 10^{10}\,\cms$.

\fig{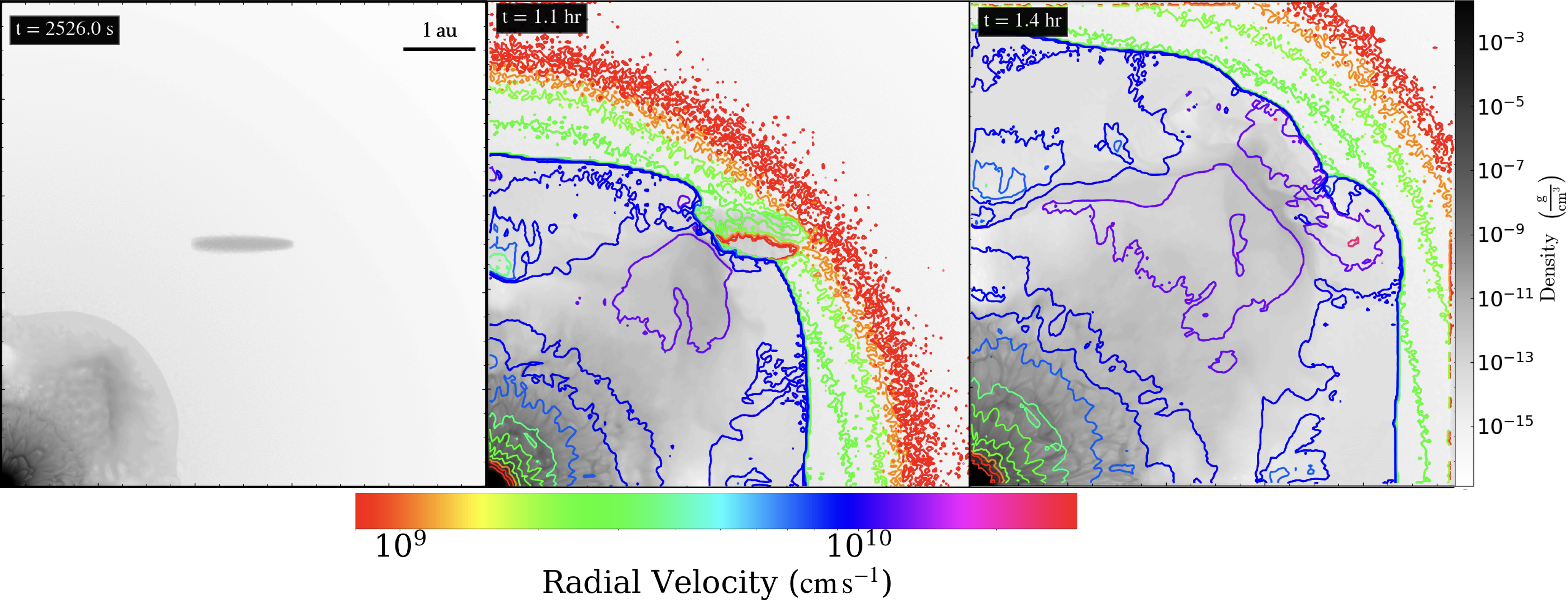}{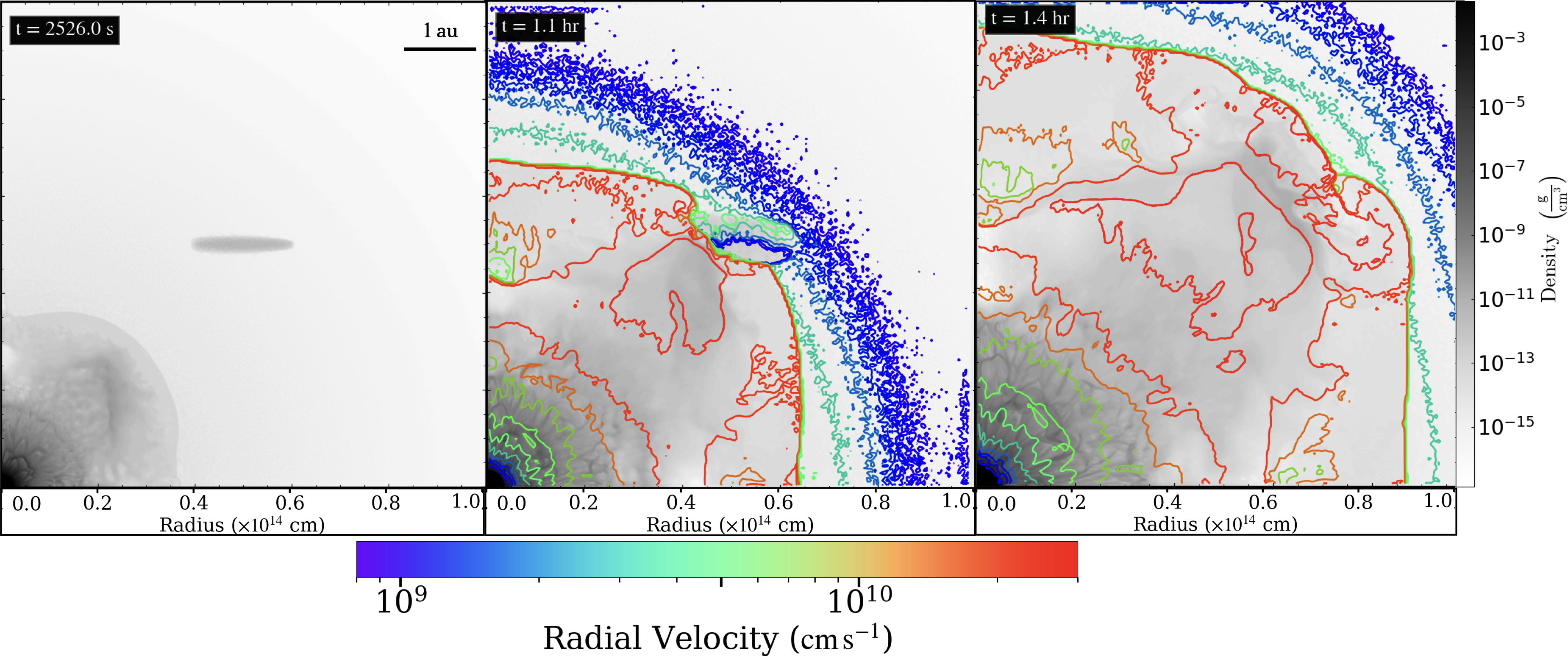}{
The density and velocity evolution of Model \texttt{R}. The middle panel reveals strong mixing around the reverse shock and shock front layer around the collision site of $r \sim (4-7)\times 10^{13}$ cm. Finally, the ring-like CSM is completely destroyed by the shock, as shown in the right panel.}{1.00}

To better evaluate the gas dynamics, we present the 1D density and velocity profiles of the right panel of \Fig{gas_ring} in \Fig{line_ring}. Due to the hindrance of the ring at $45^{\circ}$, the forward shock along $45^{\circ}$ travels to $8\times10^{13}$ cm, shorter than those in the directions of $30^{\circ}$ and $60^{\circ}$. The impinging gas flow passing around the shocked ring develop Kelvin-Helmholtz (KH) \citep{1868helmholtz, 1871Kelvin}, and Richtmyer–Meshkov (RM) instabilities \citep{1960Richtmyer,1972Meshkov}.

\sfig{line_ring}{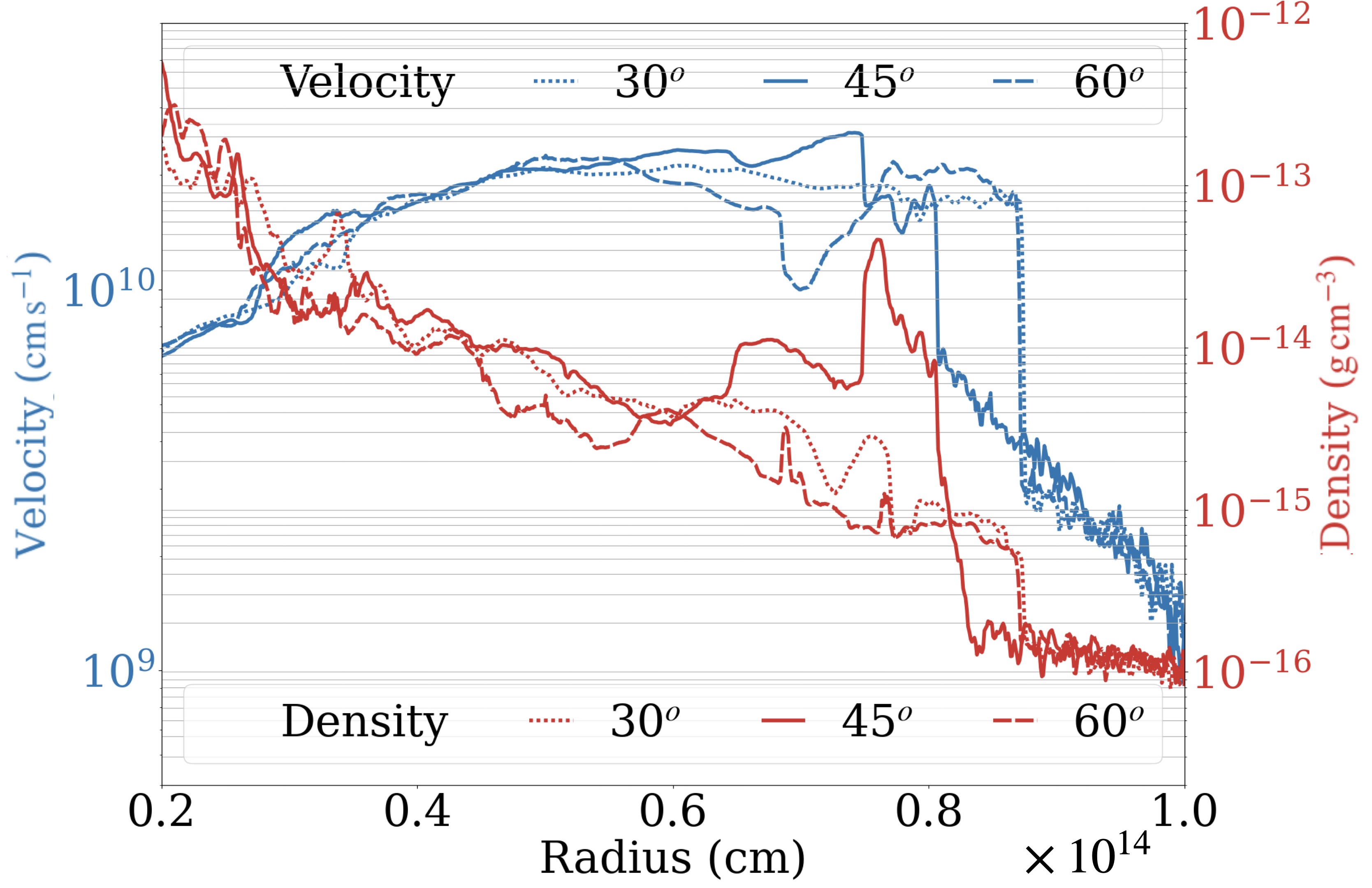}{The velocity and density profile from three different viewing angles in Model \texttt{R}. The velocity jumps at distances around $8.7\times 10^{13}$ cm and $8\times 10^{13}$ cm are the shock fronts. For the velocity profile of $45^{\circ}$, A reverse shock forms at $7.5\times10^{13}$ cm from the collision of the ring.}{0.7}

\subsection{Light Curves}
We present LCs from different viewing angles in \Fig{LC1_ring} and the corresponding multi-color LCs at a viewing angle of $45^{\circ}$ in \Fig{LC2_ring}. Based on \Fig{LC1_ring}, the LC duration is $\approx 1.45\pm0.06$ hour with a peak luminosity of $\approx (3.50\pm0.28) \times 10^{46}\,\ergs$ among viewing angles. In the rising phase of \Fig{LC1_ring}, the radiation precursor interacts with the ring first and shows luminosity fluctuations before breakout. During the breakout phase, the ring attenuates the emission of band VIII between $30^{\circ}$ and $60^{\circ}$ around $1.3$ hours, which leads to a significant deviation in LCs among viewing angles. The shock collides with the ring at $\sim 1.5$ hours that deforms the shock front and drives strong mixing, as shown in \Fig{line_ring}. This collision results in large luminosity fluctuations at approximately $1.5$ and $2$ hours in LCs for all viewing angles in \Fig{LC1_ring}. These fluctuations can be seen in bands V and VI around $1.8$ hours and in band VII around $2.0$ hours in \Fig{LC2_ring}. By comparing the RO LCs with smoothed LCs from data binning, the luminosity fluctuation is $\sim 18\%$ before the peak. After the shock collides with the ring, the luminosity fluctuation increases to $\sim 30\%$ for band VIII and $\sim 52\%$ for band VI, showing that the post-shock mixing regions significantly affect the emission of longer wavelengths. Although the ring fails to survive after the shock collision, the multi-color LCs of shock breakout still shed light on the pre-explosion environment with a ring or any confined dense CSM.
\sfig{LC1_ring}{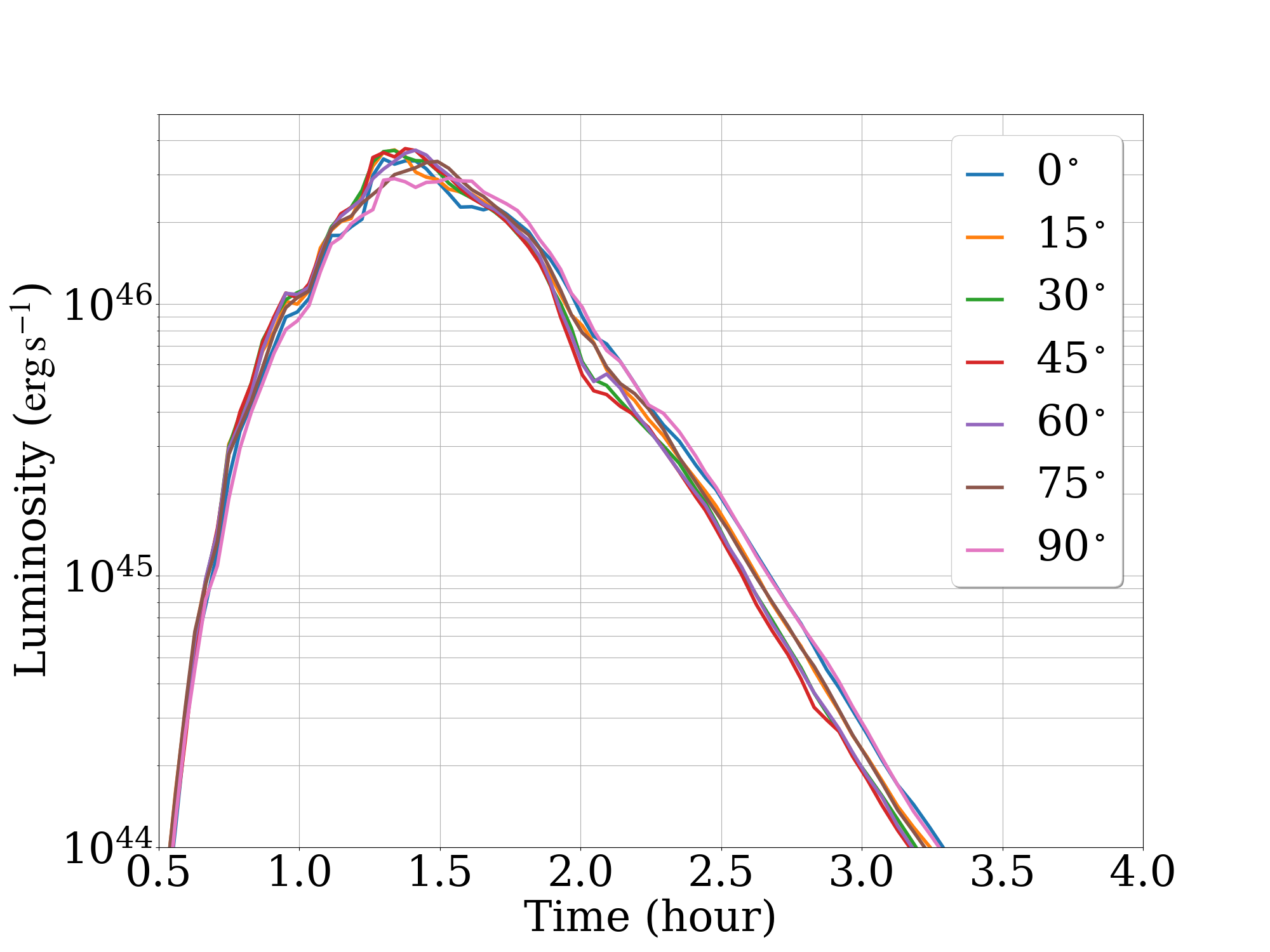}{
LCs of band VIII in Model \texttt{R}. The LC duration is $\sim1.45$ hours with peak luminosity at $\sim 3.5 \times 10^{46}\,\ergs$. The LCs fluctuate due to radiation precursor interaction with the ring.}{0.700}
\sfig{LC2_ring}{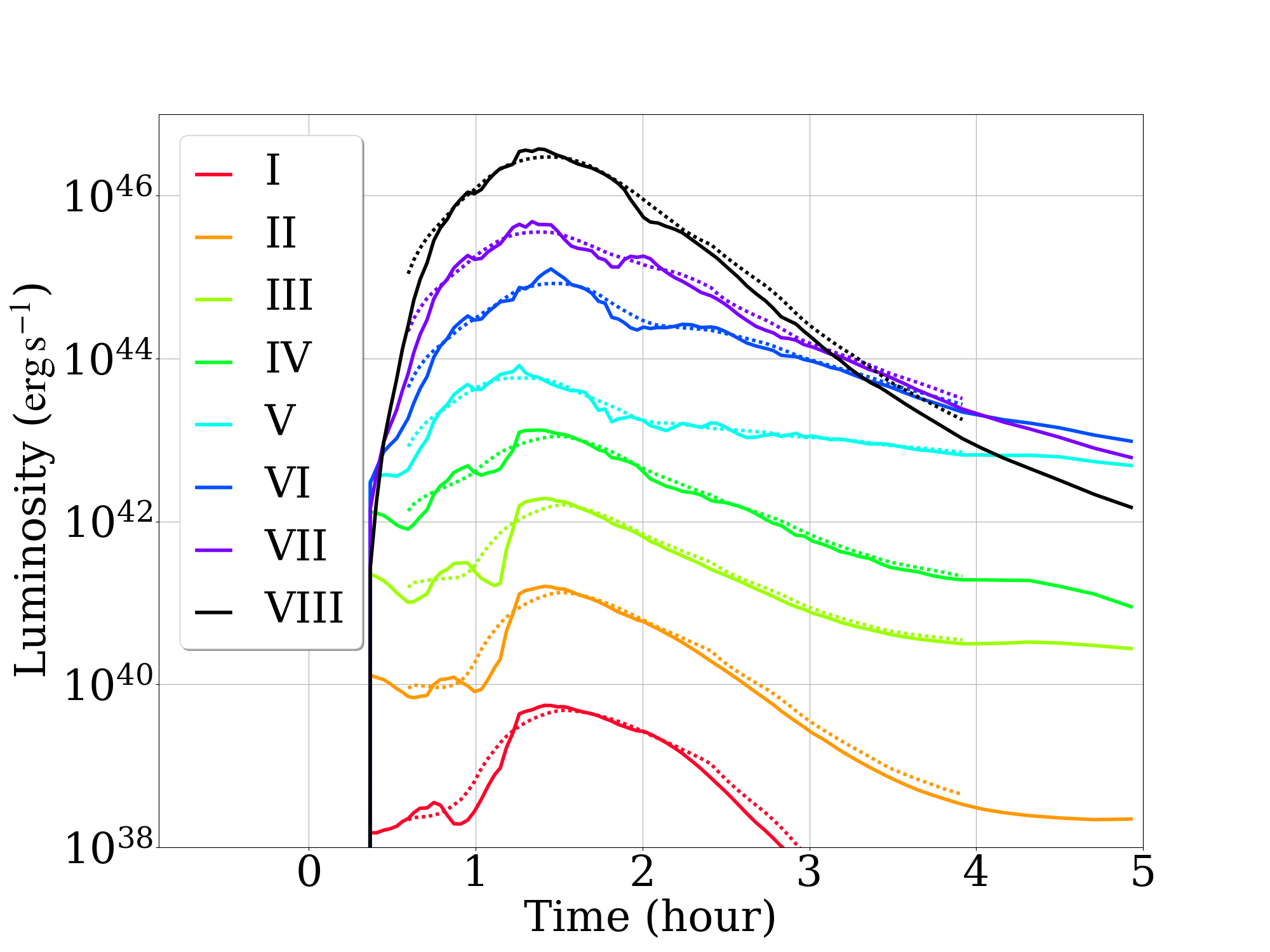}{
Color LCs of Model \texttt{R}, the dotted curves are smoothed LCs by data binning to better locate the peak. Before the luminosity peaks, the radiation precursor gradually heats the ring and creates small bumps from band I at 0.7 hours to around 1.1 hours for bands IV, V, and VI. During the breakout phase, the ring absorbs the short wavelength radiation and re-emits it into longer wavelengths. The dominant band of emission transits from VIII to VII after $\sim$3 hours of the shock breakout.}{0.700}
In \Fig{LC2_ring}, the peak luminosity of bands I to VIII is: $5.5\times 10^{39},\, 1.6\times 10^{41},\, 1.9\times 10^{42},\, 1.3\times 10^{43},\, 8.2\times 10^{43},\, 1.3\times 10^{45},\, 4.8\times 10^{45},\, \mathrm{and}\, 3.7\times 10^{46}$ \ergs, respectively. Their LC durations are 1.4, 1.3, 2.2, 1.9, 2.9, 2.0, 1.7, and 1.4 hours, respectively. Despite different peaks and durations, the characteristic LC shape is similar for all wavelengths. Due to the cooling of shock and stellar envelope, the dominant photon energy switches from high to low at $3.0$ hours. After the luminosity reaches the peak, it starts to decline at a rate of $\dot{L}=3.5\,\mathrm{mag}\,\mathrm{hour}^{-1}$ for band VIII. For band V, the rate is $\dot{L}=1.5\,\mathrm{mag}\,\mathrm{hour}^{-1}$. 

\section{Shock Breakout with a Companion Star}
\label{sec:bi}
\subsection{Gas Dynamics of Collisions}

We show the density evolution of the shock breakout with a companion star (\texttt{B}) in \Fig{gas_bi}. The collision between the SN shock and the companion star enhances the mixing via fluid instabilities in the contact discontinuity regions, and the surviving companion continues driving the post-shock turbulence through its evaporating envelope.

In the left panel of \Fig{gas_bi}, the shock front passes through the companion star at $2.1\times10^{13}$ cm with a radial velocity of $\sim 10^{9}\,\cms$. Although the shock front is generally spherical, the companion star continues to drive the fluid instabilities. The SN ejecta has traveled to $10^{13}$ cm to the surroundings of the companion star that carves a low-density region along the direction of $45^{\circ}$.

To analyze the interactions between the blast wave and the companion star, we present 1D profiles of different viewing angles at $\sim 1900$ s in \Fig{line_bi}. The forward shock of the blast wave is at $2.1\times10^{13}$ cm, and the reverse shock from the colliding companion is at $\sim$ (1.8--1.9) $\times 10^{13}$ cm. Comparing different angle profiles, a more substantial reverse shock appears in $45^{\circ}$ direction where the companion star is located.

In the middle panel of \Fig{gas_bi}, the shock front passes the companion star and leaves a strong mixing region at $\sim 2300$ s. The robustness of the companion star blocks the gas flow towards its direction. It creates a low-density region beyond the companion star, creating fluid instabilities that turn into turbulence and fill the ambient space in the shadow regions. In the right panel of \Fig{gas_bi}, the shear velocity between the companion star and SN ejecta creates KH instability; the collision also favors the formation of RM instabilities, and they mix the post-shock gas. The companion slightly expands due to the compressible heating by the shock.

We compare the density profiles before and after the shock collision with the companion in \Fig{Bi_ML}. The radius of the companion star expands, and it loses around $25\%$ of its initial mass, which is consistent with the analytic result based on \citep{1975Wheeler}.
\fig{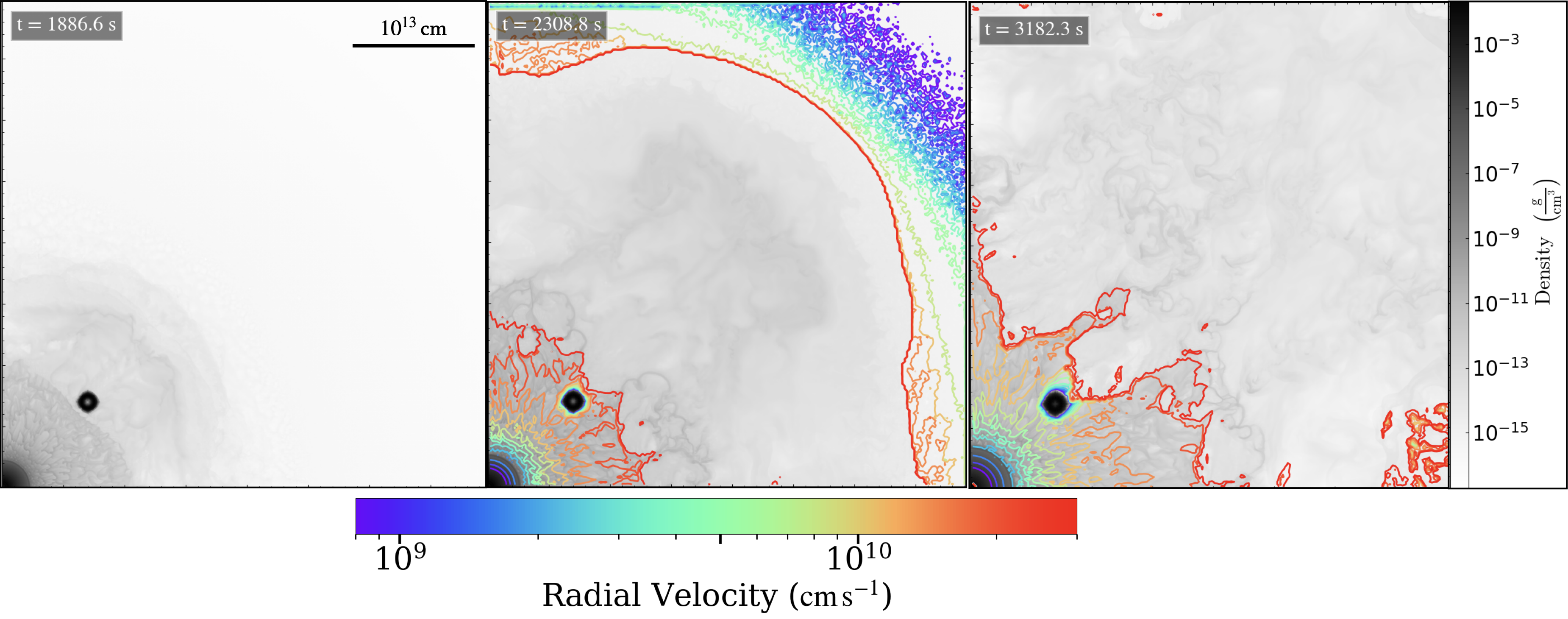}{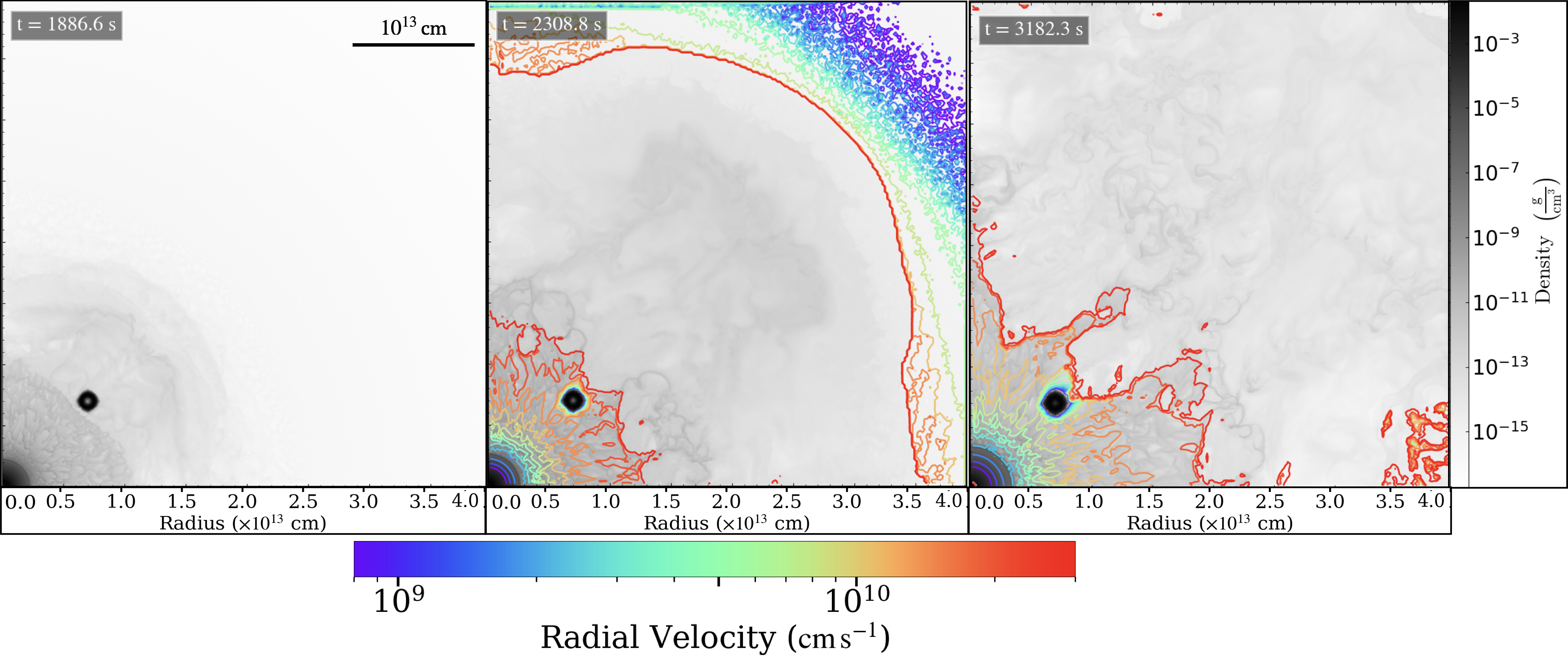}{The density and velocity evolution of Model \texttt{B}. The companion star can hinder the shock and drive the turbulence through KH instabilities when the SN ejecta and shocked CSM passes.}{1.00}
\sfig{line_bi}{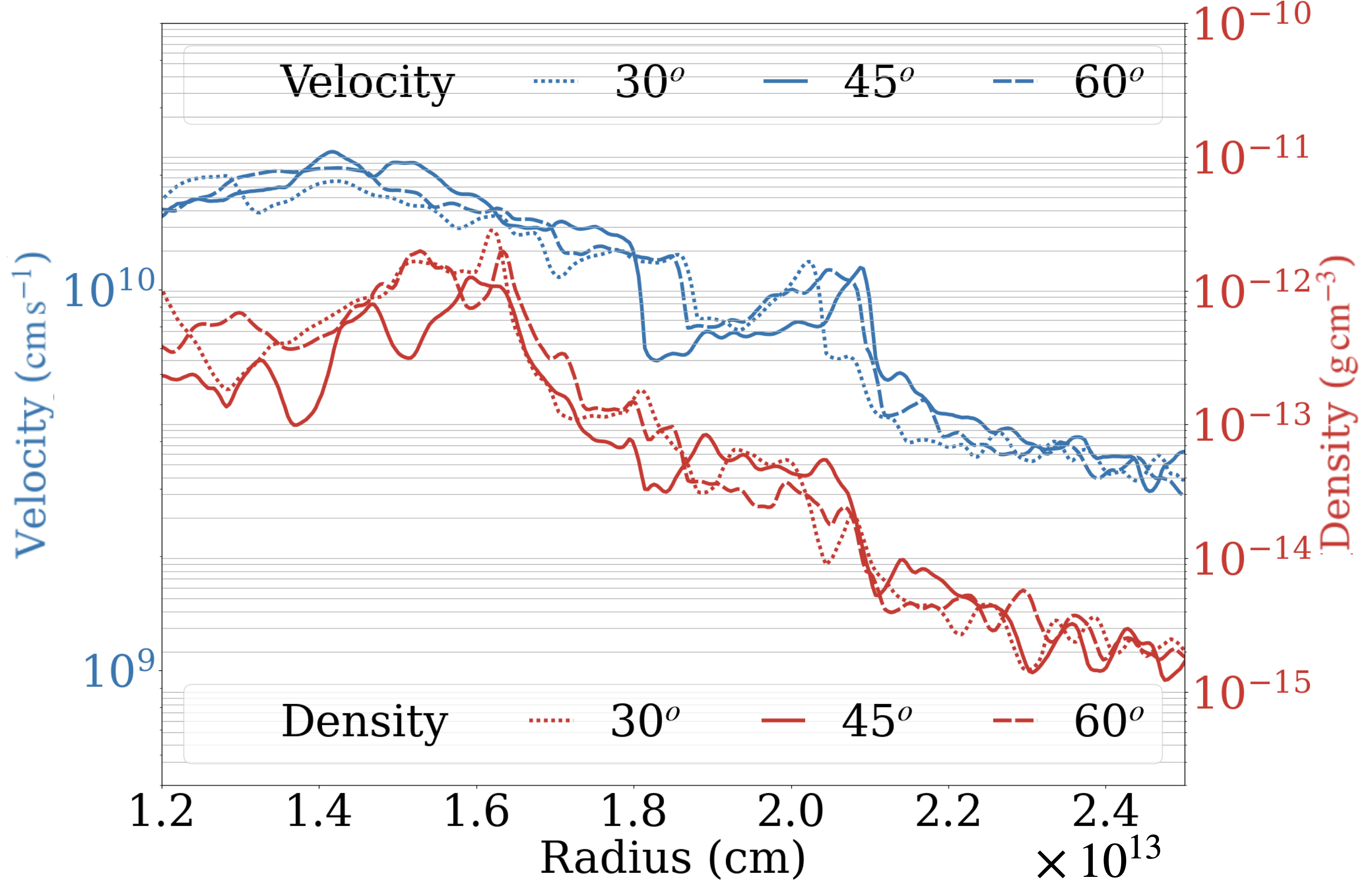}{The velocity and density profiles of Model \texttt{B} at $\sim 1890$ s. At this time, the shock reaches $2.1\times 10^{13}$ cm, and the collision between the shock and the companion forms a strong reverse shock along $45^{\circ}$ at $1.8\times 10^{13}$.}{0.7}
\sfig{Bi_ML}{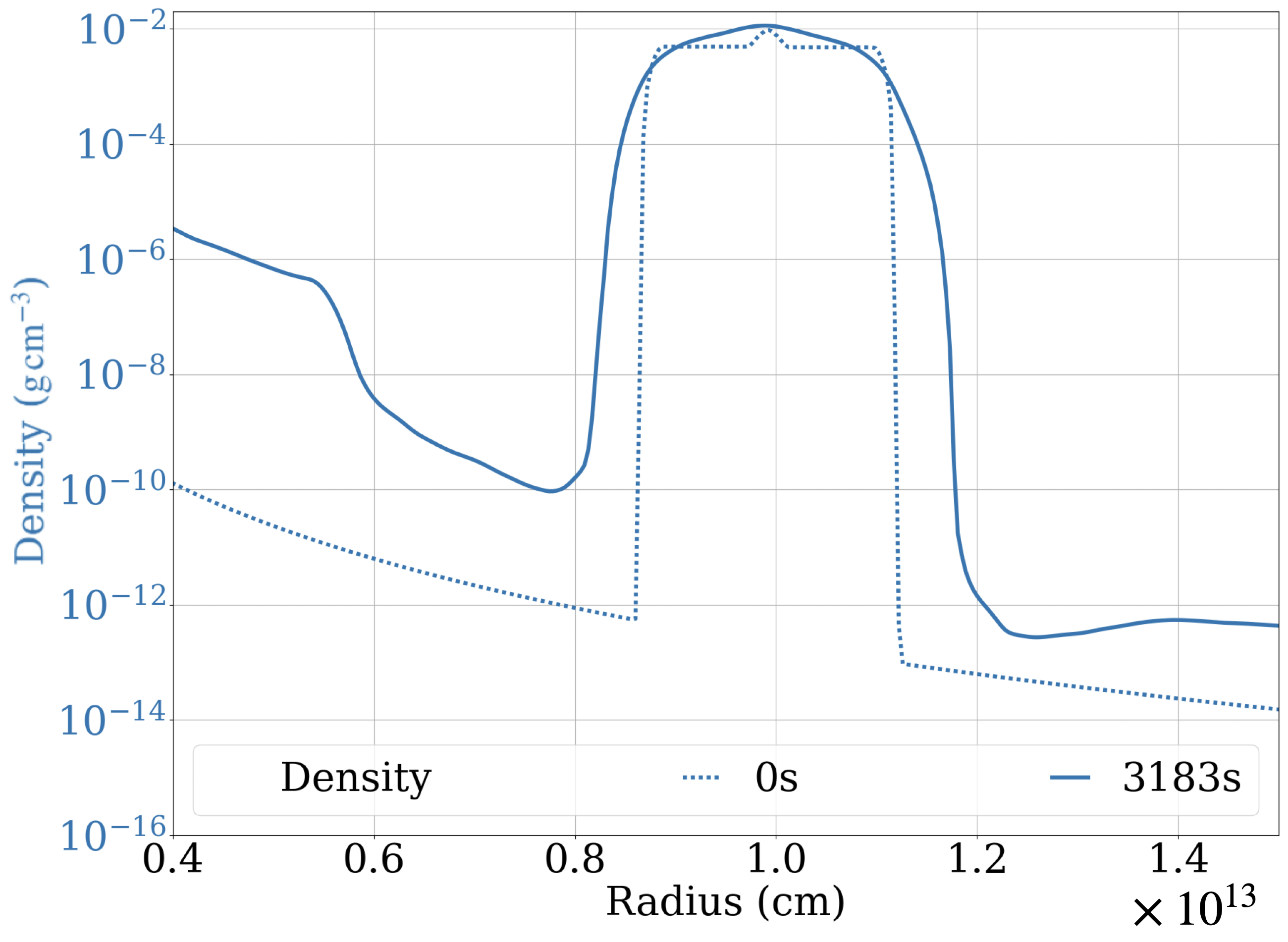}{Density profiles of Model \texttt{B} before and after the shock collision. Due to the shock heating, the companion expands around $25\%$ in size, and the heated envelope of the companion partly escapes through ablation and stripping.}{0.7}

\subsection{Light Curves}
We show the resulting LCs in \Fig{SHt3LC} and \Fig{SHt3LC2}. 
The LC of band VIII has a peak luminosity of ($1.95\pm 0.18) \times 10^{46}\,\ergs$ with an LC duration of $0.84\pm0.04$ hours. Due to the smaller box size of the binary simulations, a smaller collecting sphere for RO LC calculations is applied. We have to compensate the light travel time of $\sim 1580$ s to the binary as moving to the reference point of $8\times 10^{13}$ cm applied for other models.
So we can compare the LCs with other models.

The fluctuations in LCs depend on a blocking effect by the companion star and the shock propagation due to the collision with the companion star. The densities around the companion star become a bit higher after the shock sweeps through it, as shown in the right panel of \Fig{gas_bi}. This higher-density region attenuates the passing radiation and creates smaller rising luminosity at the viewing angles of $30^\circ$ and $60^\circ$. However, photons propagating inside the higher density regions preferentially start to escape to the shadow regions and create an initial sharp rise at $45^{\circ}$. The radiation and shock propagating along $45^{\circ}$ is absorbed by the companion star and drives the mass loss from the companion star. Such mass loss rate is huge, and it can fluctuate LCs significantly and produce multiple minor peaks at $\approx (1.2-1.4)$ hours. The shock collision makes denser regions surrounding the companion star, stripping part of the companion envelope and being disturbed by instabilities, which may later be heated by the photons originally propagating along $45^{\circ}$ to contribute to the secondary peaks.

As seen in \Fig{SHt3LC2}, the peak luminosities for various bands (from I to VIII) are $1.2\times10^{40}$, $1.8\times10^{41}$, $1.4\times10^{42}$, $7.7\times10^{42}$, $5.9\times10^{44}$, $7.4\times10^{44}$, $3.7\times10^{45}$, and $2.7\times10^{46}$ \ergs, respectively. The corresponding LC durations are 0.84, 0.85, 0.85, 0.80, 0.70, 0.80, 0.81, and 0.88 hours, respectively. The overall patterns show higher fluctuations in longer wavelengths, and the luminosity can vary up to $\sim 27\%$ around the peak for band VI. On the other hand, the fluctuation is minimized to $\sim 12\%$ for band VIII.

The peak luminosity appears at $\sim 1$ hour for all wavelengths, while the second peaks occur at $1.24$ hours for band I to band V and 1.26, 1.31, and 1.40 hours for band VI to VIII. Along $45^\circ$, where the companion mostly shadows the ejecta, the secondary peak arises from the envelope heating of the companion star by absorbing high energy radiation flux. For example, around $1.31$ hours in \Fig{SHt3LC2}, we observe the luminosity drop in band VIII and the secondary luminosity spikes among bands IV to VII. 

The companion star continues to stir the system, resulting in a faster luminosity drop rate of $\dot{L}=4.5\,\mathrm{mag}\,\mathrm{hour}^{-1}$ after 0.8 hours of the breakout peak luminosity. The luminosity of band VI increases with $\dot{L}=-1\,\mathrm{mag}\,\mathrm{hour}^{-1}$ after 1.5 hours of the explosion. If the simulation continues, the highest energy band will shift to band VI at around $2.5$ hours because the expansion of the photosphere cools down radiation temperature, similar to Model \texttt{T}, \texttt{P}, and \texttt{R}.
\sfig{SHt3LC}{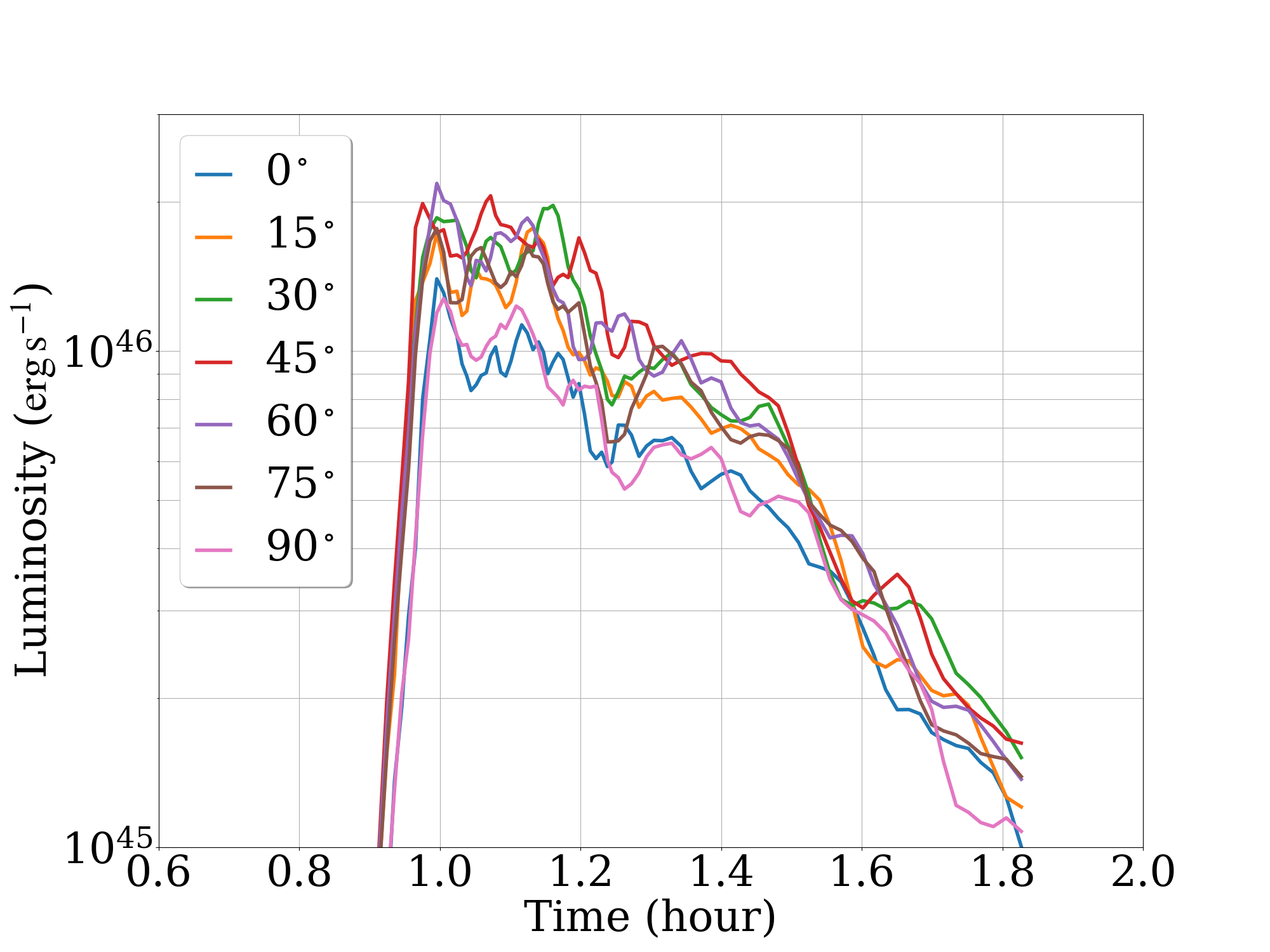}{LCs of band VIII in Model \texttt{B}. Small fluctuations appear in the rising phase, then soon grow into large variations because of the strong KH and RM instabilities driven by the shock collision with the companion star.}{0.700}
\sfig{SHt3LC2}{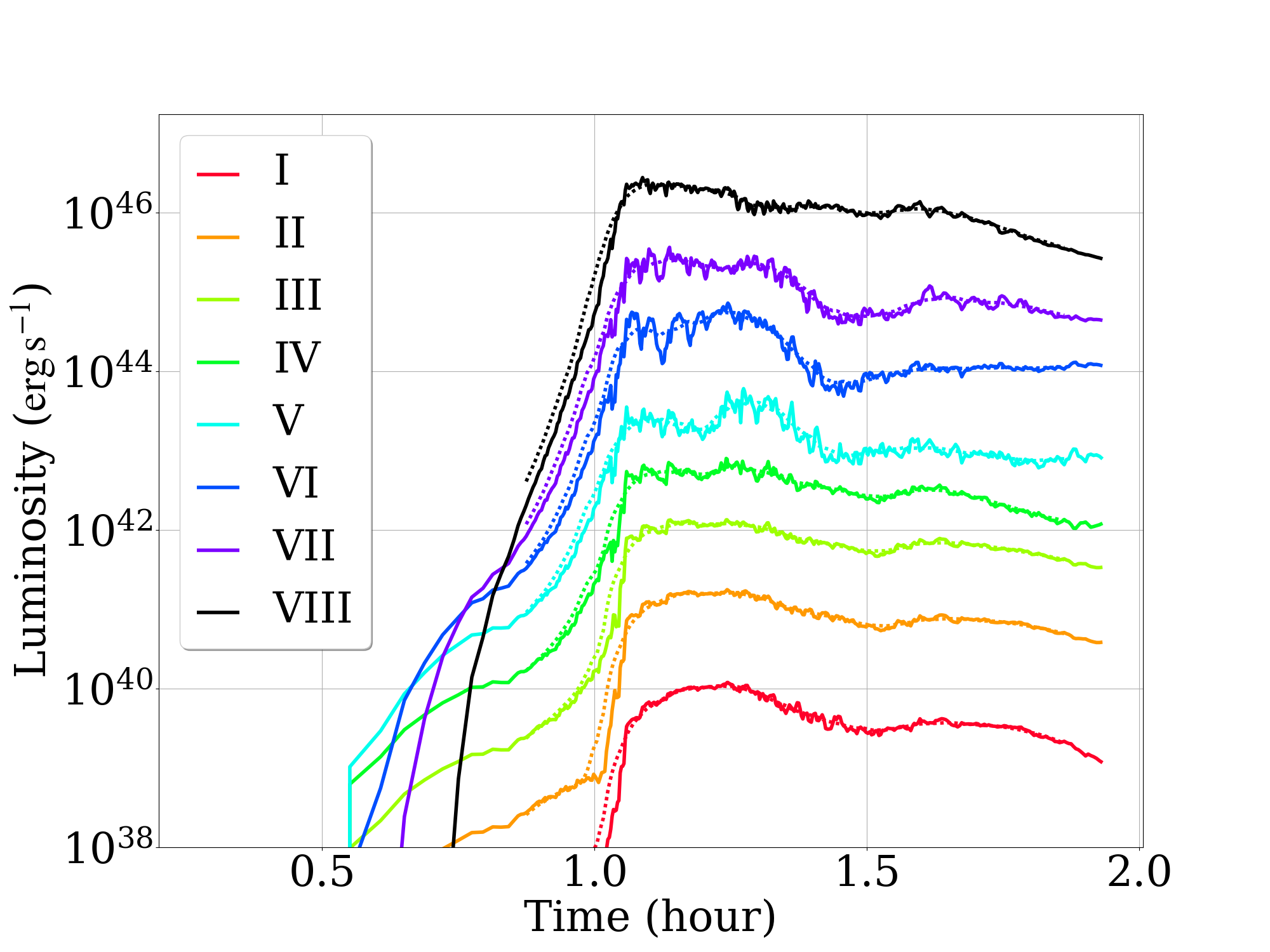}{Color LCs of Model \texttt{B}, the dotted curves are smoothed LCs from data binning. The companion star absorbs and blocks the high energy photons, then releases the emission of longer wavelengths at $1.25$ hours. A double-bump feature appears for longer wavelengths (Band I-IV).}{0.700}
\begin{deluxetable*}{clcccc}
\tabletypesize{\scriptsize}
\tablewidth{0pt}
\tablecaption{LC durations and Peak Luminosities of our Simulations\label{tab:value}}
\tablehead{
\colhead{Dimension} & \colhead{Model} &\colhead{Box Size (cm)} & \colhead{Resolution} & \colhead{FW10\%M (hour)} & \colhead{Peak Luminosity ($10^{46}\,\ergs$)}
} 
\colnumbers
\startdata 
{    1D }&   \texttt{N}$^*$  &    $10^{14}$  &     102400  & 0.20  &1.25 \\
{       }&   \texttt{N}   & $10^{14}$ & 1024 & 0.33  &0.47 \\
\hline
{       }& \texttt{T}       &$10^{14}$&$1024^2$&$1.48\pm0.04$  &$3.69\pm0.14$\\
{       }& \texttt{P}$^*$  &$10^{14}$&$1024^2$&$1.68\pm0.06$  &$3.23\pm0.24$\\
{   2D  }& \texttt{P}      &$10^{14}$&$1024^2$&$1.42\pm0.04$  &$3.83\pm0.15$\\
{       }& \texttt{R}     &$10^{14}$&$1024^2$&$1.45\pm0.06$  &$3.50\pm0.28$\\
{       }& \texttt{B}       &$4\times10^{13}$&$2048^2$& $0.84\pm0.04$ &$1.95\pm 0.18$\\
\enddata
\tablecomments{The mean LC duration and peak luminosity with standard deviation across viewing angles of our models based on band VIII. The detailed model setup can be found in Table \ref{tab:name}. Our 1D Models \texttt{N} agree with the previous 1D results from \cite{2017ApJ...845..103L}. Comparisons between models are discussed in \S \ref{sec:Dis}. }
\end{deluxetable*}

\section{Discussion}
\label{sec:Dis}
\subsection{Multi-Dimensional Effect}

To evaluate the impact of multidimensional effects on the shock breakout of SN~1987A progenitor, we compare all LCs from our simulations in \Fig{ComparisonLC} and summarize their characteristics in Table~\ref{tab:value}. The LCs from our multidimensional simulations have a longer duration than those in the previous 1D calculations \citep{1992ApJ...393..742E,1999ApJ...510..379M,2017ApJ...845..103L}. Previous 1D simulations are known for forming nonphysical thin shells near the forward shock due to limited degrees of freedom and inefficient formation of fluid instabilities, shrinking the LC duration.

As shown in \Fig{ComparisonLC}, the 1D peak occurs earlier and its peak luminosity is lower compared to 2D models. Our 2D LC duration is around three times longer, and the peak luminosity is $3-6$ times brighter than the previous 1D models. The main differences between 1D and 2D LCs are due to the dimensional effects discussed below.

\Fig{1D2Devo} shows the evolution of the density profiles for 1D and 2D models. The figure shows that the forward shock (the region with a steep density jump) in 1D keeps its steepness during the evolution. On the other hand, in 2D, the density profile around the shocked structure is smoothed out over time because the shock front is diffused out partly due to fluid instabilities such as RT instability. Before the initial rise of the luminosity, the gases in shocked regions are unstable against the RT instability. RT instability may dredge up the inner denser material to outer regions and smooth the density profile. The smoothed density profile reduces the optical depth gradients, extends the location of the photosphere where optical depth is $\tau\sim\frac{2}{3}$, and facilitates photon diffusion from the shock breakout. On the other hand, the high optical depth barrier created by the density steepness in 1D traps some of the outgoing photons, and their radiation energy is used to accelerate the gas instead of emission as shown in \Fig{1D2Devo}. As shown in \Fig{EE}, the finger-like structures developed due to RT instabilities evolve into larger structures between the photosphere and the forward shock before the initial rise of the luminosity. The regions for such finger-like structures correspond to the regions of fluid instability between shock and SN ejecta as shown in the upper panels in \Fig{1D2Devo}, where the shock front is at $\sim 2\times 10^{13}$ cm, and the photosphere is at $\sim 10^{13}$ cm).

The LC duration reflects the shock structure and the breakout radius \citep{2011ApJ...727..104C}. In our 2D simulations, the smoothed density profiles extend the shock front and photosphere outward, which increases the diffusion time of radiation, delays the rising time, and extends the LC duration. Furthermore, smoother profiles behind the photosphere in 2D increase the radiation flux compared to 1D, in which radiation is highly trapped, and the energy is converted to the kinetic energy of the gas.

In \Fig{1D2D}, we examine the energetics of 1D and 2D runs at the time when luminosity reaches its peak: $\sim0.3$ hours for 1D and around 1 hour for 2D. By comparing Figures~\ref{fig:1D2Devo} and \ref{fig:1D2D}, we find the density and optical depth profiles in 2D are smoother, and the photosphere in 2D is also larger. From \Fig{1D2D}, the profiles both in 1D and 2D exhibit distinct features in the gradients inside the original stellar radius, $r \leq 3.2 \times 10^{12}$ cm. As seen in the middle panel, the radiation flux in 1D is trapped inside the radius and has an $r^{-2}$ tail.  The 1D gas energy density, including kinetic energy, thermal energy, and gravitational energy, is shown at the bottom of \Fig{1D2D}. The gas energy becomes higher within the original stellar radius, which implies that part of the trapped radiation energy is converted to the gas energy in 1D. To compare the energy density profiles at the peak luminosity between 1D and 2D, we define a normalized radius to the photosphere as
\begin{equation}
\xi\equiv \frac{r}{r_{\mathrm{ph}}}-1,
\end{equation}
where $r_{\mathrm{ph}}$ is the photosphere radius. \Fig{1D2Drat} shows the gas and radiation energy density across the photosphere in 2D that have shallower gradients compared with 1D. It suggests that photon diffusion is more efficient in 2D and creates a stronger radiation precursor. The inner regions in 1D have larger radiation and gas energy than that in 2D. This implies the formation of a thin shell that traps outgoing photons. In the bottom panel of \Fig{1D2Drat}, there is a significant drop in the radiation-to-gas energy ratio inside the photosphere in 1D, where the gas energy behind the photosphere is higher in 1D, due to the trapping of photons and the energy conversion from the radiation to the gas.

We summarize our findings in Figures~\ref{fig:1D2Devo}, \ref{fig:1D2D}, and \ref{fig:1D2Drat}. The effective luminosities at the photosphere are $\sim 2 \times 10^{45}\ergs$, and $\sim 6 \times 10^{46}\ergs$ for 1D and 2D, respectively.
On the other hand, the gas energy inside the photosphere is increased from the initial value of $\sim 1.27 \times 10^{51}$ erg to a higher value $\sim 1.89 \times 10^{51}$ erg in 1D and $\sim 1.69 \times 10^{51}$ erg in 2D, which suggests that the part of radiation energy is converted to the gas energy in 1D. This feature is due to photon trapping behind the photosphere from a steep density gradient in 1D. Energy converts from radiation to gas with a higher optical depth in the inner regions, corresponding to the lower regions in the bottom panel of \Fig{1D2Drat}.

Multidimensional simulations reflect the nature of radiation transport involving multi-wavelength radiation fluxes from various distances and angles \citep[][Page~13, Equation~(71)]{chandrasekhar2013radiative}. Multidimensional simulations account for the development of turbulent structures, breaking the spherical symmetry and enhancing fluid mixing. The formation of turbulence structures is sensitive to the progenitor star and its CSM profile, as shown in our multidimensional simulations. \cite{suzuki2016} studied the asymmetric shock breakout of SN~1987A using 2D ray-tracing calculations with the progenitor model from \cite{1988A&A...196..141S} and \cite{1990ApJ...360..242S}. This mass loss of this progenitor is $\sim 3\times 10^{-6}\,\Ms\mathrm{yr^{-1}}$ with a wind velocity of $330\,\mathrm{km\,s^{-1}}$ that corresponds to the steady wind of $\alpha\sim 0.5$ in our study. Due to the asymmetric shock breakout, large variations of breakout emissions among viewing angles. Their LC duration is larger than the duration found in our 1D models, but it is consistent with our 2D results. Therefore, 2D results demonstrate the importance of multidimensional RHD effects in modeling the shock breakout not only from the asymmetry in the supernova shock but also from mixing originating from fluid instabilities, as presented.
\fig{ComparisonLC}{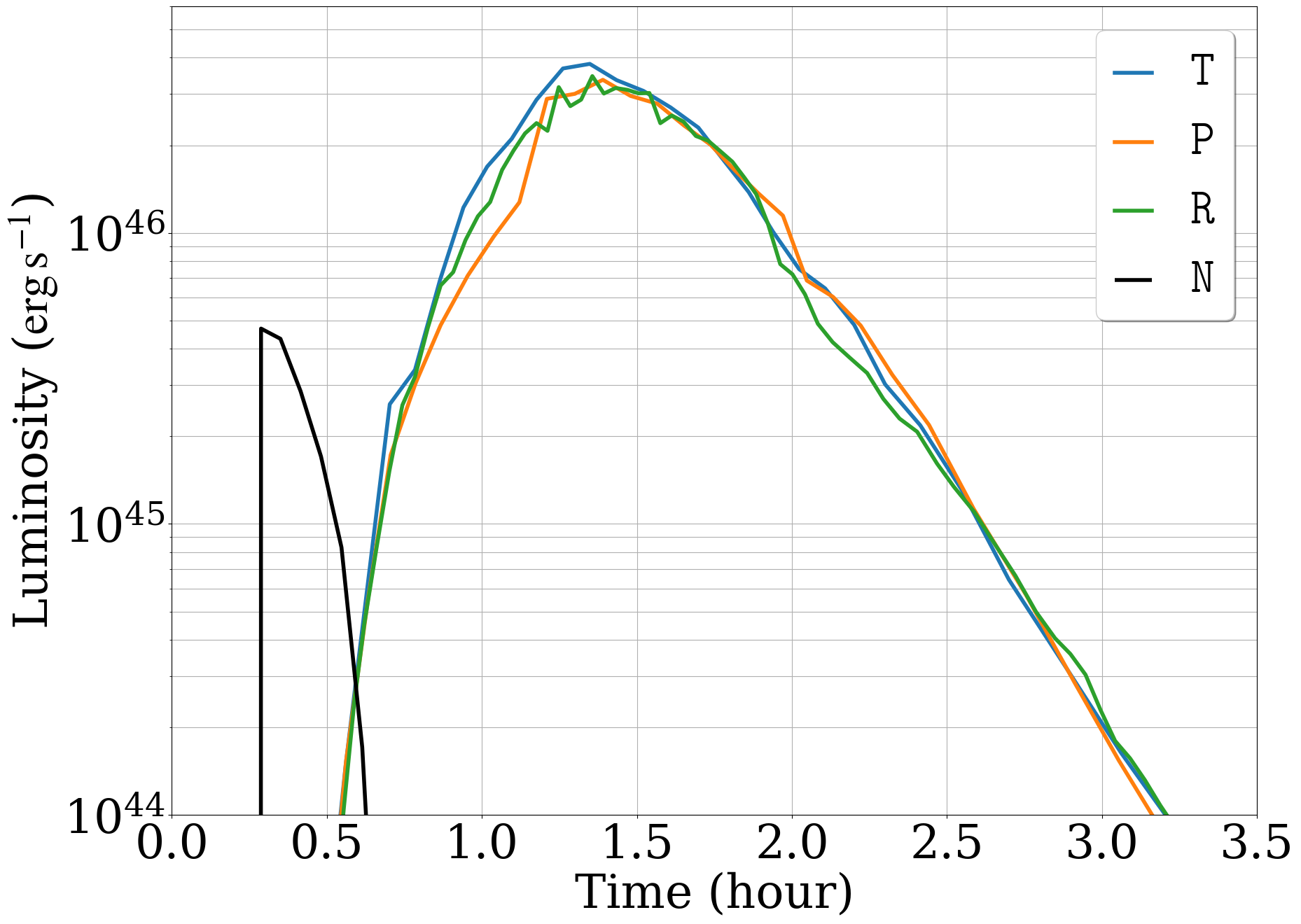}{Comparison of 1D and 2D LCs from the viewing angle of $45^\circ$. 1D and 2D LCs look different in terms of duration, peak luminosity, and the timing of the peak. Comparing among 2D models, Model \texttt{P} has a lower peak luminosity, and Model \texttt{R} shows stronger fluctuations.}{1.00}

\fig{1D2Devo}{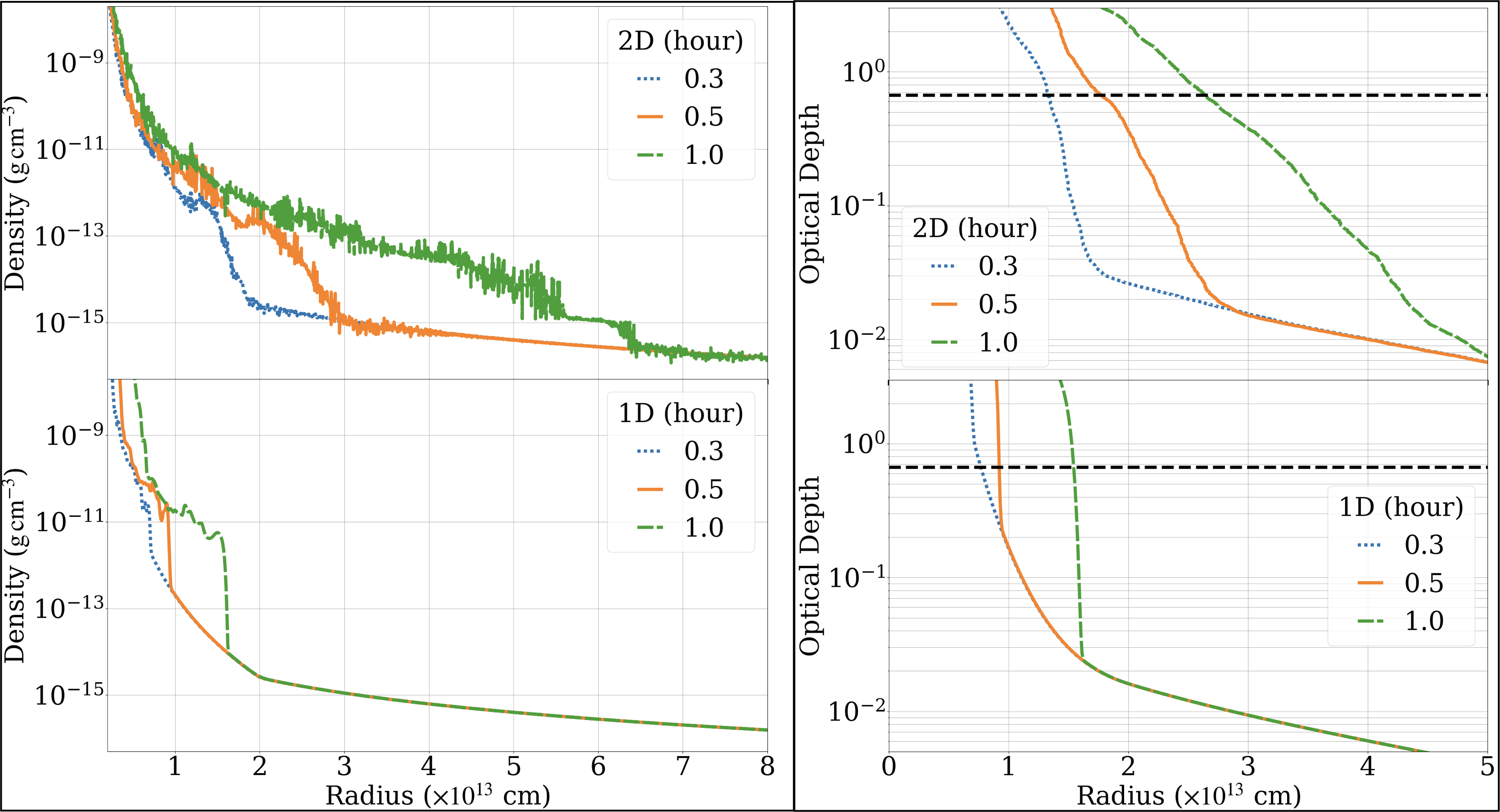}{The comparison between 1D (Model \texttt{N}) and 2D (Model \texttt{T}). The left panels show the evolution of density profiles, and the right panels show the optical depth profiles for band VIII ( Model \texttt{T}). The black horizontal lines indicate the photosphere radius. As seen in the right panel, the shock front travels further in 2D during the same time span, which enlarges the photosphere.}{1}

\fig{1D2D}{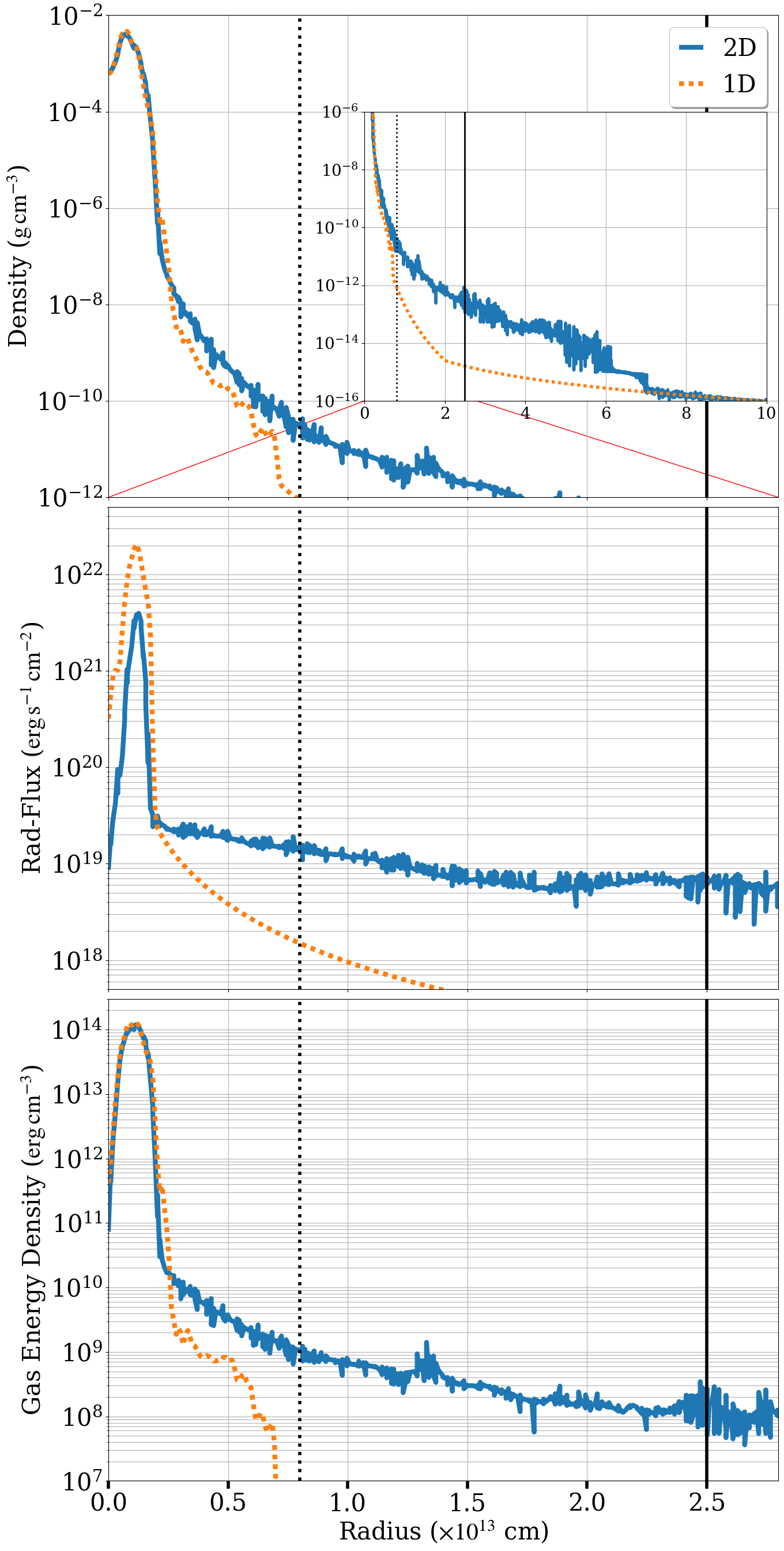}
{Density and energetics profiles of 1D (Model \texttt{N}) and 2D (Model \texttt{T}) at the peak luminosity. The \textbf{top} shows the gas mass density, \textbf{middle} shows radiation flux, and \textbf{bottom} shows the gas energy density. The vertical black lines indicate the photosphere radius for 1D (dotted) and 2D (solid). The locations of photosphere is $\sim 7\times 10^{12}$ cm in 1D and $\sim 3 \times 10^{13}$ cm in 2D. Radiation flux is also larger in 2D. Therefore, the resulting luminosity is $\sim 10^{46}\ergs$ in 2D and $\sim 10^{45}\ergs$ in 1D.}{0.5}
\fig{1D2Drat}{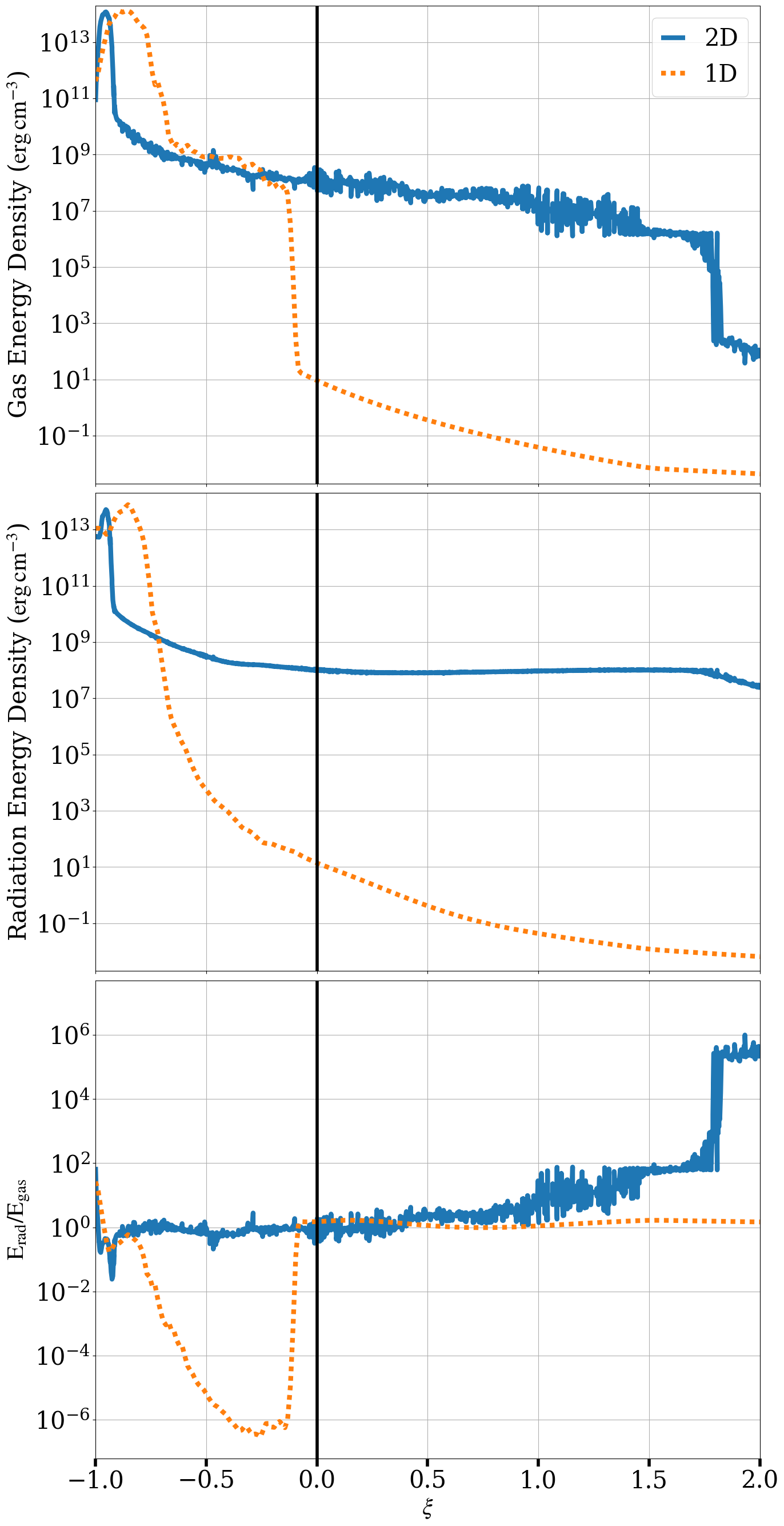}{Density and energetics profiles of 1D (Model \texttt{N}) and 2D (Model \texttt{T}) at the peak luminosity. The \textbf{top} shows the gas energy density, \textbf{middle} shows radiation energy density, and \textbf{bottom} shows the ratio of the radiation to gas energies (${E_{\rm rad}/E_{\rm gas}}$) as a function of the normalized radius to the photosphere.
The vertical black line shows the location of the photosphere at $\xi=0$.}{0.5}
\fig{EE}{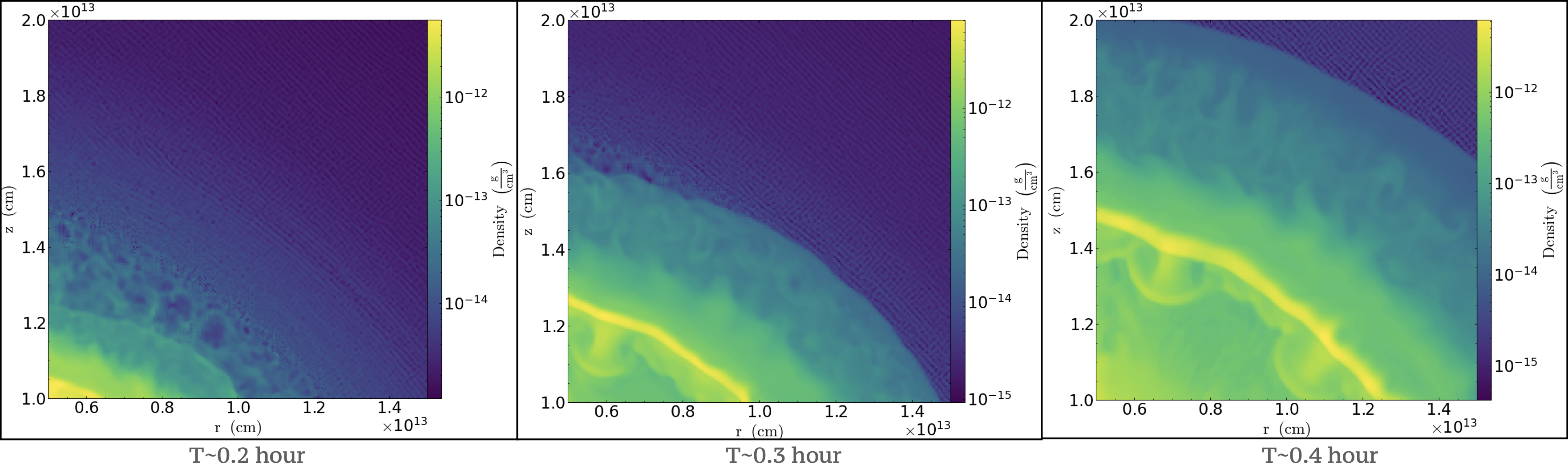}{The growth of RT Fingers during the early phase of shock breakout for Model \texttt{T} with $2048^2$. Panels from left to right show 0.2, 0.3, and 0.4 hours respectively. RT fingers start to emerge and grow in the shocked SN ejecta and CSM.}{1.0}
\fig{LE}{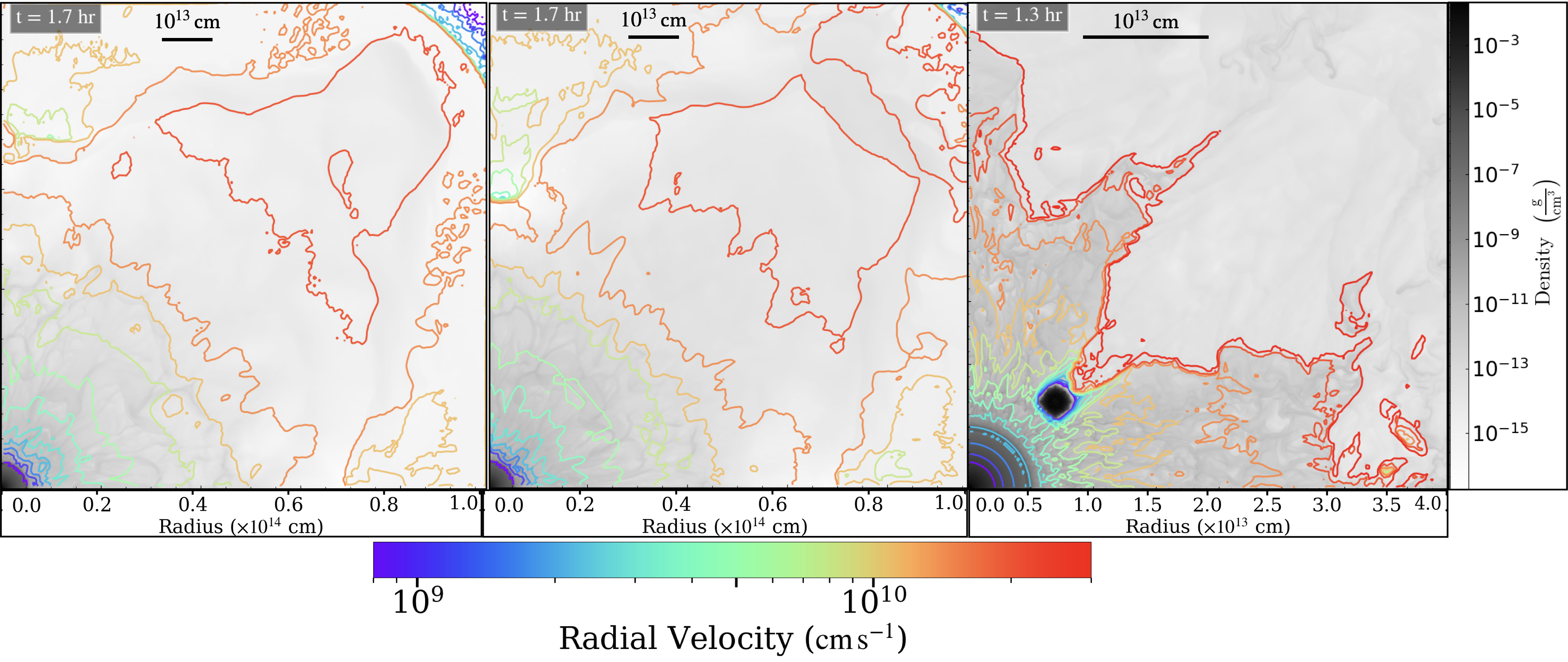}{Snapshots of density and velocity of three scenarios at the end of simulations. From left to right, the panel shows Models \texttt{P}, \texttt{R}, and \texttt{B}, respectively. The distinct differences in the density and velocity distributions among the three models reflect the impact of the CSM on the shock breakout dynamics that affect the consequent observational signatures.}{1.0}
\subsection{Comparison between 2D models}
Our simulations are more realistic than previous models because we consider the multidimensional and multi-group radiation hydrodynamics with detailed opacity. We show the development of turbulent structures under different CSM conditions in \Fig{LE} and summarize the values of LC duration and its peak luminosity for each viewing angle in Table \ref{tab:value}. \Fig{ComparisonLC} and Table \ref{tab:value} show that LCs of Models \texttt{P}, \texttt{R}, and \texttt{B} have smaller peak luminosity with shorter duration than that of Model \texttt{T}. The differences between Models \texttt{P} and \texttt{R} are minor. We discuss the possible mechanism that causes the differences in the LCs among the 2D models as below.
The perturbed CSM supplies mass around the progenitor star. The perturbed CSM from different environments can grow into different clumpy structures via fluid instabilities. These clumpy CSM structures generate large density fluctuations that affect the opacity of breakout emissions. Consequently, the optical depth along the line of sight in different CSM environments varies significantly, impacting the resulting LCs.

\subsubsection{Impact of perturbations}
In \S \ref{sec:uni}, we compare simulation Model \texttt{P} to Model \texttt{T}. We find a higher peak luminosity and a shorter LC duration in Model \texttt{P} by comparing the values in Table \ref{tab:value}.

The initial perturbations grow into turbulent mixing during the shock breakout. The turbulent mixing in the shocked CSM dissipates part of the radiation energy from emissions. The LCs of Model \texttt{P} shown in \Fig{2D325LC} and \Fig{2D325LC2} exhibit small fluctuations across viewing angles during the rising phase. 

\subsubsection{Ring-Like CSM}
Comparing to Model \texttt{T} in \Fig{ComparisonLC}, Model \texttt{R} has a shorter LC duration, lower peak luminosity, and larger fluctuations around the luminosity peak. In addition to lower luminosity at the rising phase compared to Model \texttt{T}, the radiation precursor collides with the ring earlier and leads to large luminosity fluctuations in \Fig{LC1_ring} and \Fig{LC2_ring}. 

When the shock collides with the ring-like CSM in the middle panel of \Fig{gas_ring}, it creates the impinging flows around the ring. The asymmetric geometry formed by the shock collision creates an efficient passage for radiation to escape from the emission layer that shortens the LC duration. As seen in the sudden drop in \Fig{LC2_ring}, the ring converts the radiation of short wavelengths to long wavelengths and deforms the spherical shock. The post-breakout luminosity decline rate is slightly slower than that of Model \texttt{P}. 

\subsubsection{Explosion with a Companion Star}
Based on Table \ref{tab:value} and the right panel of \Fig{LE}, Model \texttt{B} shows lower peak luminosity and a shorter LC duration due to the deformation of the shock front by collision with the companion star. The surviving binary companion continues to create larger fluctuations in LCs across viewing angles after the breakout. The LCs of long wavelength show a double-bump feature \citep[e.g.,][]{Kasen2016double}. Depending on the nature of the companion star, such as a red or blue supergiant, the emission from the surviving companion during the breakout phase can vary.

To summarize our 2D models, the evolution of shock breakout is sensitive to the physical properties of CSM and companion stars. Therefore, the breakout LCs vary across viewing angles and colors among all models. Therefore, our results can probe the nature of the pre-supernova environments unavailable in previous 1D or grey studies.

\subsection{Further Improvements and Limitations}

\subsubsection{Observational Perspective}
Although shock breakout of SN~1987A had not been directly detected, we can probe the early phase from the UV observations of SN~1987A around one day after the event \citep{1987Cassatella, 1992ApJ...393..742E}, when the radiation temperature has cooled to around $14,000$ K and continued dropping to around $10,000$ K after another 24 hours. Based on our simulations, the radiation temperature reaches $\sim 10^{7}$ K right after the breakout. The luminosity reaches its peak following a luminosity decay rate of $\sim 1.5\,\mathrm{mag}\,\mathrm{hour}^{-1}$, which takes $\sim 40$ hours for the effective temperature to drop from $10^7$ to $10^{5}$ K, this agrees well with the early observational data of 1987A and another 1987A-like SN \citep{Fransson_hour_1989ApJ,2019Avinash_1987}.
 
Searching the early emission of SN shock breakouts is more approachable with X-ray and UV observations \citep{ 2020RSG_Alex, erosita, Bayless_2022} corresponding to the band V to band VIII in our simulation. The late-time shock breakout emission can be probed by the excessive optical and infrared rays \citep{Dwek_2008, Suwa2017lensing} falling in the bands I to IV in our simulations. 
Furthermore, shock breakouts characterized by a sharp luminosity transition within short periods are more suitable for high redshift observations and can provide promising clues of the evolution of the early universe \citep{Blondin_2008}.

\subsubsection{Realistic Opacity}
We use opacities derived from \code{OPAL} depending on the gas density, temperature, and radiation frequency. Our simulations properly model the post-breakout cooling and produce correct color LCs with a rapid luminosity decline rate that cannot be done with a constant opacity. However, this opacity can be further improved by using multiple opacity tables from \cite{2005MNRAS.362L...1S,2017ApJ...835..119P} for gas temperature $\geq 10^4$ K, and \cite{2005ApJ...623..585F,2009A&A...508.1539M} for gas temperature $< 10^4$ K. These tables are operated well in the {\MESA} code \citep{2011ApJS..192....3P, 2013ApJS..208....4P, 2015ApJS..220...15P, 2018ApJS..234...34P,2019ApJS..243...10P, 2023ApJS..265...15J} that cover wider range of density, temperature, and metallicity, especially taking into account individual line opacities contributed from $\mathrm{^{56}Ni}$, carbon, and oxygen. Although the total opacity may not change dramatically, it may reflect in line emissions and absorption of SN spectra.
\subsubsection{Towards 3D Simulations}

In this study, we only show the results from 2D simulations. However, the 3D can be more realistic than 2D due to the nature of turbulence. 3D simulation allows $(r,\theta,\phi)$ motions that can further break cylindrical symmetry \citep{2022ApJJiang}. For our Model \texttt{B}, the 2D Cartesian slice is difficult to capture the detailed companion impacts, 3D coordinates better describe the geometry, rotation, and asymmetry of the binary interaction \citep{Liu2012}. High-velocity bulbs in 2D explosion shock as shown in \Fig{2D325}, \Fig{gas_ring}, and \Fig{gas_bi} evolve into 3D bubbles with \text{Rayleigh-B\'enard} convection that the radiative heating post-shocked regions mix with radiative cooling shock front. This large-scale circulation affects differently on the energy transport of shock in 3D \citep{2015arXiv150101827V}. Nevertheless, our 2D simulations generally have an advantage in following later evolution with higher resolution. Furthermore, stellar convection and its penetration depth show no significant difference in the stellar simulations between 2D and 3D \citep{2020A&A...638A..15P}.


\subsubsection{Realistic Circumstellar Medium}

We examine different CSM structures by adjusting the $\alpha$ parameter for mass loss rates of the steady wind and simulate the eruptive mass loss $\sim$ 10 days before the explosion by placing a ring-like structure along $45^\circ$ in Model \texttt{R}. CSMs may reflect time-dependent mass loss history of the progenitor stars \citep{2015ApJWesley}, with higher mass loss rates compared to normal mass-loss winds \citep{2007Natur.450..390W, 2014ApJ...792...44C, 2014ARA&A..52..487S, 2021ApJ...923...41L, Hiramatsu_2023}. The internal motion of CSM may also affect the shock breakout dynamics and its LC. Our simulation results can be used as input for post-processing the spectra of SN~1987A. 

\section{Conclusions} \label{sec:conclusions}
We perform the first 2D MGFLD simulations of the shock breakout of SN~1987A with realistic opacities using {\OPAL}. 2D structure formation in the gas decodes the nonphysical thin shell found in previous 1D models and allows us to study nonuniform pre-supernova environments such as stellar convection, eruptive mass loss wind, and a binary companion. Our multi-group radiation hydrodynamics simulations include eight radiation bands ranging from X-ray to infrared that offer the color evolution of shock breakout emissions and unveil the impact of radiative cooling on shock propagation. 

We find that the LC durations of $\sim 1\mathrm{\, hour}$ observed in our simulation result from broadened density structures around the shock front; the durations overall agree with those of asymmetric shock breakouts with 2D ray-tracing calculations by \cite{suzuki2016}. Our 2D simulations demonstrate the impacts of multidimensional mixing on the shock dynamics and LCs. The color LCs offer clues for the observational strategy for shock breakouts. Moreover, we discover that the effect of radiation precursor can lead to substantial fluctuations in LCs for a ring-like CSM and double-bump X-ray LCs for the presence of a nearby companion star. We plan to advance this study with more realistic opacities, including line emission and absorption in 3D geometry. Furthermore, we will explore the shock breakouts of RSG progenitors, which contain an extensive hydrogen envelope.

Our 2D MGFLD simulations of the SN~1987A shock breakout demonstrate the impact of the pre-breakout environments and the opacity on the emission and dynamics of the shock propagation. This work can extend to provide solid predictions for future searches of SN shock breakouts and deepen our understanding of the physics of supernovae and their exploding environments.

\begin{acknowledgments}
We thank Po-Sheng Ou and Sung-Han Tsai for the useful discussion.  This research is supported by the National Science and Technology Council, Taiwan, under grant No. MOST 110-2112-M-001-068-MY3, NSTC 113-2112-M-001-028-, and the Academia Sinica, Taiwan, under a career development award under grant No. AS-CDA-111-M04. This research was supported in part by grant NSF PHY-2309135 to the Kavli Institute for Theoretical Physics (KITP) and grant NSF PHY-2210452 to the Aspen Center for Physics. Our computing resources were supported by the National Energy Research Scientific Computing Center (NERSC), a U.S. Department of Energy Office of Science User Facility operated under Contract No. DE-AC02-05CH11231 and the TIARA Cluster at the Academia Sinica Institute of Astronomy and Astrophysics (ASIAA).
\end{acknowledgments}

\bibliography{sample631}{}

\begin{thebibliography}{}
\expandafter\ifx\csname natexlab\endcsname\relax\def\natexlab#1{#1}\fi
\providecommand{\url}[1]{\href{#1}{#1}}
\providecommand{\dodoi}[1]{doi:~\href{http://doi.org/#1}{\nolinkurl{#1}}}
\providecommand{\doeprint}[1]{\href{http://ascl.net/#1}{\nolinkurl{http://ascl.net/#1}}}
\providecommand{\doarXiv}[1]{\href{https://arxiv.org/abs/#1}{\nolinkurl{https://arxiv.org/abs/#1}}}

\bibitem[{Almgren {et~al.}(2020)Almgren, Sazo, Bell, Harpole, Katz, Sexton,
  Willcox, Zhang, \& Zingale}]{Almgren2020}
Almgren, A., Sazo, M.~B., Bell, J., {et~al.} 2020, Journal of Open Source
  Software, 5, 2513, \dodoi{10.21105/joss.02513}

\bibitem[{{Almgren} {et~al.}(2010){Almgren}, {Beckner}, {Bell}, {Day},
  {Howell}, {Joggerst}, {Lijewski}, {Nonaka}, {Singer}, \&
  {Zingale}}]{2010ApJ...715.1221A}
{Almgren}, A.~S., {Beckner}, V.~E., {Bell}, J.~B., {et~al.} 2010, \apj, 715,
  1221, \dodoi{10.1088/0004-637X/715/2/1221}

\bibitem[{{Alp} {et~al.}(2019){Alp}, {Larsson}, {Maeda}, {Fransson},
  {Wongwathanarat}, {Gabler}, {Janka}, {Jerkstrand}, {Heger}, \&
  {Menon}}]{2019ApJ...882...22A}
{Alp}, D., {Larsson}, J., {Maeda}, K., {et~al.} 2019, \apj, 882, 22,
  \dodoi{10.3847/1538-4357/ab3395}

\bibitem[{{Arnett} {et~al.}(1989){Arnett}, {Bahcall}, {Kirshner}, \&
  {Woosley}}]{1989ARA&A..27..629A}
{Arnett}, W.~D., {Bahcall}, J.~N., {Kirshner}, R.~P., \& {Woosley}, S.~E. 1989,
  ¥araa, 27, 629, \dodoi{10.1146/annurev.aa.27.090189.003213}

\bibitem[{{Arnett} \& {Meakin}(2011)}]{2011ApJ...741...33A}
{Arnett}, W.~D., \& {Meakin}, C. 2011, \apj, 741, 33,
  \dodoi{10.1088/0004-637X/741/1/33}

\bibitem[{{Bayless} {et~al.}(2015){Bayless}, {Even}, {Frey}, {Fryer}, {Roming},
  \& {Young}}]{2015ApJWesley}
{Bayless}, A.~J., {Even}, W., {Frey}, L.~H., {et~al.} 2015, \apj, 805, 98,
  \dodoi{10.1088/0004-637X/805/2/98}

\bibitem[{Bayless {et~al.}(2022)Bayless, Fryer, Brown, Young, Roming, Davis,
  Lechner, Slocum, Echon, \& Froning}]{Bayless_2022}
Bayless, A.~J., Fryer, C., Brown, P.~J., {et~al.} 2022, The Astrophysical
  Journal, 931, 15, \dodoi{10.3847/1538-4357/ac674c}

\bibitem[{Blondin {et~al.}(2008)Blondin, Davis, Krisciunas, Schmidt, Sollerman,
  Wood-Vasey, Becker, Challis, Clocchiatti, Damke, Filippenko, Foley,
  Garnavich, Jha, Kirshner, Leibundgut, Li, Matheson, Miknaitis, Narayan,
  Pignata, Rest, Riess, Silverman, Smith, Spyromilio, Stritzinger, Stubbs,
  Suntzeff, Tonry, Tucker, \& Zenteno}]{Blondin_2008}
Blondin, S., Davis, T.~M., Krisciunas, K., {et~al.} 2008, The Astrophysical
  Journal, 682, 724, \dodoi{10.1086/589568}

\bibitem[{{Buchler} \& {Yueh}(1976)}]{1976ApJ...210..440B}
{Buchler}, J.~R., \& {Yueh}, W.~R. 1976, \apj, 210, 440, \dodoi{10.1086/154847}

\bibitem[{{Cassatella} {et~al.}(1987){Cassatella}, {Fransson}, {vant
  Santvoort}, {Gry}, {Talavera}, {Wamsteker}, \& {Panagia}}]{1987Cassatella}
{Cassatella}, A., {Fransson}, C., {vant Santvoort}, J., {et~al.} 1987, \aap,
  177, L29

\bibitem[{Chandrasekhar(2013)}]{chandrasekhar2013radiative}
Chandrasekhar, S. 2013, Radiative Transfer, Dover Books on Physics (Dover
  Publications).
\newblock \url{https://books.google.com.tw/books?id=1YHCAgAAQBAJ}

\bibitem[{Chen {et~al.}(2013)Chen, Heger, \& Almgren}]{CHEN201370}
Chen, K.-J., Heger, A., \& Almgren, A.~S. 2013, Astronomy and Computing, 3-4,
  70, \dodoi{https://doi.org/10.1016/j.ascom.2014.01.001}

\bibitem[{{Chen} {et~al.}(2014){Chen}, {Heger}, {Woosley}, {Almgren}, \&
  {Whalen}}]{2014ApJ...792...44C}
{Chen}, K.-J., {Heger}, A., {Woosley}, S., {Almgren}, A., \& {Whalen}, D.~J.
  2014, \apj, 792, 44, \dodoi{10.1088/0004-637X/792/1/44}

\bibitem[{Chen {et~al.}(2014{\natexlab{a}})Chen, Heger, Woosley, Almgren, \&
  Whalen}]{Ken2014PI}
Chen, K.-J., Heger, A., Woosley, S., Almgren, A., \& Whalen, D.~J.
  2014{\natexlab{a}}, The Astrophysical Journal, 792, 44,
  \dodoi{10.1088/0004-637X/792/1/44}

\bibitem[{{Chen} {et~al.}(2023){Chen}, {Whalen}, {Woosley}, \&
  {Zhang}}]{Ken2023ApJ}
{Chen}, K.-J., {Whalen}, D.~J., {Woosley}, S.~E., \& {Zhang}, W. 2023, \apj,
  955, 39, \dodoi{10.3847/1538-4357/ace968}

\bibitem[{Chen {et~al.}(2014{\natexlab{b}})Chen, Woosley, Heger, Almgren, \&
  Whalen}]{Chen2014PPI}
Chen, K.-J., Woosley, S., Heger, A., Almgren, A., \& Whalen, D.~J.
  2014{\natexlab{b}}, The Astrophysical Journal, 792, 28,
  \dodoi{10.1088/0004-637X/792/1/28}

\bibitem[{{Chevalier}(1976)}]{1976ApJ...207..872C}
{Chevalier}, R.~A. 1976, \apj, 207, 872, \dodoi{10.1086/154557}

\bibitem[{{Chevalier} \& {Fransson}(1987)}]{1987Natur.328...44C}
{Chevalier}, R.~A., \& {Fransson}, C. 1987, \nat, 328, 44,
  \dodoi{10.1038/328044a0}

\bibitem[{{Chevalier} \& {Irwin}(2011)}]{2011ApJ...729L...6C}
{Chevalier}, R.~A., \& {Irwin}, C.~M. 2011, \apjl, 729, L6,
  \dodoi{10.1088/2041-8205/729/1/L6}

\bibitem[{{Couch} {et~al.}(2011){Couch}, {Pooley}, {Wheeler}, \&
  {Milosavljevi{\'c}}}]{2011ApJ...727..104C}
{Couch}, S.~M., {Pooley}, D., {Wheeler}, J.~C., \& {Milosavljevi{\'c}}, M.
  2011, \apj, 727, 104, \dodoi{10.1088/0004-637X/727/2/104}

\bibitem[{{Dohi} {et~al.}(2023){Dohi}, {Greco}, {Nagataki}, {Ono}, {Miceli},
  {Orlando}, \& {Olmi}}]{2023ApJ...949...97D}
{Dohi}, A., {Greco}, E., {Nagataki}, S., {et~al.} 2023, \apj, 949, 97,
  \dodoi{10.3847/1538-4357/acce3f}

\bibitem[{Dwek \& Arendt(2008)}]{Dwek_2008}
Dwek, E., \& Arendt, R.~G. 2008, The Astrophysical Journal, 685, 976,
  \dodoi{10.1086/589988}

\bibitem[{{Ensman} \& {Burrows}(1992)}]{1992ApJ...393..742E}
{Ensman}, L., \& {Burrows}, A. 1992, \apj, 393, 742, \dodoi{10.1086/171542}

\bibitem[{{Epstein}(1981)}]{Epstein1981}
{Epstein}, R.~I. 1981, \apjl, 244, L89, \dodoi{10.1086/183486}

\bibitem[{{Farag} {et~al.}(2024){Farag}, {Fontes}, {Timmes}, {Bellinger},
  {Guzik}, {Bauer}, {Wood}, {Mussack}, {Hakel}, {Colgan}, {Kilcrease},
  {Sherrill}, {Raecke}, \& {Chidester}}]{2024Farag}
{Farag}, E., {Fontes}, C.~J., {Timmes}, F.~X., {et~al.} 2024, \apj, 968, 56,
  \dodoi{10.3847/1538-4357/ad4355}

\bibitem[{{Ferguson} {et~al.}(2005){Ferguson}, {Alexander}, {Allard}, {Barman},
  {Bodnarik}, {Hauschildt}, {Heffner-Wong}, \& {Tamanai}}]{2005ApJ...623..585F}
{Ferguson}, J.~W., {Alexander}, D.~R., {Allard}, F., {et~al.} 2005, \apj, 623,
  585, \dodoi{10.1086/428642}

\bibitem[{{Fransson} \& {Lundqvist}(1989)}]{Fransson_hour_1989ApJ}
{Fransson}, C., \& {Lundqvist}, P. 1989, \apjl, 341, L59,
  \dodoi{10.1086/185457}

\bibitem[{{Fuller} \& {Ro}(2018)}]{2018Fuller}
{Fuller}, J., \& {Ro}, S. 2018, \mnras, 476, 1853, \dodoi{10.1093/mnras/sty369}

\bibitem[{Förster {et~al.}(2018)Förster, Moriya, Maureira, Anderson,
  Blinnikov, Bufano, Cabrera-Vives, Clocchiatti, de~Jaeger, Estevez, Galbany,
  González-Gaitán, Gräfener, Hamuy, Hsiao, Huentelemu, Huijse, Kuncarayakti,
  Martínez, \& Young}]{delay2018}
Förster, F., Moriya, T., Maureira, J., {et~al.} 2018, Nature Astronomy, 2,
  \dodoi{10.1038/s41550-018-0563-4}

\bibitem[{{Gezari} {et~al.}(2015){Gezari}, {Jones}, {Sanders}, {Soderberg},
  {Hung}, {Heinis}, {Smartt}, {Rest}, {Scolnic}, {Chornock}, {Berger}, {Foley},
  {Huber}, {Price}, {Stubbs}, {Riess}, {Kirshner}, {Smith}, {Wood-Vasey},
  {Schiminovich}, {Martin}, {Burgett}, {Chambers}, {Flewelling}, {Kaiser},
  {Tonry}, \& {Wainscoat}}]{2015ApJ...804...28G}
{Gezari}, S., {Jones}, D.~O., {Sanders}, N.~E., {et~al.} 2015, \apj, 804, 28,
  \dodoi{10.1088/0004-637X/804/1/28}

\bibitem[{{Goldberg} {et~al.}(2022){Goldberg}, {Jiang}, \&
  {Bildsten}}]{2022ApJJiang}
{Goldberg}, J.~A., {Jiang}, Y.-F., \& {Bildsten}, L. 2022, \apj, 933, 164,
  \dodoi{10.3847/1538-4357/ac75e3}

\bibitem[{{Gonz\'alez-Tor\`a, G.} {et~al.}(2023){Gonz\'alez-Tor\`a, G.},
  {Wittkowski, M.}, {Davies, B.}, {Plez, B.}, \& {Kravchenko, K.}}]{rsgwind}
{Gonz\'alez-Tor\`a, G.}, {Wittkowski, M.}, {Davies, B.}, {Plez, B.}, \&
  {Kravchenko, K.} 2023, A\&A, 669, A76, \dodoi{10.1051/0004-6361/202244503}

\bibitem[{{Greco} {et~al.}(2021){Greco}, {Miceli}, {Orlando}, {Olmi},
  {Bocchino}, {Nagataki}, {Ono}, {Dohi}, \& {Peres}}]{2021ApJ...908L..45G}
{Greco}, E., {Miceli}, M., {Orlando}, S., {et~al.} 2021, \apjl, 908, L45,
  \dodoi{10.3847/2041-8213/abdf5a}

\bibitem[{{Greco} {et~al.}(2022){Greco}, {Miceli}, {Orlando}, {Olmi},
  {Bocchino}, {Nagataki}, {Sun}, {Vink}, {Sapienza}, {Ono}, {Dohi}, \&
  {Peres}}]{2022ApJ...931..132G}
---. 2022, \apj, 931, 132, \dodoi{10.3847/1538-4357/ac679d}

\bibitem[{Hiramatsu {et~al.}(2023)Hiramatsu, Tsuna, Berger, Itagaki, Goldberg,
  Gomez, De, Hosseinzadeh, Bostroem, Brown, Arcavi, Bieryla, Blanchard,
  Esquerdo, Farah, Howell, Matsumoto, McCully, Newsome, Gonzalez, Pellegrino,
  Rhee, Terreran, Vinkó, \& Wheeler}]{Hiramatsu_2023}
Hiramatsu, D., Tsuna, D., Berger, E., {et~al.} 2023, The Astrophysical Journal
  Letters, 955, L8, \dodoi{10.3847/2041-8213/acf299}

\bibitem[{{Iglesias} \& {Rogers}(1996)}]{1996ApJ...464..943I}
{Iglesias}, C.~A., \& {Rogers}, F.~J. 1996, \apj, 464, 943,
  \dodoi{10.1086/177381}

\bibitem[{{Iliev} {et~al.}(2009){Iliev}, {Whalen}, {Mellema}, {Ahn}, {Baek},
  {Gnedin}, {Kravtsov}, {Norman}, {Raicevic}, {Reynolds}, {Sato}, {Shapiro},
  {Semelin}, {Smidt}, {Susa}, {Theuns}, \& {Umemura}}]{2009MNRAS.400.1283I}
{Iliev}, I.~T., {Whalen}, D., {Mellema}, G., {et~al.} 2009, \mnras, 400, 1283,
  \dodoi{10.1111/j.1365-2966.2009.15558.x}

\bibitem[{{Imshennik} \& {Nadezhin}(1988)}]{Imshennik1988}
{Imshennik}, V.~S., \& {Nadezhin}, D.~K. 1988, Soviet Astronomy Letters, 14,
  449

\bibitem[{{Jermyn} {et~al.}(2023){Jermyn}, {Bauer}, {Schwab}, {Farmer}, {Ball},
  {Bellinger}, {Dotter}, {Joyce}, {Marchant}, {Mombarg}, {Wolf}, {Sunny Wong},
  {Cinquegrana}, {Farrell}, {Smolec}, {Thoul}, {Cantiello}, {Herwig}, {Toloza},
  {Bildsten}, {Townsend}, \& {Timmes}}]{2023ApJS..265...15J}
{Jermyn}, A.~S., {Bauer}, E.~B., {Schwab}, J., {et~al.} 2023, \apjs, 265, 15,
  \dodoi{10.3847/1538-4365/acae8d}

\bibitem[{{Kasen} {et~al.}(2016){Kasen}, {Metzger}, \&
  {Bildsten}}]{Kasen2016double}
{Kasen}, D., {Metzger}, B.~D., \& {Bildsten}, L. 2016, \apj, 821, 36,
  \dodoi{10.3847/0004-637X/821/1/36}

\bibitem[{Katz {et~al.}(2010)Katz, Budnik, \& Waxman}]{Katz_2010}
Katz, B., Budnik, R., \& Waxman, E. 2010, The Astrophysical Journal, 716, 781,
  \dodoi{10.1088/0004-637X/716/1/781}

\bibitem[{{Katz} {et~al.}(2016){Katz}, {Zingale}, {Calder}, {Swesty},
  {Almgren}, \& {Zhang}}]{2016ApJ...819...94K}
{Katz}, M.~P., {Zingale}, M., {Calder}, A.~C., {et~al.} 2016, \apj, 819, 94,
  \dodoi{10.3847/0004-637X/819/2/94}

\bibitem[{{Kozyreva} {et~al.}(2020){Kozyreva}, {Nakar}, {Waldman}, {Blinnikov},
  \& {Baklanov}}]{2020RSG_Alex}
{Kozyreva}, A., {Nakar}, E., {Waldman}, R., {Blinnikov}, S., \& {Baklanov}, P.
  2020, \mnras, 494, 3927, \dodoi{10.1093/mnras/staa924}

\bibitem[{{Leung} {et~al.}(2021){Leung}, {Wu}, \&
  {Fuller}}]{2021ApJ...923...41L}
{Leung}, S.-C., {Wu}, S., \& {Fuller}, J. 2021, \apj, 923, 41,
  \dodoi{10.3847/1538-4357/ac2c63}

\bibitem[{Levesque {et~al.}(2012)Levesque, Stringfellow, Ginsburg, Bally, \&
  Keeney}]{ring2012}
Levesque, E., Stringfellow, G., Ginsburg, A., Bally, J., \& Keeney, B. 2012,
  The Astronomical Journal, 147, \dodoi{10.1088/0004-6256/147/1/23}

\bibitem[{Levinson \& Nakar(2020)}]{LEVINSON20201}
Levinson, A., \& Nakar, E. 2020, Physics Reports, 866, 1,
  \dodoi{https://doi.org/10.1016/j.physrep.2020.04.003}

\bibitem[{{Liu, Z. W.} {et~al.}(2012){Liu, Z. W.}, {Pakmor, R.}, {Röpke, F.
  K.}, {Edelmann, P.}, {Wang, B.}, {Kromer, M.}, {Hillebrandt, W.}, \& {Han, Z.
  W.}}]{Liu2012}
{Liu, Z. W.}, {Pakmor, R.}, {Röpke, F. K.}, {et~al.} 2012, A\&A, 548, A2,
  \dodoi{10.1051/0004-6361/201219357}

\bibitem[{{Lovegrove} {et~al.}(2017){Lovegrove}, {Woosley}, \&
  {Zhang}}]{2017ApJ...845..103L}
{Lovegrove}, E., {Woosley}, S.~E., \& {Zhang}, W. 2017, \apj, 845, 103,
  \dodoi{10.3847/1538-4357/aa7b7d}

\bibitem[{Mao {et~al.}(2015)Mao, Ono, Nagataki, aki Hashimoto, Ito, Matsumoto,
  Dainotti, \& Lee}]{Mao_2015}
Mao, J., Ono, M., Nagataki, S., {et~al.} 2015, The Astrophysical Journal, 808,
  164, \dodoi{10.1088/0004-637X/808/2/164}

\bibitem[{{Marigo} \& {Aringer}(2009)}]{2009A&A...508.1539M}
{Marigo}, P., \& {Aringer}, B. 2009, \aap, 508, 1539,
  \dodoi{10.1051/0004-6361/200912598}

\bibitem[{{Matzner} \& {McKee}(1999)}]{1999ApJ...510..379M}
{Matzner}, C.~D., \& {McKee}, C.~F. 1999, \apj, 510, 379,
  \dodoi{10.1086/306571}

\bibitem[{{Meakin} \& {Arnett}(2007)}]{2007ApJ...667..448M}
{Meakin}, C.~A., \& {Arnett}, D. 2007, \apj, 667, 448, \dodoi{10.1086/520318}

\bibitem[{{Menon} \& {Heger}(2017)}]{2017MNRAS.469.4649M}
{Menon}, A., \& {Heger}, A. 2017, \mnras, 469, 4649.
\newblock \doarXiv{1703.04918}

\bibitem[{{Menon} {et~al.}(2019){Menon}, {Utrobin}, \&
  {Heger}}]{2019MNRAS.482..438M}
{Menon}, A., {Utrobin}, V., \& {Heger}, A. 2019, \mnras, 482, 438,
  \dodoi{10.1093/mnras/sty2647}

\bibitem[{{Meshkov}(1972)}]{1972Meshkov}
{Meshkov}, E.~E. 1972, Fluid Dynamics, 4, 101, \dodoi{10.1007/BF01015969}

\bibitem[{{Nakamura} {et~al.}(2022){Nakamura}, {Takiwaki}, \&
  {Kotake}}]{2022MNRAS.514.3941N}
{Nakamura}, K., {Takiwaki}, T., \& {Kotake}, K. 2022, \mnras, 514, 3941,
  \dodoi{10.1093/mnras/stac1586}

\bibitem[{{Ono} {et~al.}(2020){Ono}, {Nagataki}, {Ferrand}, {Takahashi},
  {Umeda}, {Yoshida}, {Orlando}, \& {Miceli}}]{2020ApJ...888..111O}
{Ono}, M., {Nagataki}, S., {Ferrand}, G., {et~al.} 2020, \apj, 888, 111,
  \dodoi{10.3847/1538-4357/ab5dba}

\bibitem[{{Ono} {et~al.}(2023){Ono}, {Nozawa}, {Nagataki}, {Kozyreva},
  {Orlando}, {Miceli}, \& {Chen}}]{Ono2023}
{Ono}, M., {Nozawa}, T., {Nagataki}, S., {et~al.} 2023, arXiv e-prints,
  arXiv:2305.02550, \dodoi{10.48550/arXiv.2305.02550}

\bibitem[{{Orlando} {et~al.}(2020){Orlando}, {Ono}, {Nagataki}, {Miceli},
  {Umeda}, {Ferrand}, {Bocchino}, {Petruk}, {Peres}, {Takahashi}, \&
  {Yoshida}}]{2020A&A...636A..22O}
{Orlando}, S., {Ono}, M., {Nagataki}, S., {et~al.} 2020, \aap, 636, A22,
  \dodoi{10.1051/0004-6361/201936718}

\bibitem[{Ott {et~al.}(2008)Ott, Burrows, Dessart, \& Livne}]{Ott_2008}
Ott, C.~D., Burrows, A., Dessart, L., \& Livne, E. 2008, The Astrophysical
  Journal, 685, 1069, \dodoi{10.1086/591440}

\bibitem[{{Ouchi} \& {Maeda}(2019)}]{2019Ouchi&Maeda}
{Ouchi}, R., \& {Maeda}, K. 2019, \apj, 877, 92,
  \dodoi{10.3847/1538-4357/ab1a37}

\bibitem[{Paxton {et~al.}(2010)Paxton, Bildsten, Dotter, Herwig, Lesaffre, \&
  Timmes}]{2011Paxton}
Paxton, B., Bildsten, L., Dotter, A., {et~al.} 2010, The Astrophysical Journal
  Supplement Series, 192, 3, \dodoi{10.1088/0067-0049/192/1/3}

\bibitem[{{Paxton} {et~al.}(2011){Paxton}, {Bildsten}, {Dotter}, {Herwig},
  {Lesaffre}, \& {Timmes}}]{2011ApJS..192....3P}
{Paxton}, B., {Bildsten}, L., {Dotter}, A., {et~al.} 2011, \apjs, 192, 3,
  \dodoi{10.1088/0067-0049/192/1/3}

\bibitem[{{Paxton} {et~al.}(2013){Paxton}, {Cantiello}, {Arras}, {Bildsten},
  {Brown}, {Dotter}, {Mankovich}, {Montgomery}, {Stello}, {Timmes}, \&
  {Townsend}}]{2013ApJS..208....4P}
{Paxton}, B., {Cantiello}, M., {Arras}, P., {et~al.} 2013, \apjs, 208, 4,
  \dodoi{10.1088/0067-0049/208/1/4}

\bibitem[{{Paxton} {et~al.}(2015){Paxton}, {Marchant}, {Schwab}, {Bauer},
  {Bildsten}, {Cantiello}, {Dessart}, {Farmer}, {Hu}, {Langer}, {Townsend},
  {Townsley}, \& {Timmes}}]{2015ApJS..220...15P}
{Paxton}, B., {Marchant}, P., {Schwab}, J., {et~al.} 2015, \apjs, 220, 15,
  \dodoi{10.1088/0067-0049/220/1/15}

\bibitem[{{Paxton} {et~al.}(2018){Paxton}, {Schwab}, {Bauer}, {Bildsten},
  {Blinnikov}, {Duffell}, {Farmer}, {Goldberg}, {Marchant}, {Sorokina},
  {Thoul}, {Townsend}, \& {Timmes}}]{2018ApJS..234...34P}
{Paxton}, B., {Schwab}, J., {Bauer}, E.~B., {et~al.} 2018, \apjs, 234, 34,
  \dodoi{10.3847/1538-4365/aaa5a8}

\bibitem[{{Paxton} {et~al.}(2019){Paxton}, {Smolec}, {Schwab}, {Gautschy},
  {Bildsten}, {Cantiello}, {Dotter}, {Farmer}, {Goldberg}, {Jermyn}, {Kanbur},
  {Marchant}, {Thoul}, {Townsend}, {Wolf}, {Zhang}, \&
  {Timmes}}]{2019ApJS..243...10P}
{Paxton}, B., {Smolec}, R., {Schwab}, J., {et~al.} 2019, \apjs, 243, 10,
  \dodoi{10.3847/1538-4365/ab2241}

\bibitem[{{Poutanen}(2017)}]{2017ApJ...835..119P}
{Poutanen}, J. 2017, \apj, 835, 119, \dodoi{10.3847/1538-4357/835/2/119}

\bibitem[{{Pratt} {et~al.}(2020){Pratt}, {Baraffe}, {Goffrey}, {Geroux},
  {Constantino}, {Folini}, \& {Walder}}]{2020A&A...638A..15P}
{Pratt}, J., {Baraffe}, I., {Goffrey}, T., {et~al.} 2020, \aap, 638, A15,
  \dodoi{10.1051/0004-6361/201834736}

\bibitem[{{Predehl, P.} {et~al.}(2021){Predehl, P.}, {Andritschke, R.},
  {Arefiev, V.}, {Babyshkin, V.}, {Batanov, O.}, {Becker, W.}, {B\"ohringer,
  H.}, {Bogomolov, A.}, {Boller, T.}, {Borm, K.}, {Bornemann, W.},
  {Br\"auninger, H.}, {Br\"uggen, M.}, {Brunner, H.}, {Brusa, M.}, {Bulbul,
  E.}, {Buntov, M.}, {Burwitz, V.}, {Burkert, W.}, {Clerc, N.}, {Churazov, E.},
  {Coutinho, D.}, {Dauser, T.}, {Dennerl, K.}, {Doroshenko, V.}, {Eder, J.},
  {Emberger, V.}, {Eraerds, T.}, {Finoguenov, A.}, {Freyberg, M.}, {Friedrich,
  P.}, {Friedrich, S.}, {F\"urmetz, M.}, {Georgakakis, A.}, {Gilfanov, M.},
  {Granato, S.}, {Grossberger, C.}, {Gueguen, A.}, {Gureev, P.}, {Haberl, F.},
  {H\"alker, O.}, {Hartner, G.}, {Hasinger, G.}, {Huber, H.}, {Ji, L.},
  {Kienlin, A. v.}, {Kink, W.}, {Korotkov, F.}, {Kreykenbohm, I.}, {Lamer, G.},
  {Lomakin, I.}, {Lapshov, I.}, {Liu, T.}, {Maitra, C.}, {Meidinger, N.},
  {Menz, B.}, {Merloni, A.}, {Mernik, T.}, {Mican, B.}, {Mohr, J.}, {M\"uller,
  S.}, {Nandra, K.}, {Nazarov, V.}, {Pacaud, F.}, {Pavlinsky, M.}, {Perinati,
  E.}, {Pfeffermann, E.}, {Pietschner, D.}, {Ramos-Ceja, M. E.}, {Rau, A.},
  {Reiffers, J.}, {Reiprich, T. H.}, {Robrade, J.}, {Salvato, M.}, {Sanders,
  J.}, {Santangelo, A.}, {Sasaki, M.}, {Scheuerle, H.}, {Schmid, C.}, {Schmitt,
  J.}, {Schwope, A.}, {Shirshakov, A.}, {Steinmetz, M.}, {Stewart, I.},
  {Str\"uder, L.}, {Sunyaev, R.}, {Tenzer, C.}, {Tiedemann, L.}, {Tr\"umper,
  J.}, {Voron, V.}, {Weber, P.}, {Wilms, J.}, \& {Yaroshenko, V.}}]{erosita}
{Predehl, P.}, {Andritschke, R.}, {Arefiev, V.}, {et~al.} 2021, A\&A, 647, A1,
  \dodoi{10.1051/0004-6361/202039313}

\bibitem[{{Quataert} {et~al.}(2016){Quataert}, {Fern{\'a}ndez}, {Kasen},
  {Klion}, \& {Paxton}}]{2016Quataert}
{Quataert}, E., {Fern{\'a}ndez}, R., {Kasen}, D., {Klion}, H., \& {Paxton}, B.
  2016, \mnras, 458, 1214, \dodoi{10.1093/mnras/stw365}

\bibitem[{Rayleigh(1882)}]{1882Rayleigh}
Rayleigh. 1882, Proceedings of the London Mathematical Society, s1-14, 170,
  \dodoi{https://doi.org/10.1112/plms/s1-14.1.170}

\bibitem[{Richtmyer(1960)}]{1960Richtmyer}
Richtmyer, R.~D. 1960, Communications on Pure and Applied Mathematics, 13, 297,
  \dodoi{https://doi.org/10.1002/cpa.3160130207}

\bibitem[{{Sana} {et~al.}(2012){Sana}, {de Mink}, {de Koter}, {Langer},
  {Evans}, {Gieles}, {Gosset}, {Izzard}, {Le Bouquin}, \&
  {Schneider}}]{Sana2012}
{Sana}, H., {de Mink}, S.~E., {de Koter}, A., {et~al.} 2012, Science, 337, 444,
  \dodoi{10.1126/science.1223344}

\bibitem[{{Sapienza} {et~al.}(2024){Sapienza}, {Miceli}, {Bamba}, {Orlando},
  {Lee}, {Nagataki}, {Ono}, {Katsuda}, {Mori}, {Sawada}, {Terada}, {Giuffrida},
  \& {Bocchino}}]{2024ApJ...961L...9S}
{Sapienza}, V., {Miceli}, M., {Bamba}, A., {et~al.} 2024, \apjl, 961, L9,
  \dodoi{10.3847/2041-8213/ad16e3}

\bibitem[{Schawinski {et~al.}(2008{\natexlab{a}})Schawinski, Justham, Wolf,
  Podsiadlowski, Sullivan, Steenbrugge, Bell, Röser, Walker, Astier, Balam,
  Balland, Carlberg, Conley, Fouchez, Guy, Hardin, Hook, Howell, Pain, Perrett,
  Pritchet, Regnault, \& Yi}]{doi:10.1126/science.1160456}
Schawinski, K., Justham, S., Wolf, C., {et~al.} 2008{\natexlab{a}}, Science,
  321, 223, \dodoi{10.1126/science.1160456}

\bibitem[{Schawinski {et~al.}(2008{\natexlab{b}})Schawinski, Justham, Wolf,
  Podsiadlowski, Sullivan, Steenbrugge, Bell, Roser, Walker, Astier, \&
  et~al.}]{Schawinski_2008}
---. 2008{\natexlab{b}}, Science, 321, 223–226,
  \dodoi{10.1126/science.1160456}

\bibitem[{{Seaton}(2005)}]{2005MNRAS.362L...1S}
{Seaton}, M.~J. 2005, \mnras, 362, L1, \dodoi{10.1111/j.1365-2966.2005.00019.x}

\bibitem[{Sharma {et~al.}(1987)Sharma, Ram, \&
  Sachdev}]{sharma_ram_sachdev_1987}
Sharma, V.~D., Ram, R., \& Sachdev, P.~L. 1987, Journal of Fluid Mechanics,
  185, 153–170, \dodoi{10.1017/S0022112087003124}

\bibitem[{{Shigeyama} \& {Nomoto}(1990)}]{1990ApJ...360..242S}
{Shigeyama}, T., \& {Nomoto}, K. 1990, \apj, 360, 242, \dodoi{10.1086/169114}

\bibitem[{{Shigeyama} {et~al.}(1988){Shigeyama}, {Nomoto}, \&
  {Hashimoto}}]{1988A&A...196..141S}
{Shigeyama}, T., {Nomoto}, K., \& {Hashimoto}, M. 1988, \aap, 196, 141

\bibitem[{{Singh} {et~al.}(2019){Singh}, {Sahu}, {Anupama}, {Kumar}, {Kumar},
  {Yamanaka}, {Baklanov}, {Tominaga}, {Blinnikov}, {Maeda}, {Dutta},
  {Bhalerao}, {Anche}, {Barway}, {Akitaya}, {Nakaoka}, {Kawabata}, {Kawabata},
  {Sasada}, {Takagi}, {Maehara}, {Isogai}, {Kino}, {Taguchi}, \&
  {Nagao}}]{2019Avinash_1987}
{Singh}, A., {Sahu}, D.~K., {Anupama}, G.~C., {et~al.} 2019, \apjl, 882, L15,
  \dodoi{10.3847/2041-8213/ab3d44}

\bibitem[{{Smith}(2014)}]{2014ARA&A..52..487S}
{Smith}, N. 2014, \araa, 52, 487, \dodoi{10.1146/annurev-astro-081913-040025}

\bibitem[{{Sugerman} {et~al.}(2005){Sugerman}, {Crotts}, {Kunkel}, {Heathcote},
  \& {Lawrence}}]{2005ApJ...627..888S}
{Sugerman}, B. E.~K., {Crotts}, A. P.~S., {Kunkel}, W.~E., {Heathcote}, S.~R.,
  \& {Lawrence}, S.~S. 2005, \apj, 627, 888, \dodoi{10.1086/430396}

\bibitem[{Suwa(2017)}]{Suwa2017lensing}
Suwa, Y. 2017, Monthly Notices of the Royal Astronomical Society, 474, 2612,
  \dodoi{10.1093/mnras/stx2953}

\bibitem[{{Suzuki} \& {Maeda}(2021)}]{suzuki2021}
{Suzuki}, A., \& {Maeda}, K. 2021, \apj, 908, 217,
  \dodoi{10.3847/1538-4357/abd54c}

\bibitem[{Suzuki {et~al.}(2016)Suzuki, Maeda, \& Shigeyama}]{suzuki2016}
Suzuki, A., Maeda, K., \& Shigeyama, T. 2016, Astrophysical Journal, 825,
  \dodoi{10.3847/0004-637X/825/2/92}

\bibitem[{Taylor(1950)}]{1950Taylor}
Taylor, G.~I. 1950, Proceedings of the Royal Society of London. Series A.
  Mathematical and Physical Sciences, 201, 192, \dodoi{10.1098/rspa.1950.0052}

\bibitem[{Thomson(1871)}]{1871Kelvin}
Thomson, W. 1871, The London, Edinburgh, and Dublin Philosophical Magazine and
  Journal of Science, 42, 362, \dodoi{10.1080/14786447108640585}

\bibitem[{Tolstov {et~al.}(2013)Tolstov, Blinnikov, \&
  Nadyozhin}]{10.1093/mnras/sts577}
Tolstov, A.~G., Blinnikov, S.~I., \& Nadyozhin, D.~K. 2013, Monthly Notices of
  the Royal Astronomical Society, 429, 3181, \dodoi{10.1093/mnras/sts577}

\bibitem[{Tsai {et~al.}(2023)Tsai, Chen, Whalen, Ou, \& Woods}]{Tsai_2023}
Tsai, S.-H., Chen, K.-J., Whalen, D., Ou, P.-S., \& Woods, T.~E. 2023, The
  Astrophysical Journal, 951, 84, \dodoi{10.3847/1538-4357/acd936}

\bibitem[{{Urushibata} {et~al.}(2018){Urushibata}, {Takahashi}, {Umeda}, \&
  {Yoshida}}]{2018MNRAS.473L.101U}
{Urushibata}, T., {Takahashi}, K., {Umeda}, H., \& {Yoshida}, T. 2018, \mnras,
  473, L101, \dodoi{10.1093/mnrasl/slx166}

\bibitem[{{Utrobin} {et~al.}(2021){Utrobin}, {Wongwathanarat}, {Janka},
  {M{\"u}ller}, {Ertl}, {Menon}, \& {Heger}}]{Utrobin2021ApJ}
{Utrobin}, V.~P., {Wongwathanarat}, A., {Janka}, H.~T., {et~al.} 2021, \apj,
  914, 4, \dodoi{10.3847/1538-4357/abf4c5}

\bibitem[{{van der Poel} {et~al.}(2015){van der Poel}, {Stevens}, \&
  {Lohse}}]{2015arXiv150101827V}
{van der Poel}, E.~P., {Stevens}, R. J.~A.~M., \& {Lohse}, D. 2015, arXiv
  e-prints, arXiv:1501.01827, \dodoi{10.48550/arXiv.1501.01827}

\bibitem[{von Helmholtz(1868)}]{1868helmholtz}
von Helmholtz, H. 1868, Monatsberichte der Königlichen Preussische Akademie
  der Wissenschaften zu Berlin, 23, 215.
\newblock \url{https://www.biodiversitylibrary.org/item/111036}

\bibitem[{{Walborn} {et~al.}(1987){Walborn}, {Lasker}, {Laidler}, \&
  {Chu}}]{1987ApJ...321L..41W}
{Walborn}, N.~R., {Lasker}, B.~M., {Laidler}, V.~G., \& {Chu}, Y.-H. 1987,
  \apjl, 321, L41, \dodoi{10.1086/185002}

\bibitem[{{Waxman} \& {Katz}(2017)}]{2017hsn..book..967W}
{Waxman}, E., \& {Katz}, B. 2017, in Handbook of Supernovae, ed. A.~W.
  {Alsabti} \& P.~{Murdin} (Springer, Cham), 967,
  \dodoi{10.1007/978-3-319-21846-5_33}

\bibitem[{{Weaver} {et~al.}(1978){Weaver}, {Zimmerman}, \&
  {Woosley}}]{1978ApJ...225.1021W}
{Weaver}, T.~A., {Zimmerman}, G.~B., \& {Woosley}, S.~E. 1978, \apj, 225, 1021,
  \dodoi{10.1086/156569}

\bibitem[{{West} {et~al.}(1987){West}, {Lauberts}, {Jorgensen}, \&
  {Schuster}}]{1987A&A...177L...1W}
{West}, R.~M., {Lauberts}, A., {Jorgensen}, H.~E., \& {Schuster}, H.~E. 1987,
  \aap, 177, L1

\bibitem[{{Wheeler} {et~al.}(1975){Wheeler}, {Lecar}, \& {McKee}}]{1975Wheeler}
{Wheeler}, J.~C., {Lecar}, M., \& {McKee}, C.~F. 1975, \apj, 200, 145,
  \dodoi{10.1086/153771}

\bibitem[{{Woosley} {et~al.}(2007){Woosley}, {Blinnikov}, \&
  {Heger}}]{2007Natur.450..390W}
{Woosley}, S.~E., {Blinnikov}, S., \& {Heger}, A. 2007, \nat, 450, 390,
  \dodoi{10.1038/nature06333}

\bibitem[{{Zhang} {et~al.}(2011){Zhang}, {Howell}, {Almgren}, {Burrows}, \&
  {Bell}}]{2011ApJS..196...20Z}
{Zhang}, W., {Howell}, L., {Almgren}, A., {Burrows}, A., \& {Bell}, J. 2011,
  \apjs, 196, 20, \dodoi{10.1088/0067-0049/196/2/20}

\bibitem[{{Zhang} {et~al.}(2013){Zhang}, {Howell}, {Almgren}, {Burrows},
  {Dolence}, \& {Bell}}]{2013ApJS..204....7Z}
{Zhang}, W., {Howell}, L., {Almgren}, A., {et~al.} 2013, \apjs, 204, 7,
  \dodoi{10.1088/0067-0049/204/1/7}

\end{thebibliography}
\bibliographystyle{aasjournal}



\end{document}